\documentclass[a4paper,fleqn]{cas-dc}
\usepackage{siunitx}

\usepackage{amsmath}
\usepackage[numbers]{natbib}
\setcitestyle{sort&compress}

\usepackage{appendix}
\usepackage{tabularx}
\usepackage{multirow}
\usepackage{array}
\usepackage{caption}
\usepackage{subcaption}
\usepackage{overpic}
\usepackage{threeparttable}
\usepackage{graphicx}

\usepackage{natbib}

\def\tsc#1{\csdef{#1}{\textsc{\lowercase{#1}}\xspace}}
\tsc{WGM}
\tsc{QE}

\begin{document}
\let\WriteBookmarks\relax
\def\floatpagepagefraction{1}
\def\textpagefraction{.001}
\let\printorcid\relax 

\shorttitle{}    

\shortauthors{Haipeng Li et al.}

\title[mode = title]{The Symbolic Partition of Chaotic Flows Based on Ordinal Patterns}  

\author[1]{Haipeng Li}
\author[1,2]{Yueheng Lan}
\cormark[1]

\address[1]{School of Science, Beijing University of Posts and Telecommunications, Beijing 100876, China} 
\address[2]{State Key Lab of Information Photonics and Optical Communications, Beijing University of Posts and Telecommunications, Beijing 100876, China}
\cortext[1]{Corresponding author}  

\begin{abstract}	
	As a crucial tool in the field of chaotic systems study, Symbolic dynamics prompts extensive research into various methods for symbolic partitioning. The limitations of the majority of these methods are usually heuristic and empirical for partitioning the multivariate chaotic state space. Fortunately, we successfully take KA method and obtain primary coarse symbolic boundary and refine the symbolic boundary via GKA method on chaotic map in our previous studies. However, this method fails when applied to continuous chaotic flows.
	The trajectories of complex continuous chaotic flows are complex, and while their mechanisms are quite different from those of chaotic maps, they are by no means completely distinct—after all, both are governed by the same fundamental laws of chaos
	In response to the aforementioned challenges, a modified approach should be developed to overcome the failure of the existing method for continuous chaotic flows.
In this study, we extend the Koopman-analysis-based symbolic partitioning approach to continuous chaotic flows, with a focus on constructing Poincaré maps. For the periodically driven Duffing oscillator system, the return map can be constructed by exploiting the periodicity. For general chaotic flows, Koopman analysis is employed to identify suitable Poincaré sections. In this work, we construct different candidate Poincaré sections based on ordinal patterns. By combining this ordinal-pattern-based Poincaré section construction method with the Koopman-analysis-based return-map approach, we achieve effective symbolic partitioning of continuous chaotic flows. Noise perturbation tests are also conducted, demonstrating the robustness of the proposed method. This ordinal-pattern-based analysis is applied to Rössler system, Lorenz system, Lü system and Chen system. 
This study, with the aid of ordinal patterns, successfully introduces an effective symbolic partition into continuous systems, achieving a faithful transfer of the method from maps to continuous flows.

\end{abstract}

\begin{keywords}
Symbolic dynamics \sep 
Partition boundary \sep 
Poincaré map \sep
Ordinal pattern\sep
Generalized coordinate
\end{keywords}

\maketitle

\section{Introduction\label{sec:intro}}

Symbolic dynamics was first proposed and theoretically established by Morse and Hedlund in 1938 \cite{1938symbol}. This theory encodes system evolution trajectories into infinite sequences composed of discrete symbols \cite{1995An,1988chaos}, thereby enabling a coarse-grained representation of chaotic attractors.

The core advantage of symbolic dynamics lies in the fact that, as long as the partition of the state space is appropriate, the symbolic sequences can faithfully reflect the dynamical behavior of the original system—periodic orbits correspond to periodic symbol sequences, chaotic orbits correspond to aperiodic symbol sequences, and the entropy rate of the system can be precisely computed from the growth rate of the symbolic sequences.

By discretizing the continuous states of the phase space into partitioned regions and transforming complex continuous trajectories into symbol sequences composed of finite alphabets, this approach shares a certain degree of similarity with recurrence plot analysis \cite{1987Recurrence,MARWAN2007237}, while retaining the core deterministic dynamical information essential for rigorous statistical analysis \cite{2017Elements,1985Ergodic}. It holds irreplaceable value in the field of chaos research. After decades of development, symbolic dynamics has given rise to several important research directions, including topological characteristic analysis \cite{2024Dynamics}, complexity analysis of symbolic sequences \cite{KARAMANOS1999}, and classification of unstable periodic orbits \cite{2009Complex,DONG20221}. More specifically, in addition to the aforementioned major research directions, there are also various interdisciplinary research areas that combine symbolic dynamics with other fields, such as dynamic pattern identification \cite{2004Symbolic,2007Symbolic,2007A}, synchronization identification \cite{2008Symbolic}, parameter reconstruction of chaotic systems \cite{X1995Symbol}, chaotic encryption communication \cite{Hayes1993CommunicatingWC,1994Experimental}, and gene network modeling \cite{2001Symbolic}.

On this basis, the ultimate theoretical ideal of symbolic dynamics is to construct a generating partition. A generating partition refers to a partitioning scheme of the state space such that there exists a one-to-one correspondence—namely, a topological conjugacy—between symbolic sequences and the orbits of the original system. This means that for every true orbit of the system, there corresponds one and only one symbolic sequence; conversely, for every admissible symbolic sequence, there exists a unique orbit in the state space that realizes it. Once a generating partition is obtained, the complete dynamical information of the original nonlinear system—including the distribution of all periodic orbits, the topological structure of the chaotic attractor, the topological entropy, and the characterization of chaos—is fully encoded in the dynamics of the symbolic sequences, thereby equivalently transforming the analysis of the continuous system into the study of a finite-state symbolic system. For this reason, the generating partition is regarded as the most fundamental and central component of symbolic dynamics, and also the critical link that determines the validity and reliability of subsequent analyses. The rationality of its design directly governs the extent to which the symbolic sequences preserve the original dynamical information, and it stands as the fundamental issue that bridges the theory of symbolic dynamics with practical applications of chaotic time series. This is precisely the core research direction on which this paper focuses.

However, the construction of generating partitions faces fundamental difficulties. Theoretically, rigorous generating partitions have only been proven to exist for hyperbolic systems \cite{1970Markov}. For non-hyperbolic systems, high-dimensional systems, or systems with unknown types, whether generating partitions exist and how they can be constructed remain open mathematical problems. Extensive explorations have been devoted to generating partitions, yet many difficulties still await further investigation. For one-dimensional maps, owing to their piecewise smooth properties, the kneading theory can not only locate the partition boundaries—that is, the positions of all critical points \cite{2002chaos}—but also determine all admissible symbolic sequences. Moreover, for coupled map lattices based on one-dimensional maps \cite{2006Symbolic}, important dynamical information can also be extracted by constructing generating partitions for each coupled component. This indicates that generating partitions are not limited to temporal chaos; they are equally applicable to the analysis of spatiotemporal chaos.

For multidimensional maps, generating partitions are empirical in nature, and consequently the methods employed are diverse.
Existing partitioning approaches can be broadly classified into two categories: those that rely on a detailed geometric analysis of the chaotic region\cite{1985Generating,1989On,1997Structure,2021Symbolic}, and those that emphasize the uniqueness of symbolic sequences in order to yield a valid partition\cite{1982Symbolic,1983Symbolic,2003Estimating,2005Statistically,2018Empirical}.

Further exploration reveals that the spectrum of Koopman operator\cite{1931Hamiltonian,2010Dynamic,Williams2014ADA} provides a heuristic link to the symbolic dynamics.
According to the above conjecture, Zhang et al.\cite{2022Phase} carried out symbolic partitions for chaotic maps based on properly constructed eigenfunctions of a finite-dimensional approximation matrix of Koopman operator. This exploration reveals a connection between the symbolic partition and the eigenfunctions of the Koopman operators.
Nevertheless, the approach is crude and primitive in that it has a limited applicability and can only be applied to a subset of relatively simple chaotic systems and the precision accuracy of critical point positions is not high in 2-D map even in the absence of noise.
The Koopman eigenfunctions of this method only provide a global identification, which is a main reason for this issue. The extreme points of the obtained eigenfunctions are quite a lot, some of which coincide with the pre-image and after-image points instead of the boundary points themselves. Due to hardly identifying boundary points from these, this method is suitable for symbolic partition some simple 1-D chaotic system but fails to precisely partition the state space of multidimensional system which has complex local structures and multiple folding points.
In our former study\cite{2025The}, we propose a novel approach, combineing Koopman with stretching and folding mechanism of chaotic generation, to construct the symbolic boundary of chaotic regions. We construct a finite-dimensional approximate matrix of Koopman operator and calculate Koopman left eigenfunctions and focus on the functions whose eigenvalue are approximately equal to zero and positive-negative oscillations occur internally. Then, we can localize the whole region into the oscillation subregion and take it as the coarse symbolic boundary subregion. 
Based on this phenomenon, we propose Koopman Analysis (KA) method and then take symbolic partition for chaotic map.
In addition, we can refine the coarse symbolic boundary via combination affine transformation and  KA method on the coarse symbolic boundary. We refer to the modified method as Generalized Koopman Analysis (GKA). Although achieving successful partitioning for chaotic maps is of considerable value, this alone is insufficient, as chaotic maps do not encompass all chaotic systems—continuous chaotic flows are equally important.

Continuous chaotic systems are prevalent in real-world applications. These systems evolve in a smooth manner and generally demand equal-step sampling. However, the amount of dynamical information encapsulated within a single step of evolution is exceedingly scarce, rendering the extraction of meaningful dynamical characteristics a challenging task.

 However, the trajectories of complex continuous chaotic flows are complex, and while their mechanisms are quite different from those of chaotic maps, they are by no means completely distinct—after all, both are governed by the same fundamental laws of chaos
In response to the aforementioned difficult, 
in this study, we extend the Koopman-analysis-based symbolic partitioning approach to continuous chaotic flows, with a focus on constructing return maps based on Poincaré sections—also referred to as Poincaré maps.

To effectively assign symbols to continuous choatic flows, a promising partitioning scheme is to construct generating partitions on their Poincaré sections.
Therefore, we focus on constructing valid return maps based on Poincaré section, also referred to as Poincaré maps in this study. In view of the favorable heterogeneity property of ordinal patterns \cite{2002Permutation,2002Permutation,2018Bandt,Amig2007Forbidden}, we employ this effective coarse-graining technique to construct the Poincaré sections in the present study.

In this paper, we propose an ordinal-pattern-based method for constructing candidate Poincaré sections. By screening out redundant sections, we obtain the most simplified Poincaré section and establish a discrete return map. Combined with the analytical method presented in our former article~\cite{2025The}, an effective symbolic partition for continuous chaos is achieved. The key step of this procedure is to obtain the optimal and most simplified Poincaré section. Specifically, all candidate Poincaré sections are subjected to screening, equivalence analysis, and complexity analysis, thereby further eliminating redundant sections. Meanwhile, this paper investigates the use of ordinal patterns for optimizing the structure of Poincaré sections, enabling finer symbolic partitioning boundaries. The proposed method is applied to several chaotic systems, including Rössler, Lorenz, Lü, and Chen, and its robustness is demonstrated through the introduction of noise.

The rest of the paper is divided into the following sections:
Section II firstly introduces the concept of ordinal patterns and proposes candidate Poincaré sections. Meanwhile, simplified single-scroll and double-scroll models are established to study the mechanism of OP-based effective Poincaré section construction.
 Then, by examining the equivalence and coupling relationships in the stepwise evolution from one candidate Poincaré section to another, redundant sections are removed, and the optimal minimal section for the return map is established. Finally, Finally, we introduces the tool of generalized coordinates, which facilitates the reduction of Poincaré sections to a reasonably low dimension, thereby enabling the application of GKA (Generalized Koopman analysis). 
 proposed method for various types of continuous chaotic systems.
At the specific application level, this chapter conducts detailed numerical experiments and analyses on the Rössler system as a typical representative of the single-scroll model.
For multi-scroll chaotic systems, this chapter selects the Lorenz system for multi-scroll analysis.
This chapter also examines more general chaotic systems that incorporate both coupling evolution and folding evolution mechanisms—namely, the Lü and Chen systems. Effective Poincaré sections and refined symbolic partition boundaries are obtained for all of these systems.
The paper is summarized in the Section V where we evaluate the proposed method and compare it with previous methods.

\section{Methodology }
	\label{sec:koopman}

\subsection{ Poincaré map} \label{mech}

\begin{figure*}[htbp]  
	\begin{minipage}{0.48\linewidth}
		\centerline{\includegraphics[width=8cm]{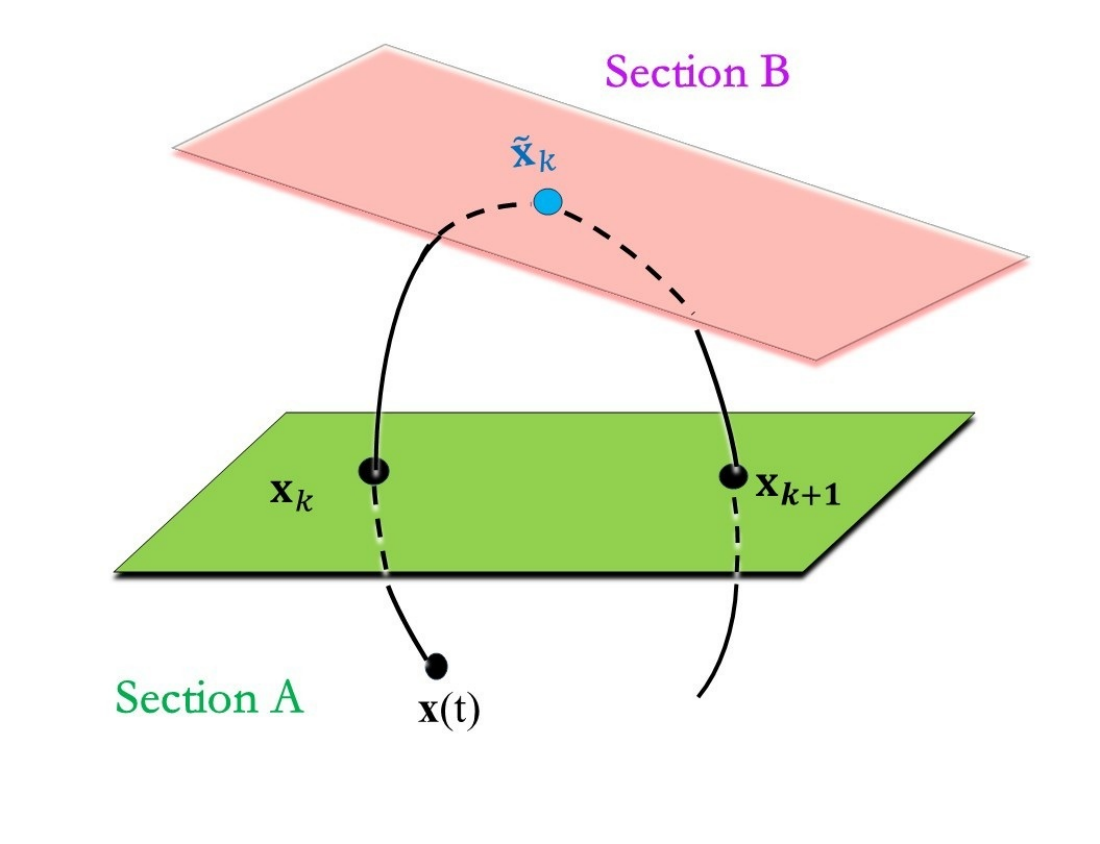}}
		\centerline{(a)}
	\end{minipage}
	\hfill
	\begin{minipage}{0.48\linewidth}
		\centerline{\includegraphics[width=8cm]{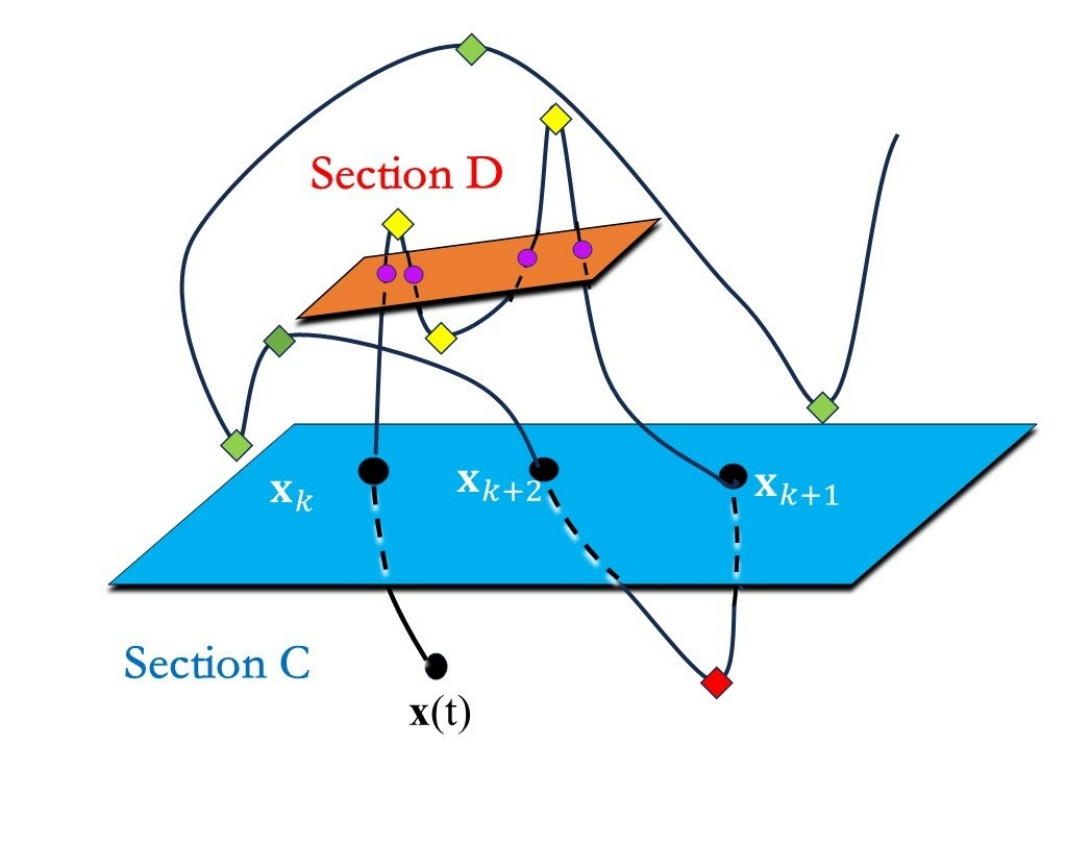}}
		\centerline{(b)}
	\end{minipage}
	\caption{Construction of Poincaré sections for continuous trajectory evolution: (a) Comparison between a valid Poincaré section (Section A) and an invalid Poincaré section of the first kind (Section B), where tangency points with the section exist; (b) An invalid Poincaré section of the second kind (Section C): important dynamical behaviors such as local oscillations occur during the nonlinear evolution of the trajectory without crossing the section.} 
	\label{fig-3-20} 
\end{figure*}

To better partitioning the choatic flow, the following auxiliary tool is adopted—namely, the Poincaré section.
A Poincaré section \cite{2004Differential} refers to a cross-section 
$\Sigma$ selected in a multidimensional phase space. By observing the intersections of the trajectory with this cross-section—referred to as Poincaré points—the continuous motion in the phase space is transformed into a mapping between a series of discrete points. This mapping is called the Poincaré map, also known as the return map.
For a general continuously evolving state variable $\mathbf{x}$, the following expression is given:
\begin{equation}
	\dot{\mathbf{x}} = \mathbf{f}(\mathbf{x}), \quad \mathbf{x} \in \mathbb{R}^d
\end{equation}
Let the initial evolution point be $\mathbf{x}(0)$
; then the evolved point at time 
t is given by 
$\mathbf{x}(t)$.
\begin{equation}
	\mathbf{x}(t) = \Phi_t(\mathbf{x}(0)) = \mathbf{x}(0) + \int_0^t \mathbf{f}(\mathbf{x}(\tau)) \, d\tau
\end{equation}
Upon constructing a Poincaré section for the system, a mapping based on forward returns to this section can be derived, which takes the following form:
\begin{equation}
	P(\mathbf{x}_k) = \mathbf{x}_{k+1} 
\end{equation}
Where $P: \Sigma \rightarrow \Sigma$
k serves as a discretization index associated with the Poincaré section. As illustrated in Fig. \ref{fig-3-20}(a), the green section denoted as Section A, which contains the intersection points 
, is a representative Poincaré section.

The correspondence between 
$\mathbf{x}_k$
and the point 
$\mathbf{x}(t)$
 on the original continuous flow is:

\begin{equation}
	\mathbf{x}_k = \mathbf{x}(t_k), \quad t_{k+1} = t_k + T(\mathbf{x}_k)
\end{equation}

Therefore, if the Poincaré section is properly constructed, the evolution of the original continuous flow can be simplified and represented in the form of a discrete mapping. The key to realizing this representation lies in determining whether the construction of the Poincaré section is appropriate.

A suitable Poincaré section must first ensure that it is not tangent to the trajectory evolution of the points; otherwise, the trajectory may merely graze the section without actually crossing it, leading to an ill-defined mapping or the appearance of singularities. Second, the section must preserve the core nonlinear dynamical properties of the system. Although ensuring non-tangency with the trajectories is a relatively easy condition to satisfy, constructing a discrete mapping that can fully retain the essential nonlinear dynamics of a complex chaotic continuous system on this basis presents significant difficulties. As shown in Fig. \ref{fig-3-20}(a), Section A is a valid Poincaré section, while the pink surface Section B is an invalid one. Taking the trajectory starting from 
$\mathbf{x}(t)$as an example, if the intersection point 
$\mathbf{x}_{k+1}$
with Section A is regarded as the departure point, then for a valid section, there must necessarily be a return point 
$\mathbf{x}_{k+1}$
to this section. If every point on the trajectory has a one-to-one correspondence between departure and return, the section can be considered valid. However, for the same trajectory starting from 
$\mathbf{x}(t)$, its intersection with Section B consists of only a single point 
$\tilde{\mathbf{x}}_{k}$
, which is a tangency point. With only one point, it can neither represent departure nor return, and thus cannot fully reflect the system's evolution. Therefore, this section cannot serve as an effective discretizing section for the continuous system.

In addition, for complex chaotic systems, the trajectories are intricate. Generally speaking, constructing an 

$(n-1)$dimensional section in one direction can ensure that there are no tangency points in that direction. However, the section may only intercept a local region; due to the complexity of chaotic trajectories, which rotate in different directions, the returns of the sampling points become overly coarse. As shown in Fig. \ref{fig-3-20}(b), on the blue Poincaré section Section C, 
$x_{k+1}$
to 
$x_{k+2}$
exhibit a clear pair of entry and return intersection points, while the trajectory segments from 
$x_{k}$
to $x_{k+1}$
and from 
$x_{k+2}$
to the next Poincaré point both lose dynamical information. Between the three yellow points on the segment from 
$x_{k}$
to $x_{k+1}$
, an additional section needs to be constructed to capture the entry and return processes in the local region. As shown by the orange section Section D in the figure, four new Poincaré intersection points are obtained, presented as purple dots, indicating that there can be three returns between 
$x_{k}$
and 
$x_{k+1}$
, rather than just one. Clearly, using only Section C as the section for this trajectory segment is inadequate. Similarly, along the trajectory from 
$x_{k+3}$
to the next Poincaré point, at least one additional section analogous to Section D can be established based on the positions of the green points. Therefore, for chaotic systems, constructing an effective Poincaré section is rather difficult.

Based on the above analysis, the core principles for constructing a Poincaré section are summarized as follows:

First, transverse intersection. The section must intersect the system's dynamical flow transversely at every point, with no trajectory being parallel to or tangent to the section. This ensures that any trajectory can steadily and repeatedly cross the section to yield continuous sampling points. If any tangency exists, it constitutes an invalid Poincaré section of the first kind.

Second, unidirectional crossing. By constraining the crossing direction of trajectories, repeated sampling and point-set confusion caused by back-and-forth crossings of the same trajectory are avoided, ensuring that the evolution direction of the Poincaré map is consistent with that of the original continuous system.

Third, ensuring that all departure and return trajectories have a corresponding pair of representative points on the section. Unidirectional crossing serves to characterize the departure and return trajectories near the section, thereby preserving essential dynamical features while discretizing and simplifying the continuous system.

The Poincaré section of chaotic motion exhibits a fractal structure of complex entirety. Moreover, the degree of complexity varies significantly across different local regions, with layouts and structures of different scales nested in a layered manner—this is a typical chaotic attractor structure corresponding to a discrete mapping.

For periodically driven continuous chaotic systems, such as the Duffing oscillator chaotic system \cite{1979Randomly}, the time 
t can be treated as an additional dimension. Since the evolution of the trajectory in the time component is periodic, the one-dimensional periodic motion in the 
t-dimension can be reduced by fixing the rotation period to a zero-dimensional periodic point, thereby constructing an effective Poincaré section. This is because the system's trajectory evolution in the time component is periodic, allowing truncation and dimensionality reduction at any arbitrary point to obtain a two-dimensional return map. A return map can then be constructed on this section to achieve the generating partition of the chaotic return map on the section. Finally, the symbols assigned to different points are extended to the trajectories corresponding to each point's one-step return to the section, accomplishing effective symbolization of the continuous chaotic system.

However, for general continuous chaotic systems, the components are subject to complex coupling and nonlinear evolution processes. Under such complex circumstances, constructing an effective Poincaré section is a challenging endeavor. Only by constructing an effective Poincaré section can the generating partition of the return map built upon it be used to achieve correct symbolization of the original continuous chaotic system.

	\subsection{Ordinal-Pattern-Based Construction of Effective Poincaré Sections}
	
Here we presents the role of Poincaré sections in continuous chaotic systems. However, how to construct effective Poincaré sections such that the derived Poincaré map can efficiently preserve the nonlinear characteristics of the original system remains an important research direction.
To address this challenge, this chapter attempts to construct effective Poincaré sections using the ordinal pattern (OP), a coarse-grained tool that reflects dynamic variations.
Ordinal pattern analysis is a symbolization process based on the relative positions of adjacent evolutionary points. For a time series $$x(T) = (x_1, x_2, \ldots, x_T)$$ evolved over period \(T\), a sliding window with interval $\tau$ can be defined as:
\begin{equation}
	w_t = \left( x_t, x_{t+\tau}, \ldots, x_{t+(D-2)\tau}, x_{t+(D-1)\tau} \right)
	\label{eq:one_logistic_eq}
\end{equation}
Each sampling point within the window is assigned a distinct symbol by sorting according to its relative magnitude in ascending or descending order, thereby yielding the ordinal pattern.
For each window $w_t$ at a given time $t$, the ordinal pattern consists of the permutation $$\pi_t = (r_1, r_2, \ldots, r_D)$$, which satisfies the following relation:
\begin{equation}
	x_{t+r_1-1} \le x_{t+r_2-1} \le \cdots \le x_{t+r_{D-1}-1} \le x_{t+r_D-1}
	\label{eq:order1}
\end{equation}
Continuous sorting operations eventually generate a sequence of ordinal symbols $\pi_i$, namely the symbolic sequence of ordinal patterns.
Specifically, Fig. \ref{fig-5-1}(b) illustrates the construction of a typical ordinal pattern. The sliding windows $w_1, w_2, \ldots$ are set with a dimension of $D=3$, and the permutation $\pi$ correspondingly contains three elements. The figure presents six distinct types of $\pi$, such as $132$ and $321$. Accordingly, the $T-D+1$ sliding windows constructed from \(T\) evolutionary points can be categorized by six types of $\pi$. In practical scenarios, not all six patterns appear simultaneously, and the absent ones are defined as forbidden patterns.
To intuitively and effectively validate the rationality of the OP method, this paper constructs a generalized simplified continuous chaotic model for verification. The chaotic model is processed via OP coarse-graining, and the correlation between OP and Poincaré sections is analyzed to preliminarily demonstrate the effectiveness of OP-based construction. Two simplified chaotic models are adopted in this study, as shown in Fig. \ref{fig-5-1}(a) and Fig. \ref{fig-5-1}(c). The model in Fig. \ref{fig-5-1}(a) describes a single-scroll chaotic system, while the model in Fig. \ref{fig-5-1}(c) characterizes a more general and complex multi-scroll chaotic system. In this section, the construction of Poincaré sections via the OP method is first explored based on the single-scroll chaotic model, and the construction mechanism is further extended to the multi-scroll chaotic model to derive a more general and effective principle for Poincaré section construction.
	
	\subsubsection{Poincaré Section of Single-Scroll Chaos}

	\begin{figure*}[htbp]  
		\begin{minipage}{0.48\linewidth}
			\centerline{\includegraphics[width=8cm]{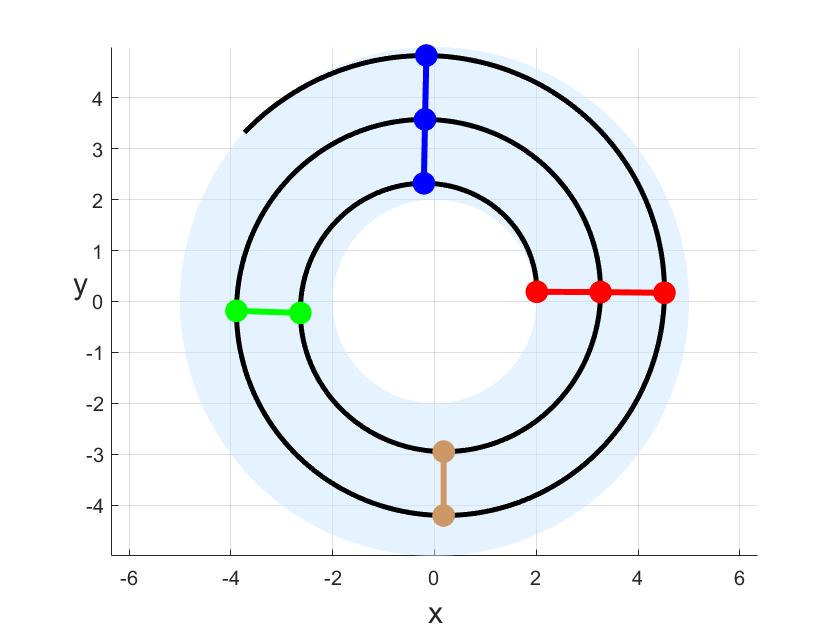}}
			\centerline{(a)}
		\end{minipage}
		\hfill
		\begin{minipage}{0.48\linewidth}
			\centerline{\includegraphics[width=8cm]{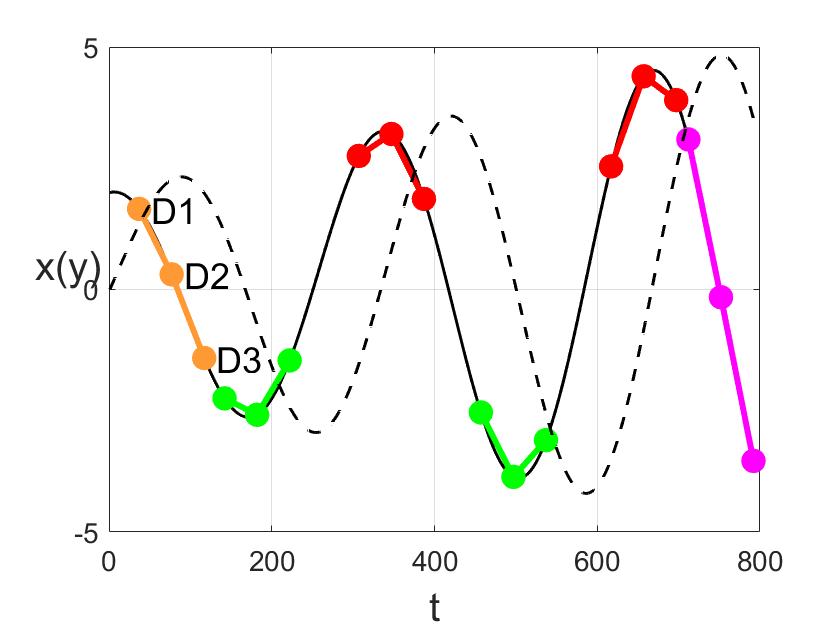}}
			\centerline{(b)}
		\end{minipage}
		\vfill
		\begin{minipage}{0.48\linewidth}
			\centerline{\includegraphics[width=8cm]{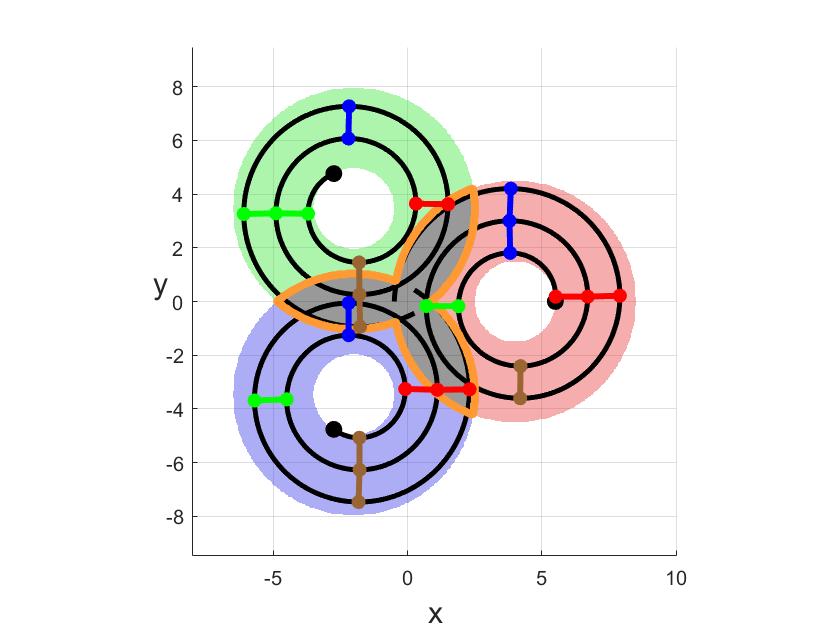}}
			\centerline{(c)}
		\end{minipage}
		\hfill
		\begin{minipage}{0.48\linewidth}
			\centerline{\includegraphics[width=8cm]{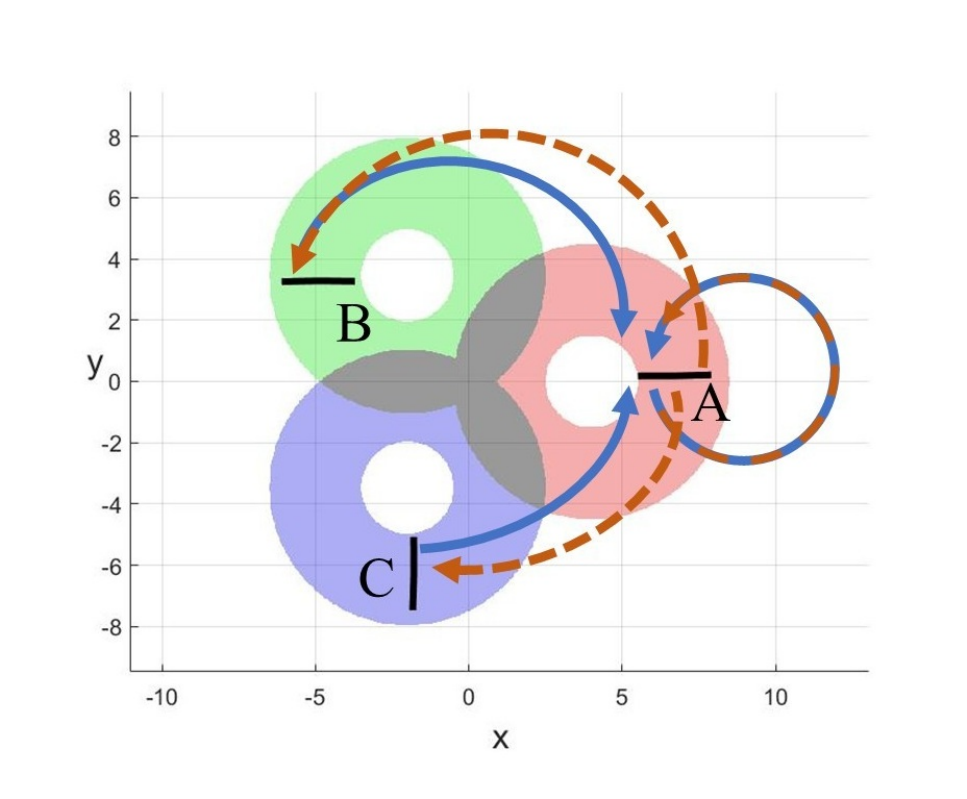}}
			\centerline{(d)}
		\end{minipage}
		\caption{Partition Mechanism of Continuous Chaos via Ordered Patterns (OP), with Local Spirals Simulating the Scrolls of Continuous Chaos and Nonlinear Processes Demonstrated through Coupling and Dispersion. (a) Spiral trajectory of single-scroll chaos; (b) Ordered pattern presentation in the x-direction; (c) Spiral trajectory of coupled multiple scrolls; (d) Schematic diagram of coupling between scrolls.
			\label{fig-5-1} }
	\end{figure*}
	
As shown in Fig. \ref{fig-5-1}(a), to intuitively characterize chaotic evolution, this paper constructs a spiral evolution trajectory to simulate the linear motion trajectory of continuous chaos. The two-dimensional spiral evolution trajectory possesses two projection directions, and ordinal patterns can be constructed in each projection direction. In this paper, time series are respectively established based on the evolution values in the two projection directions, as illustrated in Fig. \ref{fig-5-1}(b). The solid black line represents the time series in the $x$-direction, while the dashed black line denotes the time series in the $y$-direction. The OP analysis is firstly performed on the $x$-direction time series.
When constructing ordinal patterns containing extreme points using sliding windows $w$ of length $D$, it is required that $D\geq3$, which means that at least three evolutionary state points are needed to determine the relative positional relationships. Therefore, this paper processes the sequence with $$D=3$$ and sets the sampling interval of adjacent points to $\tau=50$. The three sampling points are marked as D1, D2, and D3 successively, as indicated by the orange labels in Fig. \ref{fig-5-1}(b). For continuously evolving chaotic time series obtained by equal-interval sampling, the cases of pairwise equality among the three points ($D1=D2$, $D2=D3$, and $D1=D3$) constitute the boundary of the OP region with an extremely low probability of occurrence, which are not considered in this paper. Excluding these special cases, there are six distinct combinations in total: $D1<D2<D3$, $D2<D1<D3$, $D3<D1<D2$, $D2<D3<D1$, $D1<D3<D2$, and $D3<D2<D1$. All six types of OPs exist in the $x$-direction sequence. This paper selects one typical case of each OP and presents them sequentially from left to right in Fig. \ref{fig-5-1}(b).
In the figure, the patterns $D1<D3<D2$ and $D3<D1<D2$ are marked in red. Both OPs indicate that D2 takes the maximum value, and the relative positional relationship between D1 and D3 exerts little influence on the short-term dynamics. Consequently, these two patterns share essentially identical dynamic characteristics and are merged into a unified OP: $D1<D2 \land D3<D2$. Similarly, the patterns $D2<D1<D3$ and $D2<D3<D1$ both correspond to the minimum value of D2 and are integrated into another unified OP: $D2<D1 \land D2<D3$, which is represented by green broken lines in the figure. These two merged OPs feature intermediate turning points, indicating linear rotation or nonlinear bending during evolution, and are collectively defined as non-monotonic OPs. In contrast, $D1<D2<D3$ and $D3<D2<D1$ are monotonic evolution OPs corresponding to linear growth and linear decay, which are colored blue and magenta in the figure, respectively.
Monotonic OPs represent locally linear evolutionary processes. If all projection directions exhibit monotonic variation, the corresponding local evolutionary segment is entirely linear. By contrast, non-monotonic OPs generally indicate the occurrence of nonlinear evolution processes, though they may also result from linear rotation. This paper retains the two types of nonlinear OPs (red and green) for further analysis. Given the smooth evolution of the spiral trajectory sequence, the OP sampling interval can be reduced to the interval between adjacent evolutionary points, i.e., $\tau=1$. For the OP $D1<D2 \land D3<D2$ in the $x$-direction, D2 corresponds to the local maximum within the window. This paper takes D2 as the representative point of this OP and marks all local maximum representative points in the $x$-direction in brown. Meanwhile, D2 is selected as the representative point for the OP with local minimum values satisfying $D1>D2 \land D3>D2$, and these points are highlighted in red in Fig. \ref{fig-5-1}(a). It is observed that these discrete representative points can be connected to form a sparse straight line.
This straight line is derived from a single spiral trajectory. In practical chaotic systems, apart from the quasi-periodic continuous evolution similar to spiral motion, chaotic trajectories undergo stretching, bending and overlapping nonlinear deformations, which increase the structural complexity of the straight line formed in real chaotic systems. Although chaotic trajectories continuously return and overlap, they never intersect with each other. As evolution proceeds, the line density gradually increases until the interior and adjacent outer regions of the original sparse straight line are almost fully filled.
For spiral trajectories, the evolutionary characteristics in the $y$-direction are basically consistent with those in the $x$-direction. The dashed line in Fig. \ref{fig-5-1}(b) depicts the $y$-direction time series, which shares a similar waveform with the $x$-direction series but differs in phase offset and oscillation amplitude. The phase offset is a typical feature of circular motion, while the amplitude discrepancy originates from the varying radius of the spiral trajectory. Similarly, two types of representative points corresponding to non-monotonic OPs are extracted from the $y$-direction sequence, forming two additional sparse linear regions marked in cyan and green in Fig. \ref{fig-5-1}(a). In total, four line segments can be obtained by integrating the features of the two projection directions. Each segment reduces the dimension of the original two-dimensional annular region by one, which satisfies the basic requirements for constructing Poincaré sections.
The four line segments contain redundant information, which necessitates simplification and redundant region elimination. In the absence of nonlinear stretching and folding, each line segment can be overlapped with the line segment of the next evolutionary stage via affine transformation. Since affine transformation does not involve nonlinear deformation, redundant line segments can be removed for simplification, which is defined as an equivalence process. In this simplified chaotic model, the four linear regions are pairwise equivalent before and after evolution.
However, practical chaotic systems inevitably generate stretching and folding nonlinear effects, leading to nonlinear evolution of the line segments. In this case, the post-evolutionary regions cannot be obtained by affine transformation of the pre-evolutionary regions. Nevertheless, the evolution still follows a one-to-one mapping from one line segment to another without divergence into multiple segments. Such one-to-one evolutionary correspondence of global Poincaré sub-regions, even accompanied by nonlinear processes, can still be regarded as an effective equivalence process as long as strict one-to-one correspondence between cross-sections is guaranteed.
	
	\subsubsection{Poincaré Section of Multi-Scroll Chaos}
	
However, for multi-scroll chaotic systems, the nonlinear evolutionary processes involve coupling effects, which are far more complex than the internal stretching and folding of a single scroll. Fig. \ref{fig-5-1}(c) presents the chaotic evolution of three coupled scrolls, which are distinguished by different colors in the figure. Each scroll contains self-circulating spiral trajectories depicted by black curves. It is evident that coupling interactions occur among the three scrolls within the black central region enclosed by the orange boundary. The local self-circulating regions without coupling can still be simplified via linear equivalence and generalized equivalence principles, consistent with the simplification strategy adopted for the single-scroll system in Fig. \ref{fig-5-1}(a).
Fig. \ref{fig-5-1}(c) illustrates the sub-regional straight lines composed of all non-monotonic OP representative points. After simplification, three linear regions are obtained, as shown in Fig. \ref{fig-5-1}(d). These three mutually coupled lines are denoted as A, B, and C. During evolution, partial trajectories of the three regions converge into a single common region, or partial trajectories of two regions converge into another individual region; such behavior is defined as a coupling process. As demonstrated in Fig. \ref{fig-5-1}(d), the three solid blue arrows represent the three-region coupling process, in which partial trajectories from regions A, B, and C converge into region A. The overlapping of sub-regions during coupling can be regarded as folding in a generalized sense. Moreover, different local trajectories originating from a single region often participate in distinct coupling processes and eventually evolve into different target regions. Such trajectory divergence within an individual region corresponds to stretching in a generalized sense. Similarly, the three orange dashed arrows in Fig. \ref{fig-5-1}(d) indicate three different evolutionary trajectories derived from region A, which separately evolve back to region A itself, region B, and region C.
Regions B and C exhibit similar divergence behaviors and also serve as target regions for multi-region coupled evolution. The divergence and coupling evolution among multiple scroll sub-regions essentially reflect generalized stretching and folding mechanisms, which conform to the fundamental characteristics of chaotic evolution. For the Poincaré sections of each scroll, the existence of divergence and coupling evolution breaks the one-to-one equivalent evolutionary correspondence, enabling one section to evolve into multiple sections. Therefore, further simplification of these sections is no longer feasible. Such one-to-many and many-to-one evolutionary behaviors are defined as non-equivalence processes.
It is found that continuous chaotic systems differ from discrete chaotic mappings in that they generally involve multi-region coupling in addition to inherent folding behaviors. In practical complex chaotic systems, internal scroll stretching-folding and multi-scroll coupling coexist simultaneously. The OP-based generalized Poincaré section construction can effectively characterize the return mapping of continuous chaos, identify the symbolic partition boundaries based on the generalized Poincaré sections, and realize reasonable symbolic assignment for continuous chaotic systems.
To further intuitively elaborate the construction principle of OP-based Poincaré sections and the subsequent symbolic partition mechanism, this chapter verifies the proposed method via typical practical chaotic systems. The Rössler system is adopted to illustrate the symbolization process of single-scroll chaos, and the Lorenz system is used for multi-scroll chaos. Furthermore, the proposed OP Poincaré section construction method is applied to process complex evolutionary behaviors of the Lu system, Chen system, and high-dimensional hyperchaotic systems, which verifies the excellent generalization capability of the proposed method.


	\section{Result and Discussion}
	\label{result00}
	
	In this section, we will apply the method to achieve symbolic partition in several typical examples in chaotic series originating from different choatic flows. The following examples are categorized into univariate and multivariate series. We first investigates single-scroll chaos, and then extends the analysis to multi-scroll chaos with different levels of complexity.
	Then several typical examples of complex chaotic flow are partitioned by necesary  OP pretreatment, generalized coordinate and the GKA method.

	\subsection{Rössler system}
	
To further intuitively explore the rationality and interpretability of Poincaré section construction and symbolic partitioning schemes, this study selects single-scroll chaotic systems with prominent chaotic characteristics and simple topological structures as the research prototype for mechanism analysis.
The Rössler system, proposed by the German scholar Otto E. Rössler in 1976 \cite{1976An}, is one of the most classical single-scroll continuous chaotic systems in chaotic dynamics. Inspired by the Lorenz system, Rössler devoted to exploring the simplest nonlinear flow that can generate chaotic behaviors and finally established this landmark dynamical system. The most prominent advantage of the Rössler system is that it contains only a single nonlinear term but can produce abundant and complex chaotic dynamical responses, rendering it an ideal minimal model for fundamental research on chaos theory.
	
	\subsubsection{Initial Poincaré Section Construction Using the Ordinal Pattern (OP) Method}
	
Based on this ideal simplest chaotic model, this paper will reasonably and thoroughly introduce the Poincaré section construction method of the OP method from abstraction to concreteness, and further illustrate the symbolic partitioning mechanism for specific chaotic systems.

The evolution equations of the Rössler chaos are as follows:
	
	\begin{equation}
		\begin{cases}
			\dot{x} = -y - z \\
			\dot{y} = x + a y \\
			\dot{z} = b + z(x - c)
		\end{cases}
		\label{eq:5_1}
	\end{equation}
	Here, $(a, b, c)$ denote the system parameters. In this paper, classic chaotic attractors are generated by setting $a = 0.2, b = 0.2, c = 5.7$.

\begin{figure*}{*}[htbp] 
		
		\begin{minipage}{0.48\linewidth}
			\centerline{\includegraphics[width=8cm]{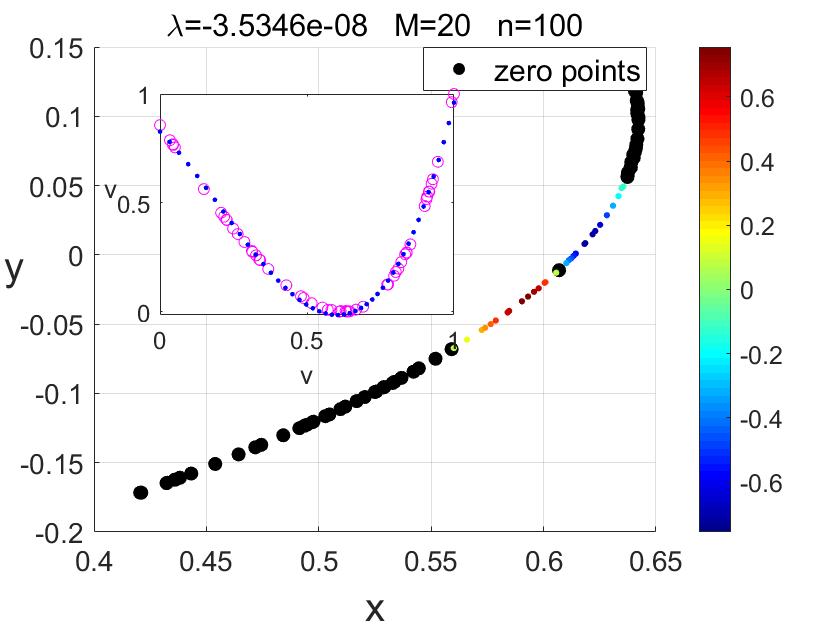}}
			\centerline{(a)}
		\end{minipage}
		\hfill
		\begin{minipage}{0.48\linewidth}
			\centerline{\includegraphics[width=8cm]{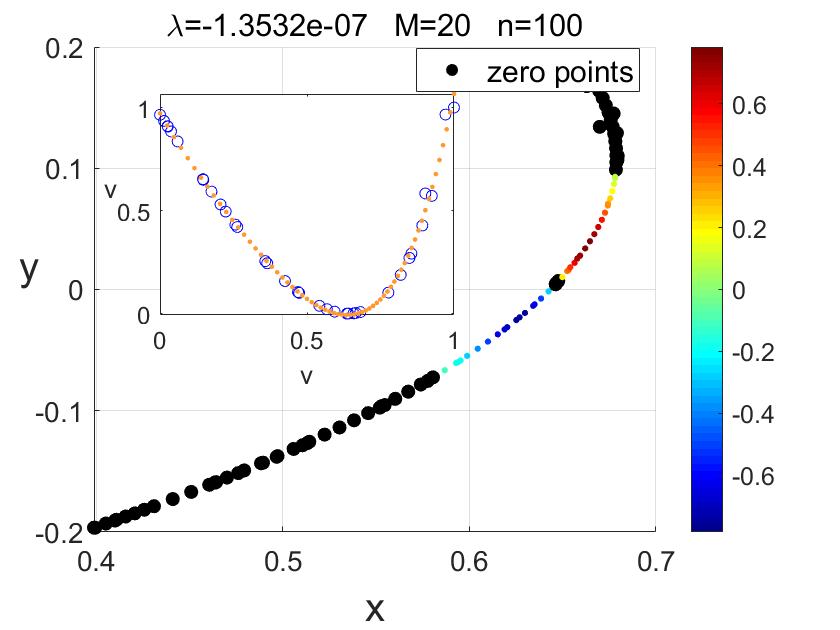}}
			\centerline{(b)}
		\end{minipage}
		\vfill
		\begin{minipage}{0.48\linewidth}
			\centerline{\includegraphics[width=8cm]{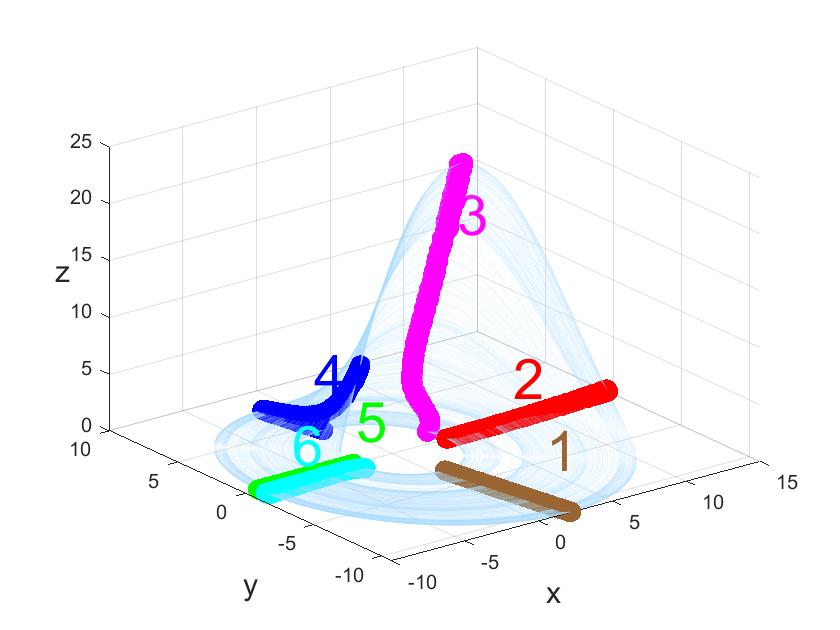}}
			\centerline{(c)}
		\end{minipage}
		\hfill
		\begin{minipage}{0.48\linewidth}
			\centerline{\includegraphics[width=8cm]{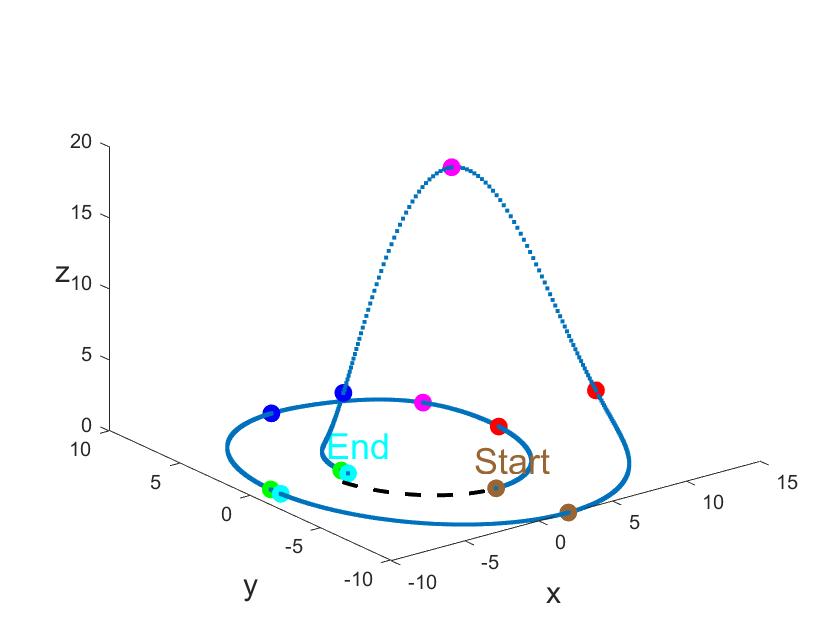}}
			\centerline{(d)}
		\end{minipage}
		
		\caption{Construction  of generalized Poincaré sections and subregion classification. (a)-(b) Construction of ordinal patterns. The time interval is set as $\tau=0.01$, and two non-monotonic OPs in the $x$-direction and $z$-direction within the time span $t\in[0,1]$ are selected and presented. (c) The generalized Poincaré section composed of six OP based constructed subregions. (d) The initial evolution trajectory, which indicates the evolutionary direction of the subregions of the generalized Poincaré section.} 
		\label{fig-5-2} 
	\end{figure*}

In this study, stationary time series are extracted and uniformly sampled at equal intervals with a sampling step of $0.0$. The first $1000$ transient points are discarded, and a total of $300000$ sampling points are selected to construct the chaotic attractor. Timing starts from the $1001$-th point, such that $t\in[0,3000]$. For three-dimensional chaotic systems, nonlinear OP regions corresponding to all three dimensions must be constructed. For continuous chaotic systems, the OP method adopts continuous non-overlapping sampling with a fixed time interval $\tau=0.01$. Only three sampling points are required for OP calculation to extract non-monotonic OPs, which are denoted as D1, D2, and D3 in sequence. The middle sampling point D2 is taken as the representative point. The OP regions corresponding to local maxima and minima are marked separately, which always correspond to distinct local positions during continuous rotational evolution.
Fig. \ref{fig-5-2}(a) presents two typical OPs in the $x$-direction within the time interval $t\in[0,1]$. Although rotational evolution induces stretching and compression and thus varying oscillation amplitudes, the two OPs are clearly distributed on opposite sides. Similarly, the two nonlinear OPs obtained in the $y$-direction exhibit a phase offset relative to the $x$-direction counterparts, which is consistent with the rotational characteristics of single-scroll spiral trajectories. Fig. \ref{fig-5-2}(b) shows two OPs in the $z$-direction for $t\in[0,1]$. Despite minor oscillations, the dominant dynamic feature lies in the stretching behavior of the magenta sub-region along the $z$-axis. In contrast, the other OP region remains nearly unchanged, indicating that this local structure is mainly governed by the stretching process within the stretching-folding mechanism of chaotic evolution.
By integrating all valid OPs extracted from the $x$, $y$, and $z$ directions, a generalized Poincaré section consisting of six OP sub-regions is obtained, as illustrated in Fig. \ref{fig-5-2}(c). Similar to the generalized Poincaré section of single-scroll chaos, the proposed multi-dimensional section can be simplified according to the inherent evolutionary rules. First, the overall evolutionary direction is determined by tracing system trajectories. The trajectory is evolved from the initial point until each sub-region is traversed at least twice, as shown in Fig. \ref{fig-5-2}(d). The dashed line represents the trajectory from the initial point to the first sub-region, followed by successive passages through all sub-regions.
It is observed that each sub-region is exactly traversed twice, demonstrating that the initial point of each sub-region satisfies the generalized rotation rule with a consistent evolutionary direction from the black region to the red region. If all points within other sub-regions follow the same rotational pattern illustrated in Fig. \ref{fig-5-2}(d), the entire section undergoes generalized rotation with a unified evolutionary orientation. The first sub-region is defined as Region 1, and the remaining sub-regions are numbered sequentially as 2, 3, 4, 5, and 6 along the evolutionary direction, as marked in Fig. \ref{fig-5-2}(c). The overall evolutionary order follows $1\rightarrow2\rightarrow3\rightarrow4\rightarrow5\rightarrow6\rightarrow1\rightarrow\cdots$, which is determined based on the initial trajectory characteristics in Fig. \ref{fig-5-2}(d).
Ideally, one-to-one correspondence among OP points implies identical point counts across all sub-regions. However, the actual number of sampling points distributed in the six sub-regions is 513, 513, 512, 512, 512, and 512, respectively. Regions 1 and 2 contain one extra point compared with the other four sub-regions. This discrepancy originates from insufficient evolution of the tail trajectory. Specifically, the last point of Region 6 further evolves and passes through Regions 1 and 2, resulting in the extra sampling points in these two regions. After eliminating the incompletely evolved tail segment, all sub-regions possess identical point quantities, ensuring strict one-to-one correspondence of evolutionary points across the entire Poincaré section.

	According to the above analysis, each section satisfies a one-to-one evolutionary correspondence, which is consistent with the OP section evolution mechanism of the simplified single-scroll model. The evolution between adjacent sections belongs to an equivalence process. Through equivalence simplification, redundant sections can be eliminated to obtain a unique effective Poincaré section.
	In this paper, the multiple sections prior to equivalence judgment are defined as ordinal pattern-based Candidate Poincaré Sections (CPS).
	Different from the simplified model, the practical chaotic system considers the complex stretching and folding mechanism of chaos, where local regions exhibit distinct magnitudes of stretching and folding. This leads to significant differences in structural complexity among different local sections. Instead of simple straight line segments, the practical sections are thick curves with varying degrees of bending. Therefore, the screening of redundant sections and the reservation of a unique effective section must take structural complexity into account. Specifically, structurally complex sections are regarded as redundant, while structurally simple and equivalent sections are retained as valid ones.
	Among all structurally simple valid sections, the one with the lowest complexity is selected as the final Poincaré section, which conforms to the simplification principle and facilitates subsequent symbolic partition analysis. The critical indicators for effective section screening are presented in the following complexity analysis section.
	
	\subsubsection{complexity analysis on CPS}
	
	Since the fractal dimension of the system in this case is 2.01, the local regions of trajectory evolution are approximately two-dimensional. As Candidate Poincaré Sections (CPSs) are constructed based on the extreme points of local dynamic regions, each CPS has one fewer dimension than its corresponding local region, rendering the obtained sections approximately one-dimensional. Nevertheless, intense stretching and folding during chaotic evolution leads to heterogeneous structural complexity across different local regions, which further results in distinct complexity levels for the CPSs extracted from individual regions. This paper conducts a complexity analysis on all CPSs to eliminate subregions with excessive folding degrees.
	As low-dimensional structures embedded in high-dimensional phase space, CPSs require dimension reduction into valid low-dimensional manifolds for reliable structural complexity evaluation. Two typical dimension reduction algorithms are adopted in this study: Isometric Mapping (Isomap) \cite{2000AX} and Principal Component Analysis (PCA) \cite{1901LIII,Hotelling1933Analysis}.
	In this paper, Isomap is firstly applied to all CPSs for complexity quantification. For local regions undergoing severe stretching and folding, structural bending generally occurs along the third dimension, making such local regions approximately three-dimensional and the corresponding CPSs two-dimensional. Intensive folding dramatically increases structural complexity, and a typical characteristic is the elevation of the effective embedding dimension. Therefore, such highly complex CPSs need to be excluded. The Isomap algorithm can project structurally complex datasets onto valid low-dimensional spaces, enabling the screening of complex CPSs by judging their effective embedding dimensions.

	\begin{figure*}[htbp]  
		\begin{minipage}{0.48\linewidth}
			\centerline{\includegraphics[width=8cm]{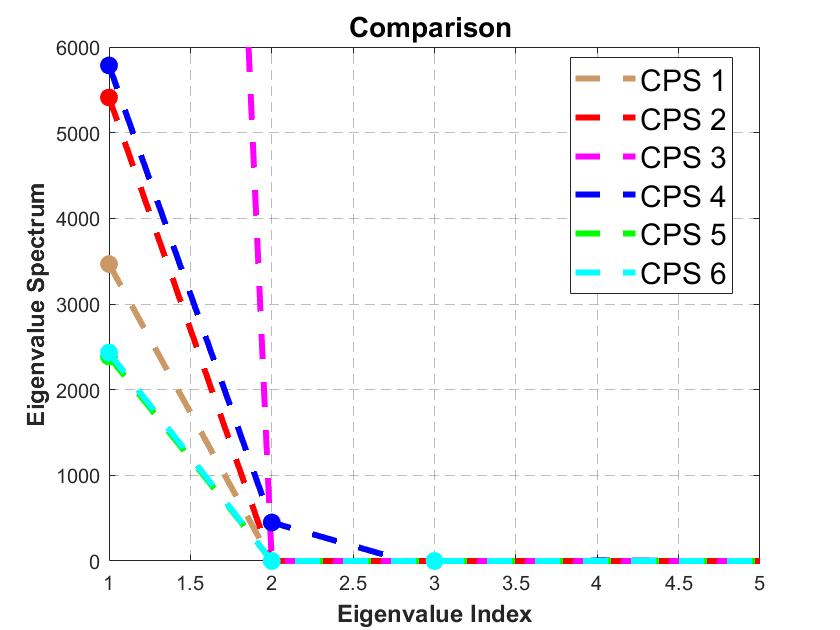}}
			\centerline{(a)}
		\end{minipage}
		\hfill
		\begin{minipage}{0.48\linewidth}
			\centerline{\includegraphics[width=8cm]{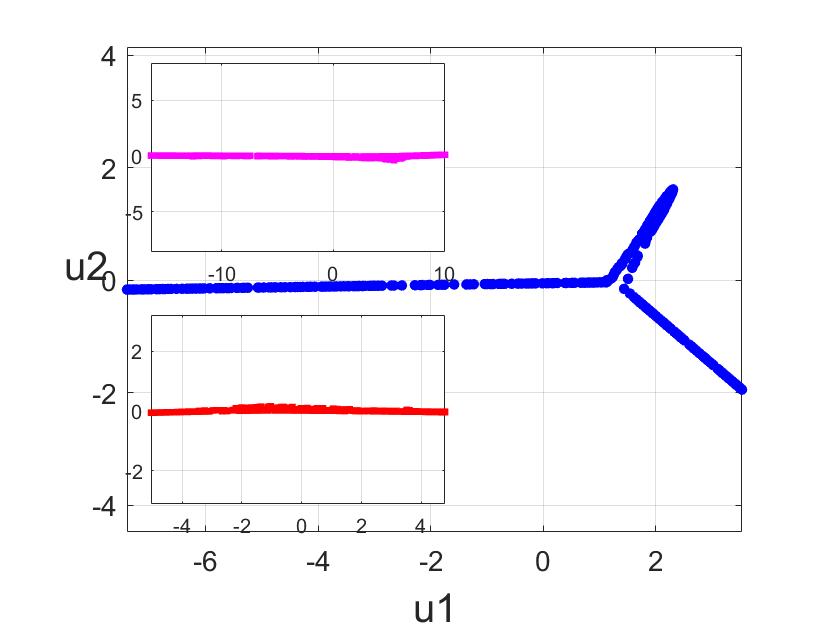}}
			\centerline{(b)}
		\end{minipage}
		\vfill
		\begin{minipage}{0.48\linewidth}
			\centerline{\includegraphics[width=8cm]{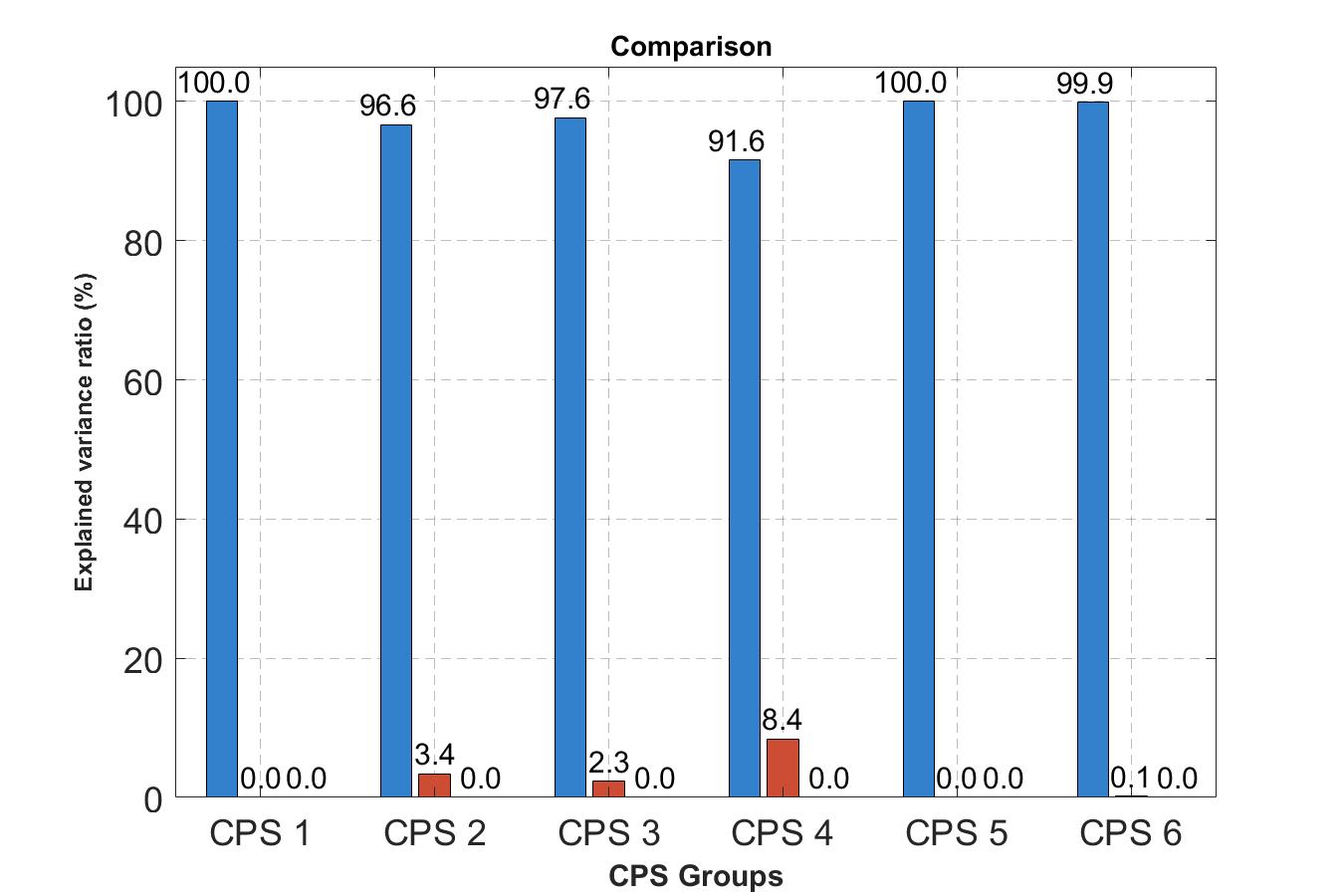}}
			\centerline{(c)}
		\end{minipage}
		\hfill
		\begin{minipage}{0.48\linewidth}
			\centerline{\includegraphics[width=8cm]{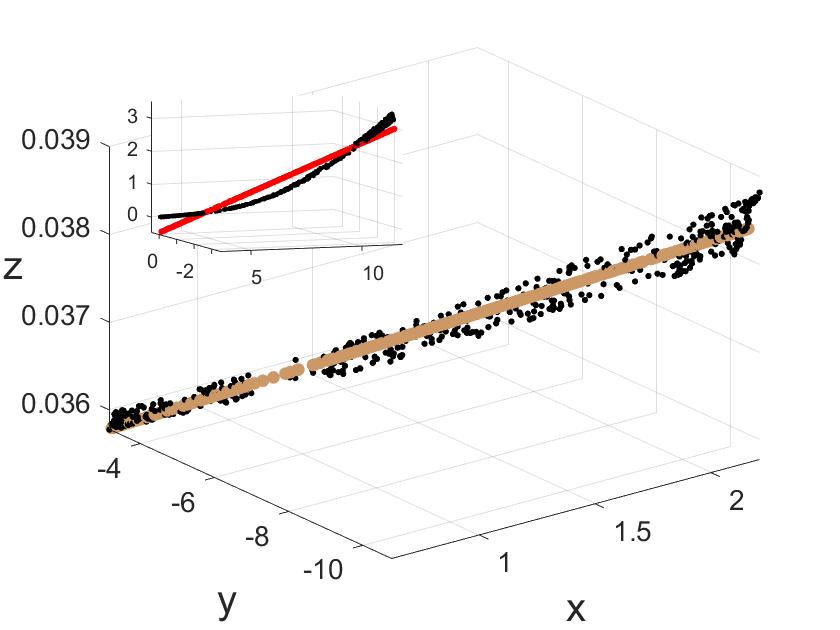}}
			\centerline{(d)}
		\end{minipage}
		\caption{Complexity analysis of CPSs via PCA and Isomap algorithms. (a) Comparison of eigenvalues of the multidimensional scaling (MDS) distance matrix obtained from Isomap analysis for different CPSs. (b) Two-dimensional Isomap embedding manifolds of CPS2, CPS3, and CPS4. (c) Single principal component variance contribution rates of all valid CPSs obtained by PCA. (d) Comparison between the reconstructed sections based on effective principal components and the original sections for CPS1 and CPS2.} 
		\label{fig-5-4} 
	\end{figure*}

Isomap is an unsupervised nonlinear manifold dimensionality reduction algorithm. Different from PCA, which adopts Euclidean distance, Isomap characterizes sample similarity via geodesic distances on the manifold surface. The algorithm proceeds as follows. First, a neighborhood parameter $k$ is set to construct a $k$-nearest-neighbor topological graph. Second, the Dijkstra algorithm is utilized to calculate the shortest path between any two samples, so as to establish the geodesic distance matrix $D_G$. Finally, multidimensional scaling (MDS) is performed on $D_G$ to implement eigenvalue decomposition:
\begin{equation}
	-\frac{1}{2}J D_G^2 J = V\Lambda V^\text{T}
	\label{eq:5_6}
\end{equation}
where $J$ denotes the centering matrix, and $\Lambda=\text{diag}(\lambda_1,\lambda_2,\dots,\lambda_r)$ represents the Isomap eigenvalue matrix. The eigenvalue $\lambda_i$ quantifies the inherent structural energy of each manifold dimension. Obtained by eigenvalue decomposition of the geodesic distance matrix, such eigenvalues reflect the intrinsic degrees of freedom of the manifold.
In this paper, the discriminant criterion for effective projection dimensions is defined as follows. The Isomap eigenvalue decay curve is plotted, and the intrinsic dimension of the point set is determined according to the abrupt inflection point of eigenvalue attenuation. The corresponding projection components are finally reserved as valid low-dimensional projections.
In this case study, the neighborhood parameter is set as $k=20$, accounting for approximately $4\%$ of the total sampling points. Based on the above algorithm, the eigenvalues and projection regions of each CPS are solved. Fig. \ref{fig-5-4}(a) presents the Isomap eigenvalue decay curves of all CPSs. It can be observed that all CPSs except CPS4 possess only one non-zero eigenvalue. In contrast, CPS4 has two non-zero eigenvalues, while the third eigenvalue approaches zero. This indicates that CPS4 requires two embedding dimensions to represent its valid low-dimensional structure, as illustrated by the blue region in Fig. \ref{fig-5-4}(b), which is consistent with the structural characteristics in Fig. \ref{fig-5-2}(a).
The upper and lower subgraphs in Fig. \ref{fig-5-2}(a) show the low-dimensional projections of CPS2 and CPS3 constructed by the first two dimensions. It is obvious that the second dimension can be discarded, which verifies that their second eigenvalues are equal to zero. Nevertheless, the first eigenvalue of CPS3 in Fig. \ref{fig-5-4}(a) is significantly larger than those of other CPSs, which is caused by the excessive length of its projected region. By comparing the two subgraphs in Fig. \ref{fig-5-4}(b), the main projection length of CPS3 is approximately three times that of CPS2. Since eigenvalues are proportional to the square of the projection length, the first eigenvalue of CPS3 is nearly nine times larger than those of the remaining CPSs, demonstrating that CPS3 is located in the local region with the most intense stretching.

	According to the Isomap analysis, CPS3 and CPS4 are identified as complex candidate Poincaré sections and eliminated, while the remaining four sections are retained as valid CPSs.
	Nevertheless, structural complexity varies among these valid regions. This is because the stretching and folding of chaotic systems is a global evolutionary behavior rather than a purely local one; such nonlinear effects are intensified in certain local regions and inconspicuous in others. To select the simplest structural region from the four valid CPSs, the PCA method is further adopted for secondary screening. Although all valid CPSs possess a one-dimensional effective dimension, the nonlinear low-dimensional projection of Isomap neglects mild bending structures. In ideal scroll chaotic models, valid CPSs present standard linear shapes. In practice, curved CPS segments suffer stronger stretching and folding interference than linear ones, resulting in higher structural complexity. Therefore, the PCA dimensionality reduction method is employed to screen optimal linear low-dimensional sections embedded in high-dimensional phase space.
	To quantitatively characterize the principal component structure of each CPS subregion, PCA dimensionality reduction is implemented for all valid subregions in this study, and the detailed algorithm procedures are presented as follows.
	First, only centralization is performed on the original curved point sets to provide standardized input for PCA decomposition. The centralization formula is defined as:
	\begin{equation}
		x_c = x-\mu
		\label{eq:5_2}
	\end{equation}
	where $\mu$ denotes the mean value of the features. In this paper, the PCA algorithm only conducts data centralization without variance normalization, aiming to preserve the original amplitude differences and variance distribution characteristics of the raw data.
	Subsequently, linear dimensionality reduction is performed on the centralized point sets via PCA. The PCA implementation is based on the singular value decomposition (SVD) of the centralized matrix. For the original feature matrix $X\in \mathbb{R}^{n\times p}$, the centralized matrix $X_c$ is obtained and decomposed as follows:
	\begin{equation}
		X_c = U\Sigma V^\text{T}
		\label{eq:5_3}
	\end{equation}
	where $\Sigma=\text{diag}(\sigma_1,\sigma_2,\dots,\sigma_p)$ represents the singular value matrix, and the square of each singular value $\sigma_i^2$ characterizes the variance contribution of the corresponding principal component. The principal component score matrix is calculated as:
	\begin{equation}
		Y = U\Sigma
		\label{eq:5_4}
	\end{equation}
	Each column of matrix $Y$ corresponds to an independent principal component, and the significance of principal components decreases sequentially. The principal components are denoted as $Y_1,Y_2,\dots,Y_p$ in descending order of importance. In this paper, the cumulative variance contribution rate ${Cont}(r)$ reaching $98\%$ is set as the screening threshold for valid principal components.
	
	\begin{equation}
		\text{Cont}(r)=\frac{\sum_{i=1}^{r}\sigma_i^2}{\sum_{i=1}^{p}\sigma_i^2} \ge 98\%
		\label{eq:5_5}
	\end{equation}

The first $r$-order principal components that satisfy the threshold are selected to form the low-dimensional feature subset. If $r=1$, the point set exhibits a strong linear distribution characteristic, and nonlinear dimensionality reduction is unnecessary; the one-dimensional principal component can be directly adopted as the valid projection result. If $r>1$, the first $r$ principal components are retained and further imported into the Isomap algorithm for nonlinear manifold analysis.
Fig. \ref{fig-5-4}(a) presents the PCA principal component variance contribution rates of the six CPSs. The results indicate that the variance contribution rates of the first principal components of CPS2, CPS3, and CPS4 are lower than $98\%$. Meanwhile, the distance comparison of the three principal components of different CPSs in Fig. \ref{fig-5-4}(b) demonstrates that the second principal component $Y_2$ of these three CPSs possesses a considerably large spatial distance, further verifying that two principal components are required to construct their valid low-dimensional projections.
The low-dimensional projection results of two typical subregions, CPS3 and CPS4, are illustrated in Fig. \ref{fig-5-4}(c). The main blue region corresponds to CPS4, and the magenta subgraph represents CPS3. The figure shows the two-dimensional projection regions constructed by $Y_1$ and $Y_2$, whose coordinate system $(u_1,u_2,\dots,u_p)$ differs from the original $(x,y,z)$ coordinate system of the chaotic system. It can be observed that both CPS3 and CPS4 exhibit obvious structural bending, and the bending degree of CPS4 is significantly higher than that of CPS3. This reveals that the folding behavior of chaotic evolution is mainly concentrated in the regions of CPS3, CPS4, and their adjacent areas.
If only the first principal direction $Y_1$ is adopted for structural evaluation, the projection degenerates into a horizontal line along the $u_1$ axis, which fails to reflect the inherent folding trend and leads to incorrect topological detection. CPS2 presents slight bending similar to CPS3, indicating that weak folding effects exist from CPS1 to CPS3, though such effects are far less intense than those occurring from CPS3 to CPS4.
In addition, Fig. \ref{fig-5-4}(b) shows that the distance proportions of the three principal directions vary distinctly among different CPSs. This phenomenon indicates that the stretching and folding behaviors globally permeate the entire chaotic attractor, while the most severe folding effects are always concentrated in specific local subregions.

	\subsubsection{Judgment of Linear Equivalence in CPS Evolution via Generalized Coordinate Method}
	
\begin{figure*}[htbp]  
		\begin{minipage}{0.48\linewidth}
			\centerline{\includegraphics[width=8cm]{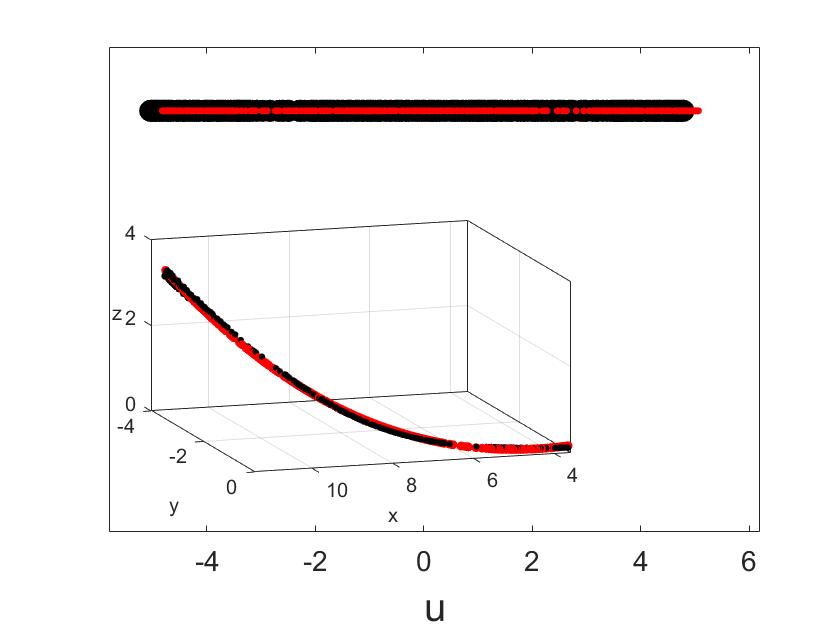}}
			\centerline{(a)}
		\end{minipage}
		\hfill
		\begin{minipage}{0.48\linewidth}
			\centerline{\includegraphics[width=8cm]{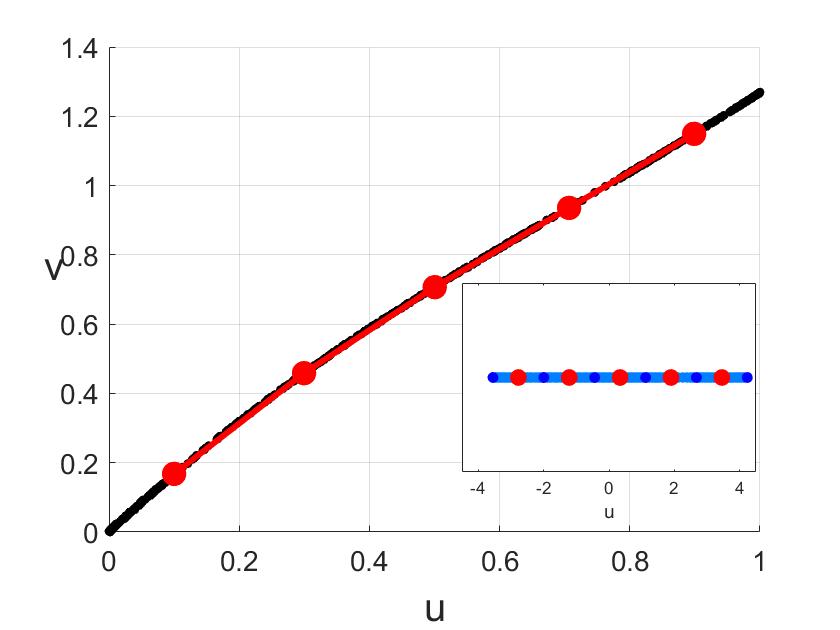}}
			\centerline{(b)}
		\end{minipage}
		\vfill
		\begin{minipage}{0.48\linewidth}
			\centerline{\includegraphics[width=8cm]{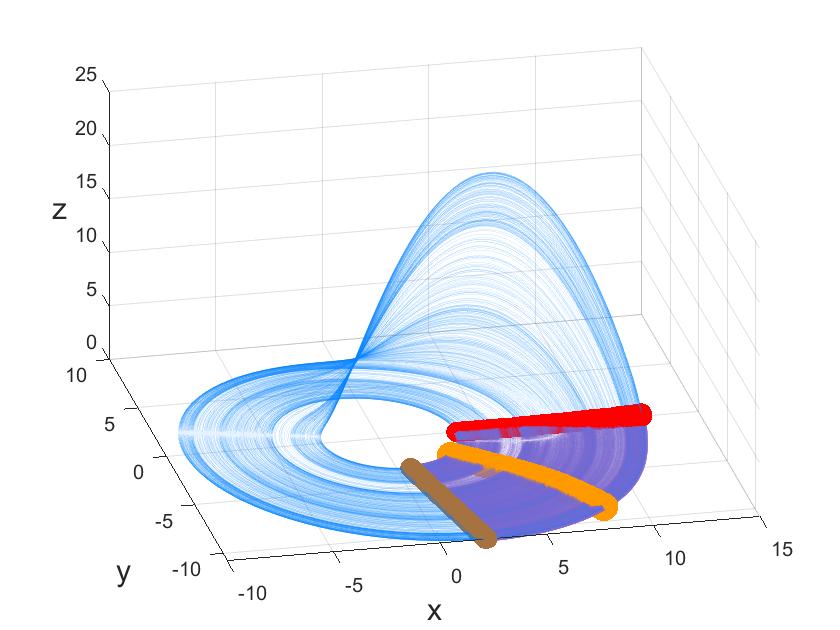}}
			\centerline{(c)}
		\end{minipage}
		\hfill
		\begin{minipage}{0.48\linewidth}
			\centerline{\includegraphics[width=8cm]{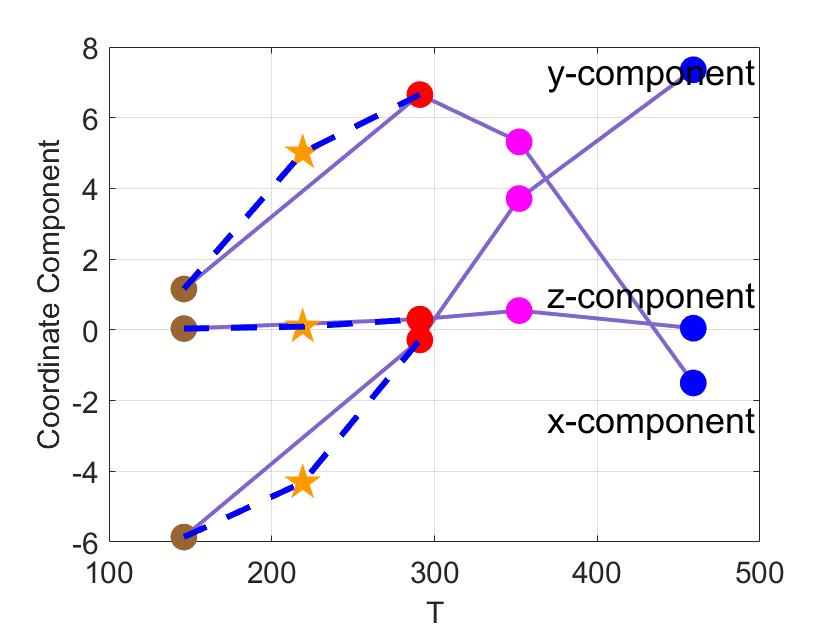}}
			\centerline{(d)}
		\end{minipage}

		\caption{Linear equivalence process analysis. (a) Reconstruction of the original region using incomplete PCA principal components combined with Isomap projection of CPS2. (b) Equivalence process under generalized coordinates. (c) Construction of linearly equivalent sections between CPS1 and CPS2. (d) OP analysis for the construction of linearly equivalent sections between CPS1 and CPS2.} 
		\label{fig-5-5} 
	\end{figure*}

Folding behaviors inevitably generate structurally complex regions and their corresponding complex CPSs, which necessitates quantitative analysis based on generalized coordinates constructed from valid low-dimensional projections. The generalized coordinate system is established on the valid low-dimensional projections of CPSs to construct generalized evolution equations between adjacent evolving CPSs. Such valid low-dimensional projections are derived from the PCA principal component projections and Isomap low-dimensional effective manifolds obtained in the previous analysis.
In this study, generalized coordinate testing is first performed on CPS1 and CPS2. Fig. \ref{fig-5-5}(a) illustrates the valid projection of CPS1, which corresponds to the linear principal component region obtained by PCA dimensionality reduction. The principal component curve is uniformly partitioned at equal intervals, as marked by blue points in the figure. The center point of each interval is extracted and labeled in red as the representative point of the corresponding local region. Notably, each red representative point is selected as the nearest real sampling point to the interval center. This selection strategy enables real trajectory points to evolve and match the low-dimensional projection positions of CPS2, thereby revealing the coarse-grained mapping relationship between adjacent cross-sections.
The subgraph in Fig. \ref{fig-5-5}(a) presents the back-projection of red representative points to the original CPS1 region. The color gradient distributed over the original region corresponds to the magnitude variation of the valid projection components in the main graph, which exhibits a strictly monotonic trend. This indicates a linear one-to-one correspondence between the original CPS region and its valid projection component, verifying the rationality of the constructed generalized coordinate system, which is defined as coordinate \(u\).
Similarly, the color distribution over the subgraph region in Fig. \ref{fig-5-5}(b) reflects the magnitude variation of valid projection components of CPS2, further demonstrating the feasibility of generalized coordinate construction, and the corresponding generalized coordinate is defined as coordinate \(v\). The main graph of Fig. \ref{fig-5-5}(b) shows a monotonically varying evolution process, indicating that the regional variation is approximately linear without folding deformation. The black curve characterizes the continuous variation relationship established by the coordinate values of the two valid projections, while the red region describes the discrete evolution process of representative points, which represents a coarse-grained mapping consistent with the black curve. The representative point-based evolution characterization is applicable to scenarios with blurred main projection boundaries or noise interference in subregions.
	
This process is defined as the linear equivalence process, which can occur between two non-adjacent subregions. All intermediate regions between the two non-adjacent regions can be proven to be equivalent to both boundary regions. As illustrated in Fig. \ref{fig-5-5}(c), the orange region denotes the intermediate subregion \($$CPS_{mid}$$\) constructed by sampling midpoints along the evolutionary trajectories between the two valid subregions CPS1 and CPS2. Fig. \ref{fig-5-5}(d) presents the time series of the \(x\), \(y\), and \(z\) components for a segment of evolutionary trajectory from CPS1 to CPS4, where orange stars mark the sampling points of \($$CPS_{mid}$$\). It can be observed that the points on \($$CPS_{mid}$$\) together with the points on CPS1 and CPS2 form monotonic OPs in all three component directions, which verifies the linear nature of the evolutionary transition.
Furthermore, arbitrary displacement of \($$CPS_{mid}$$\) within the interval between CPS1 and CPS2 still enables the construction of monotonic OPs, demonstrating the internal equivalence of the entire evolutionary interval. Consistent with the structural characteristics of the two boundary regions, the low-dimensional projection of \($$CPS_{mid}$$\) (the orange region in the figure) qualifies \($$CPS_{mid}$$\) as a valid Poincaré section. In this case, generalized coordinates established for \($$CPS_{mid}$$\) yield identical results to those presented in Fig. \ref{fig-5-5}(b) for CPS1 and CPS2. Similarly, the evolutionary process from CPS2 to CPS3 also satisfies linear equivalence. Accordingly, the entire regional layout spanning from CPS1 to CPS3 belongs to a global linear equivalence process.
	
	\begin{figure*}[htbp]  
		\begin{minipage}{0.48\linewidth}
			\centerline{\includegraphics[width=8cm]{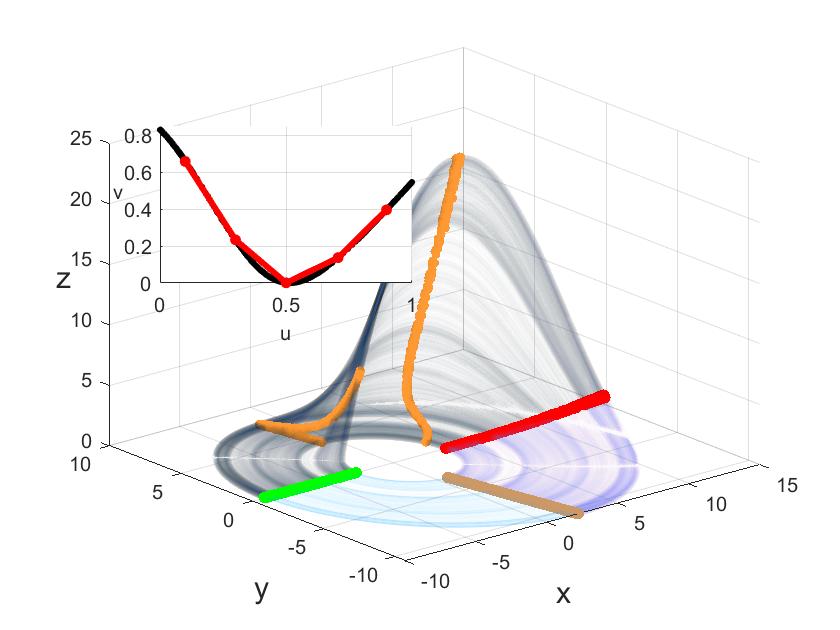}}
			\centerline{(a)}
		\end{minipage}
		\hfill
		\begin{minipage}{0.48\linewidth}
			\centerline{\includegraphics[width=8cm]{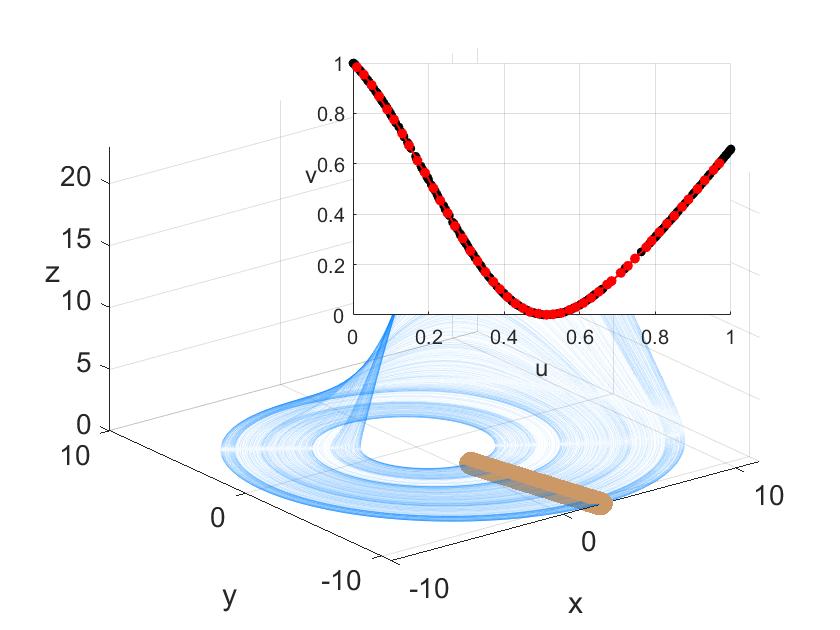}}
			\centerline{(b)}
		\end{minipage}
		\caption{Local nonlinear feature extraction of CPS via generalized coordinate construction. (a) Generalized coordinate construction from CPS1 to CPS5; (b) Self-mapping generalized coordinate construction of CPS1.} 
		\label{fig-5-6} 
	\end{figure*}

Generalized coordinates can effectively verify linearly equivalent regions among different CPSs. On this basis, this study further applies this powerful tool to the black-box local regions of chaotic attractors to identify the occurrence of nonlinear evolutionary processes.
Fig. \ref{fig-5-6}(a) illustrates the evolutionary region spanning from CPS3 to CPS5. Due to the complex topological structure of CPS4, it is infeasible to construct valid projection-based generalized coordinates for this subregion. In the subgraph of Fig. \ref{fig-5-6}(b), the displayed spatial region corresponds to CPS3, while the color gradient denotes the valid projection values of CPS5. The color distribution on the high-dimensional curved manifold of CPS3 fails to present continuous and monotonic variation, but exhibits an aggregated gradient region, which preliminarily indicates the breakdown of linear equivalence between the two subregions.
By constructing generalized coordinates based on the valid projections from CPS3 to CPS5, an evolutionary equation similar to the logistic mapping can be clearly obtained with distinct folding points, demonstrating that an obvious folding behavior occurs within this local region. Fig. \ref{fig-5-6}(c) presents the evolutionary process from CPS5 to CPS6 and further to CPS1. This paper directly verifies the linear equivalence between CPS5 and CPS1, since CPS6 serves as the intermediate transition region without adjacent evolutionary points between CPS5 and CPS1. If linear equivalence holds between CPS5 and CPS1, CPS6 is also linearly equivalent to both subregions.
The subgraph in Fig. \ref{fig-5-6}(c) shows the generalized coordinate evolution diagram based on low-dimensional projections of CPS5 and CPS1. Consistent with the linear monotonic evolutionary characteristics in the main graph of Fig. \ref{fig-5-5}(b), the results verify the linear equivalence of these two regions. Moreover, monotonic OPs can be constructed for CPS5, CPS6 and CPS1, as illustrated in Fig. \ref{fig-5-6}(d).

	\begin{figure*}[htbp]  
		\begin{minipage}{0.48\linewidth}
			\centerline{\includegraphics[width=8cm]{figure/pg6a.jpg}}
			\centerline{(a)}
		\end{minipage}
		\hfill
		\begin{minipage}{0.48\linewidth}
			\centerline{\includegraphics[width=8cm]{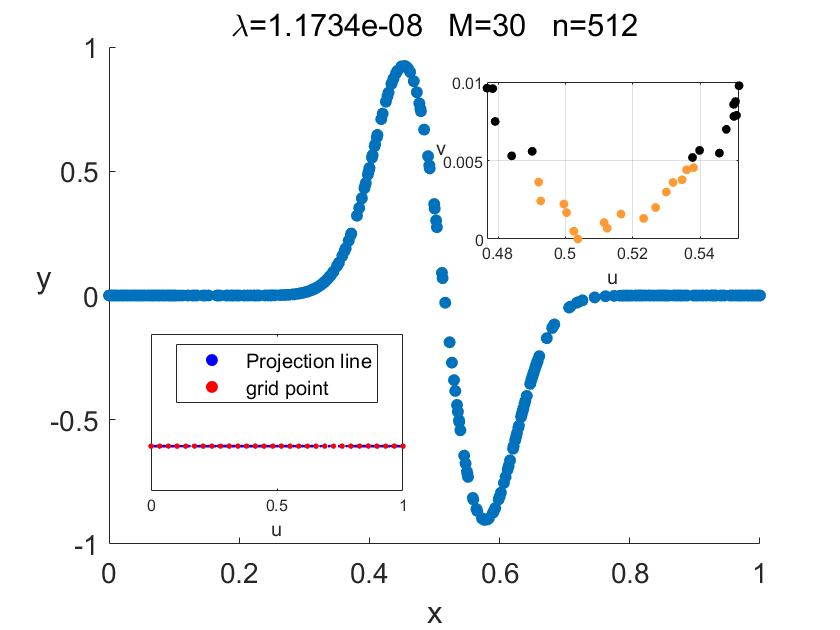}}
			\centerline{(b)}
		\end{minipage}
		\vfill
		\begin{minipage}{0.48\linewidth}
			\centerline{\includegraphics[width=8cm]{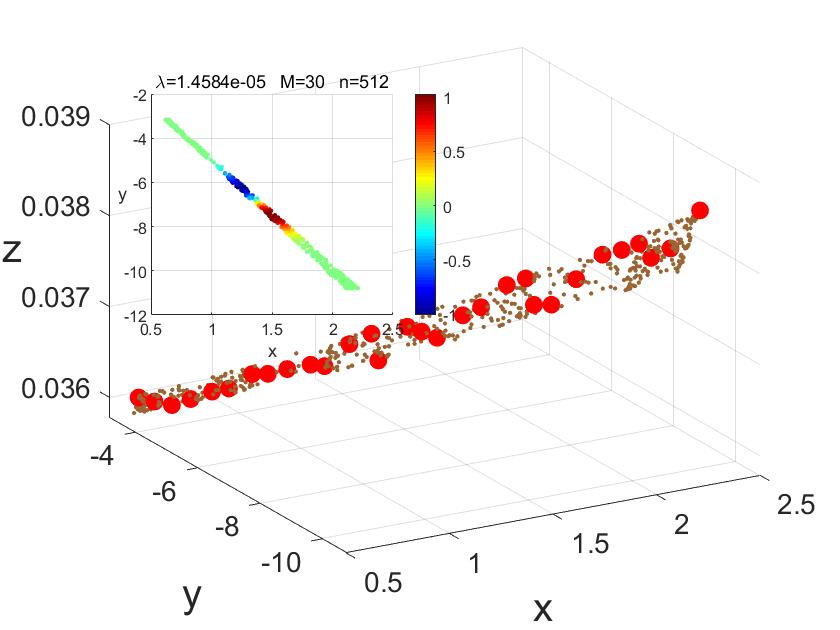}}
			\centerline{(c)}
		\end{minipage}
		\hfill
		\begin{minipage}{0.48\linewidth}
			\centerline{\includegraphics[width=8cm]{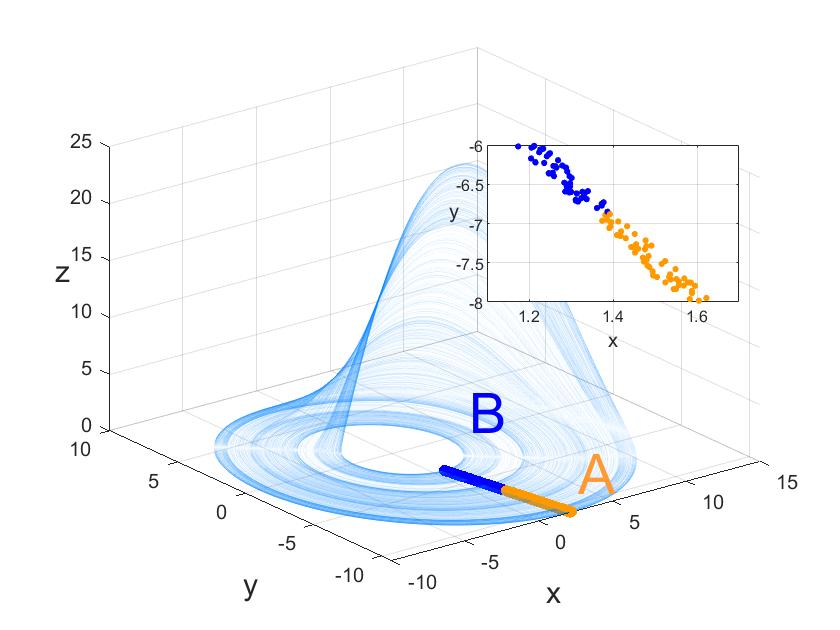}}
			\centerline{(d)}
		\end{minipage}
		\caption{Symbol partition based on the optimal OP Poincaré section. (a) Construction of generalized coordinates to identify nonlinear processes. Once folding behavior is detected, Koopman analysis is performed on the corresponding region, and the boundaries for symbol partition are located via the VLEZ method. (b) Koopman analysis of the valid projection region. (c) Koopman analysis of the original section region. (d) Acquisition of the final symbol partition results.} 
		\label{fig-5-7} 
	\end{figure*}
	
	\subsubsection{Symbolic Partitioning of the Final Section via KA Analysis}
	
The above analysis demonstrates that the investigated chaotic case contains only one local folding region, while the other two local regions exhibit purely linear evolutionary characteristics. Theoretically, the effective CPS undergoes only one folding event after traversing and evolving across the entire attractor region. In this paper, the optimal OP Poincaré section is selected according to the first principal component contribution rate of PCA. Combined with the distance ratio between the first and second principal components presented in Fig. \ref{fig-5-2}(b), CPS1 is finally determined as the optimal OP section for further analysis.
As shown in Fig. \ref{fig-5-7}(a), the generalized Poincaré section is simplified to CPS1. The subgraph displays the generalized coordinate after the effective projection evolves and returns to its initial state once. The obtained curve exhibits obvious consistency with the results in Fig. \ref{fig-5-6}(b), confirming that only one folding behavior occurs during the whole evolution process, which enables the allocation of two symbolic states for the entire system.
Unlike the one-dimensional logistic chaotic mapping, symbol partition boundaries cannot be determined via critical points in this study. This is because the point distribution near the critical position of the generalized coordinate evolution equation is unsmooth and extremely rough. To observe the local structural characteristics, the regional interval $[0.2, 0.7]$ in the evolution equation of the subgraph in Fig. \ref{fig-5-7}(a) is locally enlarged, as shown in the upper-right subgraph of Fig. \ref{fig-5-7}(b). A coarse-scale neighborhood region (marked in orange) can be clearly observed. However, this region is excessively broad for accurate symbol boundary localization. Therefore, Koopman analysis (KA) is implemented on the projected region to achieve precise symbol partition.
First, a one-dimensional Gaussian basis function is adopted to construct the basis function set for Koopman analysis. The specific implementation steps are as follows. Equally spaced sampling is performed on the valid projection region, as illustrated in the lower-left subgraph of Fig. \ref{fig-5-7}(b). A one-dimensional Gaussian wavepacket function is employed to construct the $K$ and $L$ basis function sets for the Koopman approximation matrix, with all function grids uniformly distributed on the linear valid projection region.
In this case, the valid projection is a one-dimensional straight line with endpoints of $[-3.5512, 4.2336]$ and a total length of $7.7848$. The interval between adjacent sampling points is defined as $7.7848/(M-1)$, which is taken as the wavepacket width. The grid position of each basis function is formulated as:
\begin{equation}
	x_k = -3.5512 + (k-1)r_{w}, \quad k = 1,2,\ldots, M.
	\label{eq:5_7}
\end{equation}
Subsequently, Koopman matrix construction and eigenvalue decomposition are carried out. The initial basis number is set as $M=3$. All parameters are substituted into the Gaussian basis function formula \eqref{eq:gauss}.

\begin{equation}
	g_{GA(k)}(\mathbf{x}) = \mathbf{exp}(-\dfrac{| \mathbf{x}-\mathbf{x}^{*}_{k}|^2}{r_{w}^{2}}),k=1,2,...,M
	\label{eq:gauss}
\end{equation}

In addition, $M$ is gradually increased until eigenvalues with absolute values less than $10^{-5}$ appear. When $M=24$, the wavepacket radius is $r=0.3385$, and the grid points $x_k$ are determined by Eq. \ref{eq:5_7}. The obtained valid eigenfunction VLEZ is displayed in the main graph, where the zero-crossing position in the local oscillation region accurately corresponds to the critical point of the generalized coordinate.
Another advantage of Koopman analysis is its applicability to the original high-dimensional region, which serves as the second KA processing method proposed in this paper. Consistent with the previous procedure, a valid basis function set is first constructed. Specifically, the equally spaced grid points on the low-dimensional projection are back-mapped to the original high-dimensional region. The initial wavepacket radius and grid intervals remain consistent with those adopted for low-dimensional projection analysis, and the initial grid points are still defined by Eq. \ref{eq:5_7}.
To realize region back-mapping, the nearest actual sampling points on the projected region corresponding to each grid point are matched. Based on the one-to-one mapping relationship between projected points and original regional points, equally spaced sampling points in the original high-dimensional region are obtained, as marked by red points in Fig. \ref{fig-5-7}(c). Since the intervals between back-mapped points are non-uniform, the maximum interval is selected as the final wavepacket radius to ensure full local coverage of basis functions. All parameters are substituted into the Gaussian basis function formula \ref{eq:gauss} for subsequent KA implementation.
The KA analysis starts with $M=10$. The value of $M$ is reduced if valid VLEZ is obtained; otherwise, $M$ is increased continuously. In this case, an increase in $M$ is required. Valid VLEZ can be acquired when $M=30$, as shown in the subgraph of Fig. \ref{fig-5-7}(c). The uniform interval on the low-dimensional projection is $0.3114$, while the back-mapped intervals in the original region range from $0.2708$ to $0.3480$. To guarantee complete coverage of local dynamic characteristics by wavepacket functions, the maximum interval $0.3480$ is adopted as the wavepacket radius. The red points in the main graph represent all function grid points under $M=30$.
The zero-crossing position of VLEZ in the main graph of Fig. \ref{fig-5-7}(b) divides the low-dimensional projection region into two independent parts for symbol assignment. Benefiting from the one-to-one correspondence between low-dimensional projected points and original regional points, the symbol partition results can be mapped back to the original high-dimensional section, as illustrated in Fig. \ref{fig-5-7}(d). Meanwhile, the symbol division based on the VLEZ zero-crossing position in the subgraph of Fig. \ref{fig-5-7}(c) yields completely consistent partition results. No ambiguous partition points exist in this case, which verifies the rationality of directly implementing KA analysis on low-dimensional projection regions. Accordingly, KA analysis based on low-dimensional projections is adopted for all subsequent case studies.
After completing the symbol partition of the optimal Poincaré section CPS1, a single symbolic boundary divides the entire section into two symbolic regions, realizing the symbolic partitioning of the whole Rössler chaotic system, as presented in the main graph. Nevertheless, a certain width is observed in the local partition region of CPS1 in the subgraph of Fig. \ref{fig-5-7}(d), which is attributed to boundary errors caused by large evolutionary step sizes during critical point extraction for CPS construction. The boundary error becomes more significant after one-step evolution of the local region.
The regional errors before and after evolution are presented in Fig. \ref{fig-5-8}(a). Although the overall CPS presents a curved structure, the local regions before and after evolution (blue and cyan regions) fail to form one-dimensional curved manifolds in low-dimensional projection. The first principal component contribution rates of these regions after PCA dimensionality reduction are both lower than $98\%$.
To optimize the boundary structure of CPS, adapt the effective section to the Rössler system with refined evolutionary step sizes, facilitate localized Gaussian Koopman analysis (GKA), and achieve refined symbolic boundary partitioning, this paper proposes a novel section boundary optimization method termed the monotonic OP interpolation construction method, which enables local regions to be projected into simple topological curve structures.

	\subsection{Lorenz system}
	
	Multi-scroll continuous chaos is more prevalent than single-scroll chaos. To analyze the symbolic partitioning of multi-scroll chaos, this paper first employs the more intuitive double-scroll chaos for symbolic analysis. Meanwhile, for the sake of clarity and convenience in analyzing the underlying mechanism, the Lorenz chaos—being the most classic and topologically simpler—is selected as a case study for symbolic partitioning of this class of systems.
	
	The Lorenz system was proposed by Edward N. Lorenz \cite{1963Deterministic} in 1963 as a simplified mathematical model for studying atmospheric thermal convection. Though deceptively simple in form, this system of equations serves as a cornerstone of chaos theory, profoundly revealing the mathematical essence of the "butterfly effect"—namely, that in a deterministic system, infinitesimal differences in initial conditions may lead to vastly divergent long-term evolutionary outcomes. Its specific dynamical evolution equations are given by:
	\begin{equation}
		\begin{cases}
			\dot{x} = \sigma(y-x) \\
			\dot{y} = x(\rho-z)-y \\
			\dot{z} = xy-\beta z
		\end{cases}
		\label{eq:lorenz}
	\end{equation}
	
	Where $\sigma,\rho,\beta$ are system parameters. The classical parameter values are
	$\sigma=10,\ \rho=28,\ \beta=\dfrac{8}{3}$.
	Under this parameter set, the system exhibits strong chaotic behavior, with trajectories converging to a strange attractor of fractal structure. Therefore, this paper adopts the above parameters for symbolic partitioning analysis, facilitating comparison with previous symbolic partitioning results.

	\subsubsection{Determination of Equivalent Process and Coupled Process}
	
	For the Lorenz system, which is a multi-scroll chaotic system, the nonlinear evolution involves coupling between scrolls, making it considerably more complex than the stretching and folding processes inherent to a single scroll.
	
	\begin{figure*}[htbp]  
		\begin{minipage}{0.48\linewidth}
			\centerline{\includegraphics[width=8cm]{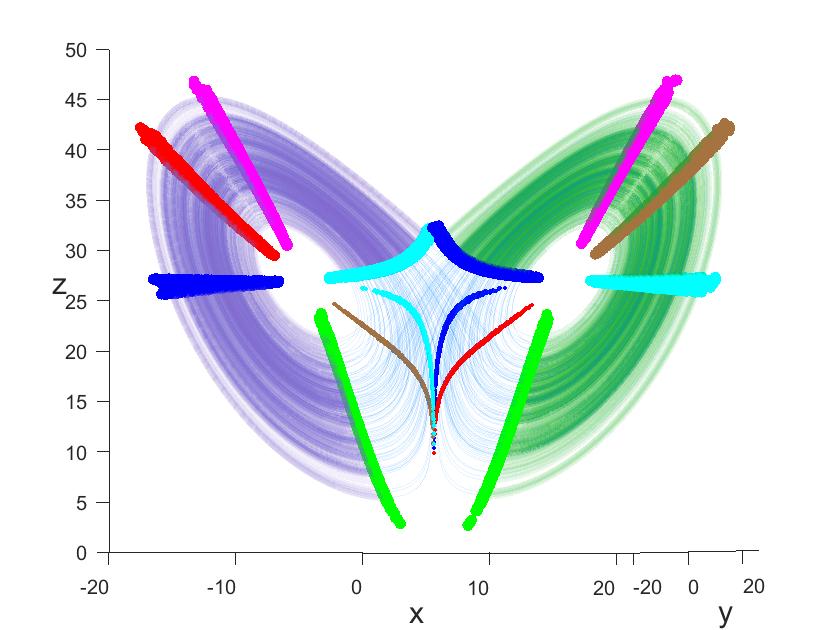}}
			\centerline{(a)}
		\end{minipage}
		\hfill
		\begin{minipage}{0.48\linewidth}
			\centerline{\includegraphics[width=8cm]{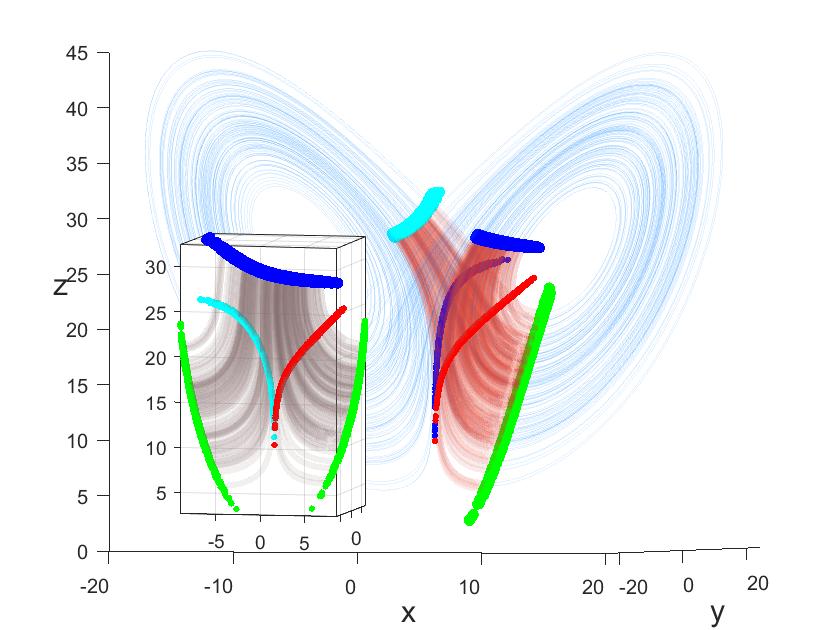}}
			\centerline{(b)}
		\end{minipage}
		\vfill
		\begin{minipage}{0.48\linewidth}
			\centerline{\includegraphics[width=8cm]{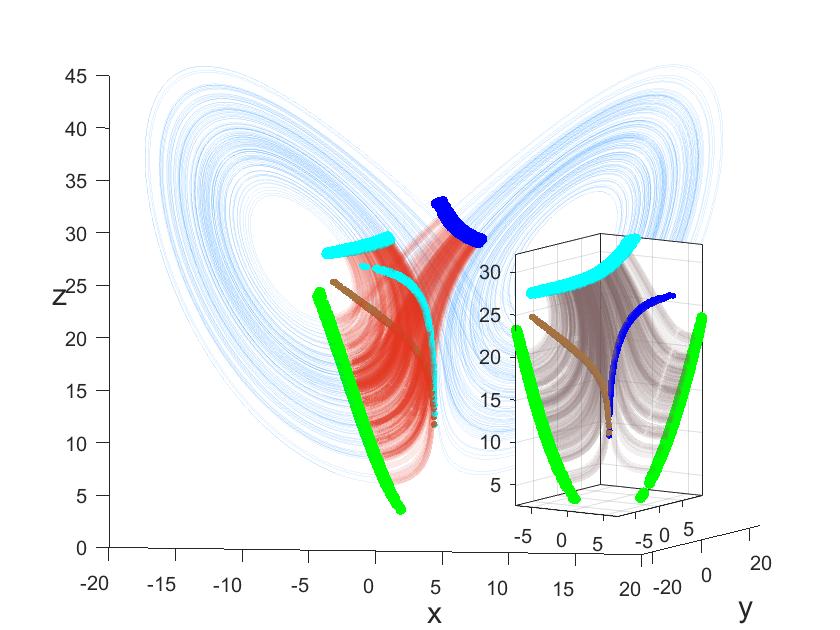}}
			\centerline{(c)}
		\end{minipage}
		\hfill
		\begin{minipage}{0.48\linewidth}
			\centerline{\includegraphics[width=8cm]{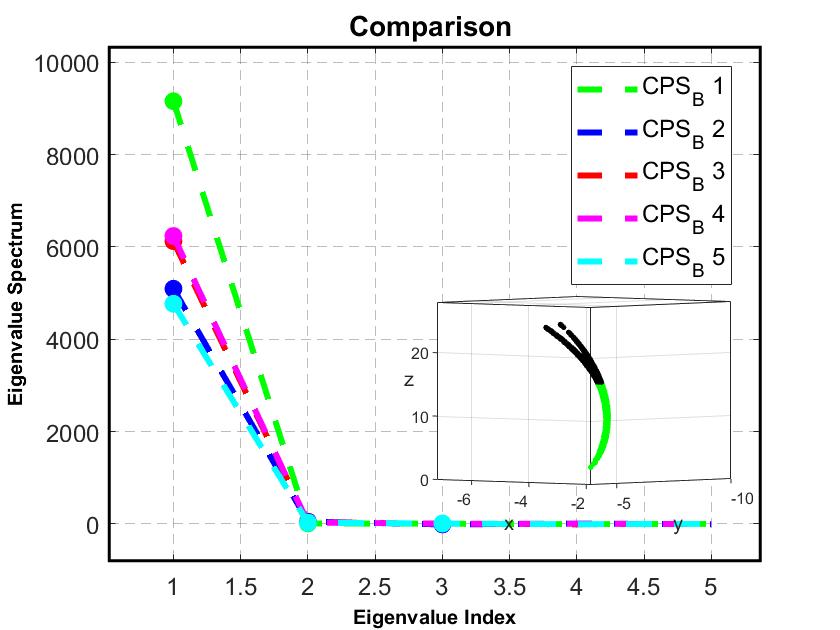}}
			\centerline{(d)}
		\end{minipage}
		\vfill
		\begin{minipage}{0.48\linewidth}
			\centerline{\includegraphics[width=8cm]{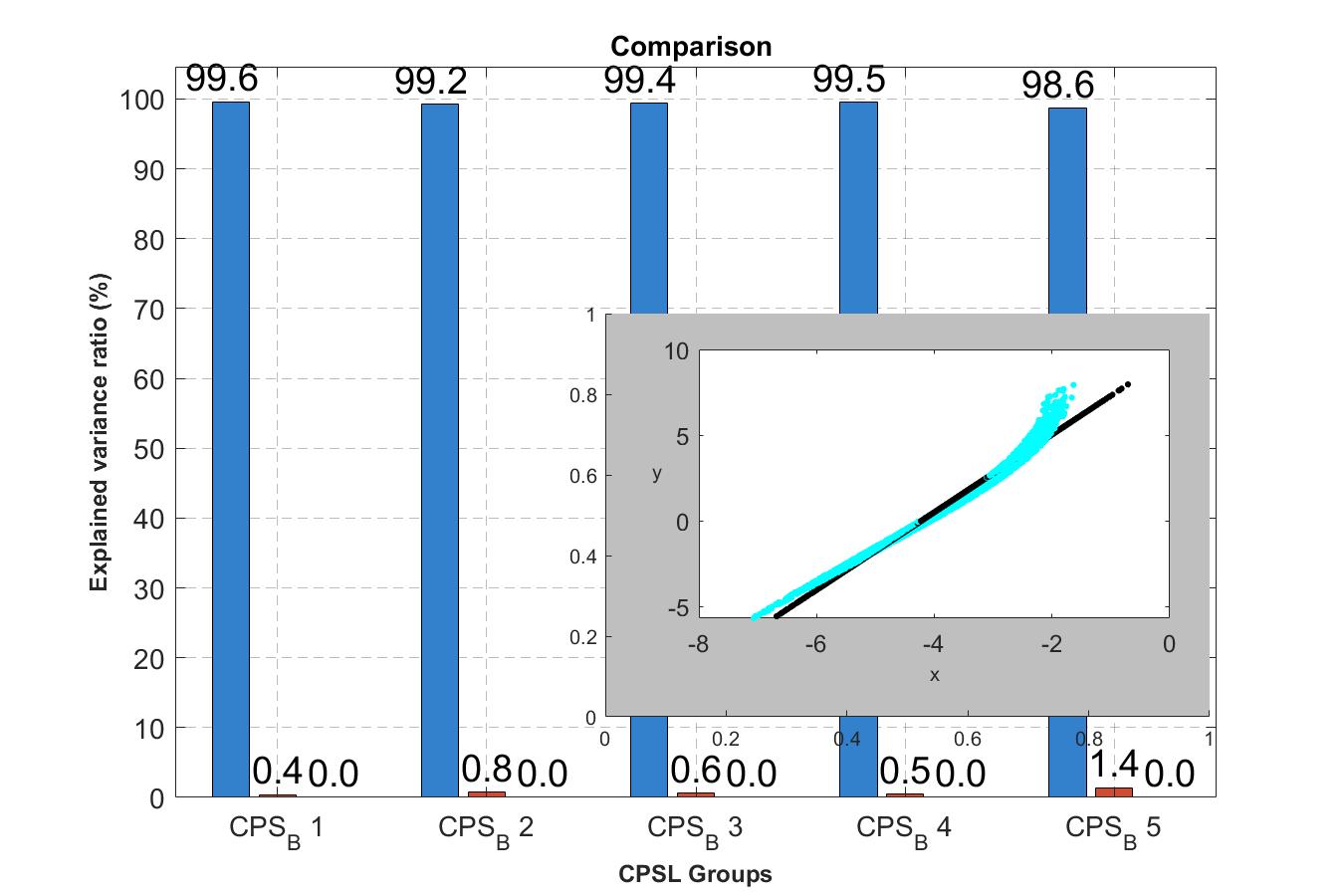}}
			\centerline{(e)}
		\end{minipage}
		\hfill
		\begin{minipage}{0.48\linewidth}
			\centerline{\includegraphics[width=8cm]{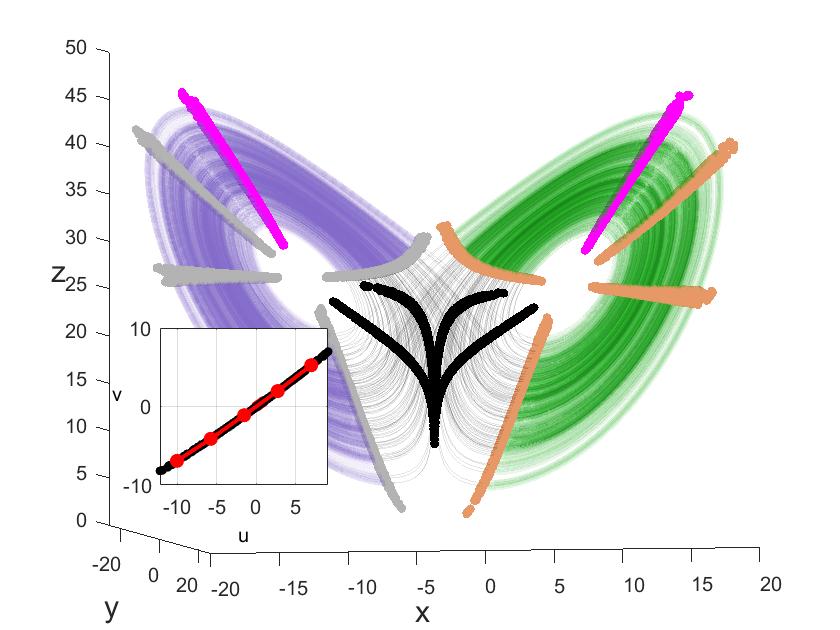}}
			\centerline{(f)}
		\end{minipage}
		\caption{CPS construction and equivalence judgment of CPS evolution for Lorenz Poincaré sections, together with optimal OP Poincaré section selection for symbol partition. (a) Initial CPS division and preliminary scroll judgment. (b) Non-equivalent divergence evolution process of the scroll corresponding to $CPS_{B}$(c)Non-equivalent divergence evolution process of the scroll corresponding to $CPS_{A}$ (d)Isomap analysis of five thick segments in 
			$CPS_{B}$
			(e)PCA analysis of five thick segments in$CPS_{B}$(f)Construction of the final Poincaré section by extracting the simplest CPS from 
			$CPS_{A}$ and $CPS_{B}$
		\label{fig-5-20} }
	\end{figure*}
	
	Fig. \ref{fig-5-20}(a) shows the structure of a chaotic attractor with two coupled scrolls, which is obtained by evolving a steady-state initial point for $800000$ steps with a time step of $0.05$. Therefore, three non-monotonic OPs in different directions can be extracted from this three-dimensional attractor. By distinguishing between local maxima and minima, six initial Candidate Poincaré Sections (CPSs) can be obtained, and these initial CPSs are marked with different colors in the figure for distinction.
	However, different from the Rössler system, the extraction of CPSs only via extreme-value OPs is obviously rough, because each initial CPS can be dispersed into at least two distinct clustered regions. In this paper, these coarse CPSs are further divided according to their clustered regions, and a total of 14 refined CPSs are obtained.
	The analysis of evolutionary order and equivalence process for these 14 CPSs is relatively complex, due to the large number of CPSs and the possible coexistence of coupling and divergence processes. Therefore, it is necessary to pre-judge whether coupling and divergence exist in this region. If such behaviors are identified, the scrolls can be classified first, and then the equivalence analysis can be carried out for each individual scroll, which can significantly simplify the equivalence process analysis. The 14 CPSs sorted by the number of sampling points are listed in the following table.
	
	\begin{table}[h]%
		
		\centering  
		\caption{Different Types of Poincaré Sections and Their Quantitative Statistics}.

		\scalebox{0.4}[0.4]{
			\resizebox{\textwidth}{!}{%
				\begin{tabular}{cccccc}
					\hline
					
					\hline
					CPS type & \multicolumn{4}{c}{number} \\ 
					\hline
					$CPS_A$ &	$2700$  & $2700$ & $2700$ & 	$2700$ & $2700$\\
					$CPS_B$ &	$2630$  & $2629$ & $2629$ & 	$2629$ & $2629$\\ 
					$CPS_C$ & $1575$  & - & - & 	- & -\\  
					$CPS_D$  &$1505$ & - & - & 	- & -\\ 
					$CPS_E$	& 	$1125$ & $1125$ & - & - & 	- \\
					
					\hline

				\end{tabular}
			}
		}
	\end{table}
	
	Based on the above results, all CPSs can be divided into five categories. Among them, $CPS_{A}$ and $CPS_{B}$ contain the largest number of CPSs (five for each category), which correspondingly correspond to two distinct chaotic scrolls. The remaining three categories comprise only four CPSs in total, with far fewer sampling points than $CPS_{A}$ and $CPS_{B}$, which are associated with the divergence and coupling processes between the two scrolls.
	Specifically, the five CPSs belonging to $CPS_{A}$ satisfy linear equivalence during system evolution. The equivalent evolutionary region of $$CPS_{A}$$ is illustrated by the green area in Fig. \ref{fig-5-20}(a), with the evolutionary sequence following green, cyan, brown, magenta, and finally blue. According to this sequence, the corresponding CPSs are named $CPS_{A}1$, $CPS_{A}2$, $CPS_{A}3$, $CPS_{A}4$, and $CPS_{A}5$. Similarly, the five CPSs of $CPS_{B}$ are mutually linearly equivalent in the evolutionary process. The equivalent region of $CPS_{B}$ is represented by the purple area in Fig. \ref{fig-5-20}(a), with the evolutionary direction following green, blue, red, magenta, and ultimately cyan. These CPSs are sequentially defined as $CPS_{B}1$, $CPS_{B}2$, $CPS_{B}3$, $CPS_{B}4$, and $CPS_{B}5$. Each of the two independent local regions presents a scroll structure similar to that of the Rössler system and remains spatially separated, which preliminarily verifies that the investigated system is a dual-scroll chaotic system.
	The four CPSs corresponding to $CPS_{C}$, $CPS_{D}$, and $CPS_{E}$ are located in the cyan transitional region in the figure. According to the distribution of sampling points, the red CPS is identified as $CPS_{C}$, the brown CPS as $CPS_{D}$, and the cyan and blue sectional regions correspond to two subsections of $CPS_{E}$, denoted as $CPS_{E}1$ and $CPS_{E}2$, respectively. One-step evolution of these four transitional CPSs leads to $CPS_{A}1$ and $CPS_{B}1$, the initial sectional regions of the two scrolls. This result excludes the existence of a third scroll and further confirms the dual-scroll characteristic of the chaotic system.
	Notably, $CPS_{B}5$, corresponding to a sampling point count of $2630$, serves as the terminal section of the local equivalent evolutionary process. The maximum-index points of $CPS_{B}5$ can achieve one-to-one evolutionary matching with the other four $CPS_{B}$ sections, while the minimum-index points of $CPS_{B}5$ are redundant and need to be eliminated. After removing redundant points, all sampling points of the five $CPS_{B}$ sections realize accurate one-to-one evolutionary correspondence.
	Consistent with the multi-scroll theoretical model proposed previously, the evolution of multi-scroll chaotic systems is inevitably accompanied by coupling and divergence processes. The four transitional sections including $CPS_{C}$, $CPS_{D}$, $CPS_{E}1$, and $CPS_{E}2$ can evolve into two independent scroll regions in one step, which motivates further investigation on the coupling mechanism between the two scrolls.
	First, backward evolution is performed on $CPS_{A}1$ (the initial section of the green region) to explore whether it originates from the coupling of multiple CPSs. As shown in Fig. \ref{fig-5-20}(b), the red area represents the trajectory region of $CPS_{A}1$ after two-step backward evolution. The one-step backward evolution of $CPS_{A}1$ converges to $CPS_{D}$ and $CPS_{E}2$, corresponding to the thin brown and blue sectional curves in the figure, while the two-step backward evolution further extends to $CPS_{A}5$ and $CPS_{B}5$, marked by the thick blue and cyan sectional curves at the upper boundary of the red region. This indicates that the scrolls corresponding to $CPS_{A}$ and $CPS_{B}$ couple into the local region of the $CPS_{B}$ scroll. The subgraph presents the two-step forward evolution of the terminal section $CPS_{A}5$, which evolves into the initial sections of both scrolls ($CPS_{A}1$ and $CPS_{B}1$), characterizing the divergence behavior accompanying the coupling process of chaotic scrolls.
	Similarly, backward evolution analysis is implemented on $CPS_{B}1$ (the initial section of the purple region). As illustrated in Fig. \ref{fig-5-20}(c), the red area denotes the two-step backward evolutionary trajectory of $CPS_{B}1$. One-step backward evolution drives $CPS_{B}1$ to $CPS_{C}$ and $CPS_{E}1$ (thin brown and cyan sectional curves), and two-step backward evolution further reaches $CPS_{A}5$ and $CPS_{B}5$ (thick blue and cyan sectional curves). This demonstrates that the two scrolls can also couple into the local region of the $CPS_{A}$ scroll. The subgraph shows that the terminal section $CPS_{B}5$ evolves forward for two steps and converges to $CPS_{A}1$ and $CPS_{B}1$, which corresponds to the typical divergence process of dual-scroll chaotic systems.
	Combining the results in Fig. \ref{fig-5-20}(b) and (c), the coupling-divergence transitional region is confirmed as the blue area in Fig. \ref{fig-5-20}(a). The four CPSs distributed in this region act as transitional coupled sections between the two scrolls and are eliminated as redundant regions. The existence of inter-scroll coupling verifies that the evolutionary process between $CPS_{A}$ and $CPS_{B}$ is non-equivalent. All non-equivalent CPSs are retained to construct the final Poincaré section, while only the structurally simplest CPSs in equivalent groups are reserved with the rest eliminated as redundant sections.
	Taking the $CPS_{B}$ group as an example, Isomap analysis is first conducted to obtain the eigenvalue distribution of the geodesic distance matrix $D_G$. The Isomap eigenvalue decay curves of five $CPS_{B}$ sections are plotted in Fig. \ref{fig-5-20}(d). The maximum eigenvalue of $CPS_{B}1$ reaches $9156$, which is significantly higher than that of other CPSs, implying obvious stretching and bending deformation in this local region. The structural morphology of $CPS_{B}1$ is displayed in the subgraph, where severe bending can be clearly observed. Specifically, the positive $z$-direction subregion (marked in black) presents a folding structure analogous to $CPS4$ in the Rössler system. Local Isomap testing verifies that this subregion exhibits a two-dimensional embedding structure similar to $CPS4$ of the Rössler system, indicating the complex structure of $CPS_{B}1$ and disqualifying it from the final valid CPS composition.
	Subsequently, PCA is performed on the remaining CPSs to extract the singular value contribution rate of each principal component, with the results presented in Fig. \ref{fig-5-20}(e). $CPS_{B}5$ exhibits a relatively low contribution rate of the first principal component. In the subgraph, the blue area represents the original $CPS_{B}5$ region, while the black area denotes the reconstruction result using only the first principal component. The poor reconstruction performance indicates that simple linear reconstruction is insufficient, and quadratic curve fitting is required. The large curvature significantly increases the structural complexity, so $CPS_{B}5$ is also excluded from the final valid CPSs.
	Among the remaining four CPSs, $CPS_{B}1$ has the lowest first principal component contribution rate, which is caused by the slender structural morphology leading to PCA misjudgment. Such misjudgment can be eliminated by reducing the evolutionary step size of adjacent OP sampling points. Therefore, two analytical methods need to be comprehensively adopted for curved CPSs with systematic errors. Finally, $CPS_{B}4$ with a high first principal component contribution rate is selected as the optimal valid component for the $CPS_{B}$ group.
	Similarly, comprehensive Isomap and PCA analysis is conducted on the $CPS_{A}$ group, and $CPS_{A}4$ with the simplest topological structure is determined as the valid component for the $CPS_{A}$ group. The final optimal Poincaré section composed of $CPS_{A}4$ and $CPS_{B}4$ is shown in magenta in Fig. \ref{fig-5-20}(f). The redundant CPSs in the $CPS_{A}$ group are marked in orange, and those in the $CPS_{B}$ group are marked in gray. Notably, although $CPS_{A}1$, $CPS_{A}5$, $CPS_{B}1$, and $CPS_{B}5$ have complex structures, they are still valid CPSs and essentially different from the redundant $CPS4$ in the Rössler system.
	Further equivalence verification is implemented by constructing generalized coordinates based on the Isomap projections of $CPS_{B}1$ and $CPS_{B}5$. The results confirm that the evolutionary process among all five $CPS_{B}$ sections satisfies linear equivalence. It is concluded that the initial and terminal sections of each equivalent CPS group possess higher structural complexity than the intermediate sections. This is because the terminal and initial sections are located at the boundaries of the coupling-divergence transitional region, suffering severe bending and distortion induced by inter-scroll coupling and divergence. The four transitional invalid CPSs in the coupling region are marked in black in the figure.

	\subsubsection{Symbolic Partitioning of the Final Section via KA Analysis}
	
\begin{figure*}[htbp]  
		\begin{minipage}{0.48\linewidth}
			\centerline{\includegraphics[width=8cm]{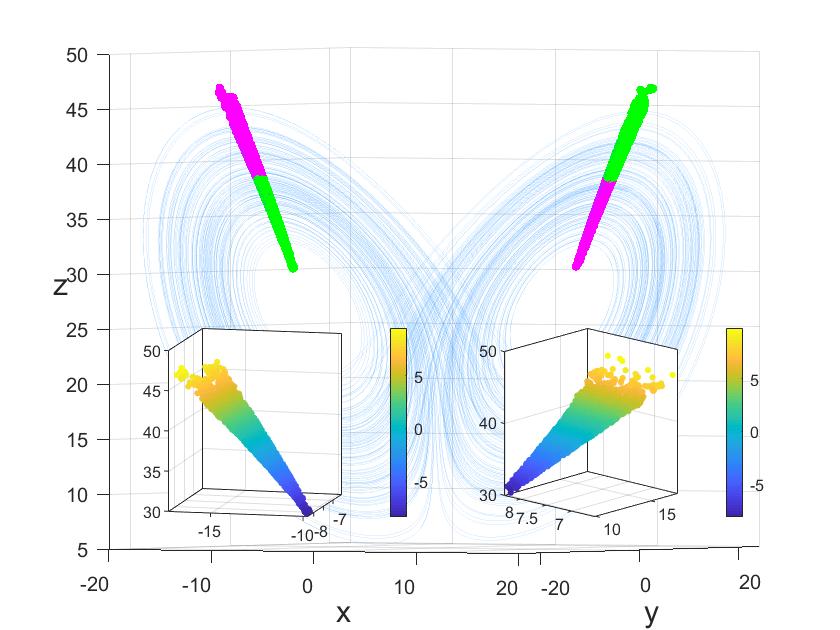}}
			\centerline{(a)}
		\end{minipage}
		\hfill
		\begin{minipage}{0.48\linewidth}
			\centerline{\includegraphics[width=8cm]{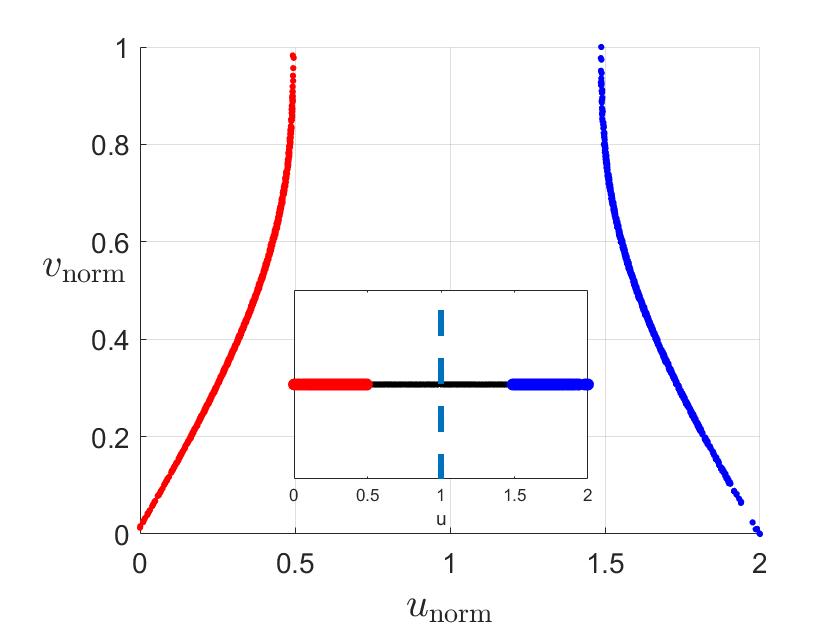}}
			\centerline{(b)}
		\end{minipage}
		\vfill
		\begin{minipage}{0.48\linewidth}
			\centerline{\includegraphics[width=8cm]{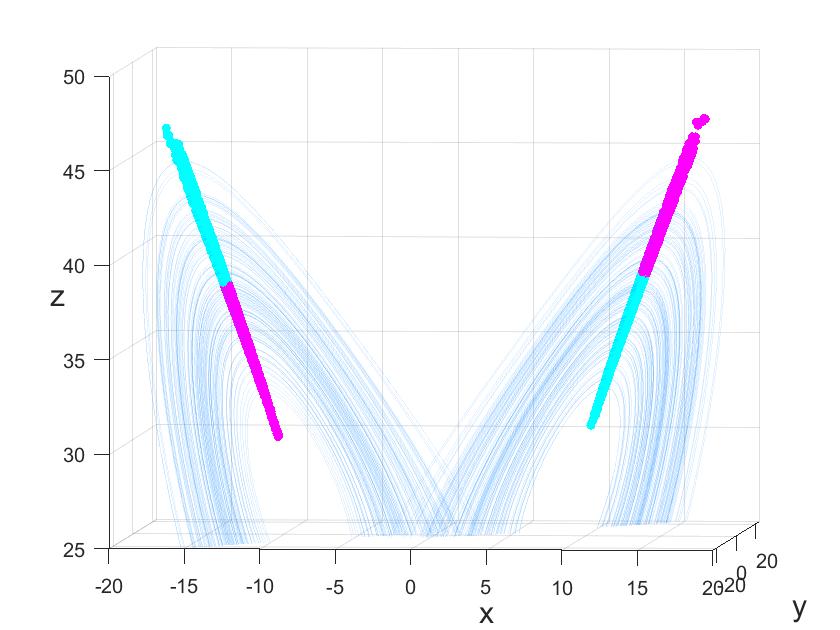}}
			\centerline{(c)}
		\end{minipage}
		\hfill
		\begin{minipage}{0.48\linewidth}
			\centerline{\includegraphics[width=8cm]{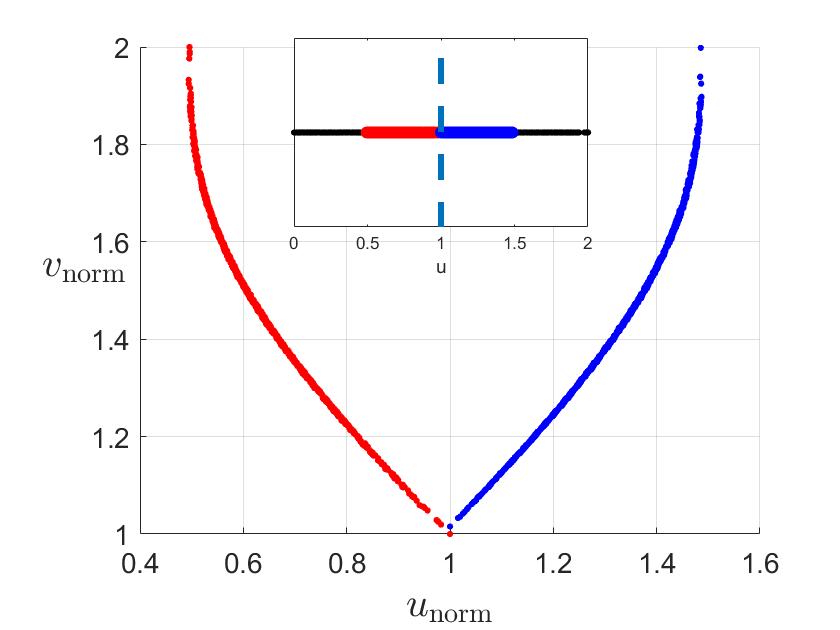}}
			\centerline{(d)}
		\end{minipage}
		\vfill
		\begin{minipage}{0.48\linewidth}
			\centerline{\includegraphics[width=8cm]{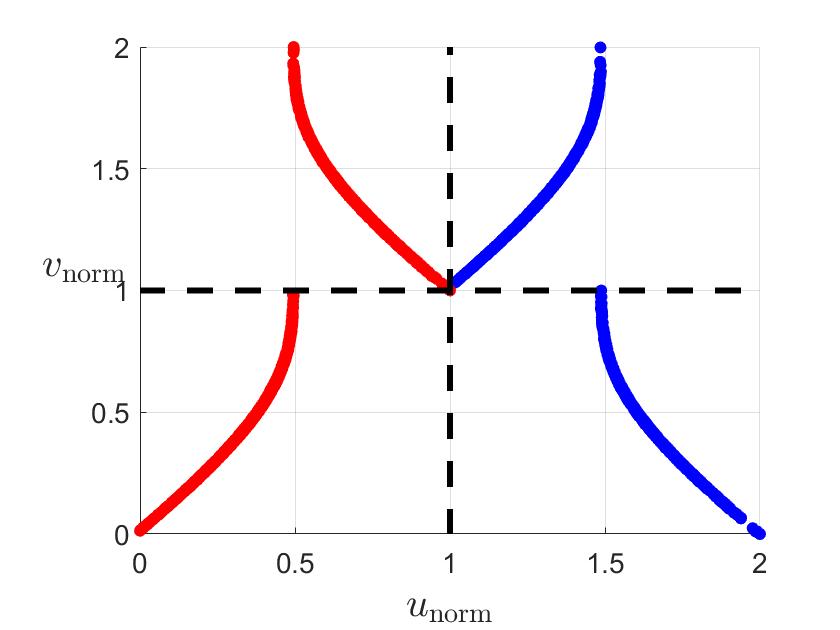}}
			\centerline{(e)}
		\end{minipage}
		\hfill
		\begin{minipage}{0.48\linewidth}
			\centerline{\includegraphics[width=8cm]{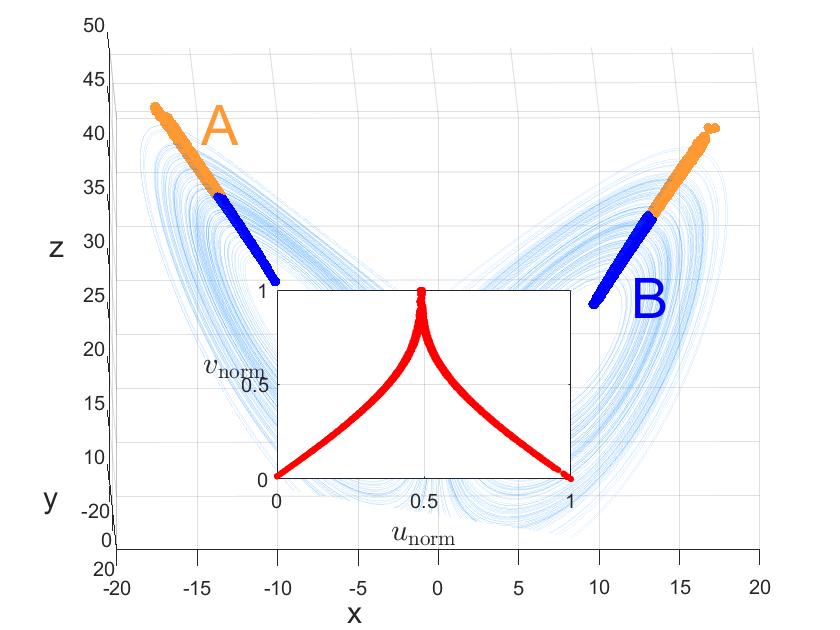}}
			\centerline{(f)}
		\end{minipage}

		\caption{Decoupling process-based low-dimensional region construction and reasonable generalized coordinate establishment for symbolic partitioning
			(a) Backward evolutionary decoupling process and low-dimensional coordinate construction of $CPS_{B}4$; (b) Generalized coordinate correspondence between the pre-image and original domain of $CPS_{B}4$; (c) Backward evolutionary decoupling process of $CPS_{A}4$; (d) Generalized coordinate correspondence between the pre-image and original domain of $CPS_{A}4$; (e) Final Poincaré generalized coordinate analysis and symbolic partition boundary determination; (f) Symmetry analysis of $CPS_{A}4$ and $CPS_{B}4$, as well as generalized coordinate simplification and symbolic partitioning of Poincaré section mapping.} 
		\label{fig-5-18} 
	\end{figure*}

	The Poincaré section structure of the studied system is more complex than that of the Rössler system. For the dual-scroll chaotic case in this work, the two sectional segments mutually couple after one-step forward evolution. However, forward evolution of a single segment only diverges into two subsections, namely $CPS_{A}4$ and $CPS_{B}4$, such that the inherent coupling process cannot be directly observed. Since the folding behavior of chaotic evolution corresponds to the inter-scroll coupling process, backward evolution is adopted in this paper to achieve decoupling and acquire the pre-image local regions of the coupled domain.
	Fig. \ref{fig-5-18}(a) presents the one-step evolutionary sections of all green local regions belonging to $CPS_{B}4$. The results show that these two local regions diverge to two terminal domains after evolution, demonstrating that the $CPS_{B}4$ component originates from the coupled evolution of the two local subregions. Both $CPS_{A}4$ and $CPS_{B}4$ can be mapped onto low-dimensional line segments of the first principal component via PCA. The two low-dimensional segments exhibit identical lengths of 17.94, with a numerical error limited to the third decimal place.
	To unify the two components within a generalized coordinate system, normalization is performed on both segments. The left segment is assigned to the generalized interval $[0,1]$, and the right segment is assigned to $[1,2]$. This generalized coordinate construction is reasonable and accurate due to the identical scaling ratio applied to both segments.
	The two subgraphs in the figure illustrate the correspondence between the original high-dimensional Poincaré sections and their low-dimensional projections, where the point sets represent the original high-dimensional sectional domains and the function values denote the corresponding one-dimensional projection results. PCA dimensionality reduction cannot characterize the projection direction from the original high-dimensional domain to the low-dimensional manifold, which leads to two possible directional solutions during the normalization of linear projected segments. To address this issue, a functional correspondence between the original sections and low-dimensional projections is established in this work. Both low-dimensional segments exhibit an approximately monotonically increasing trend with respect to the $z$-axis of the original phase space, enabling accurate mapping from the generalized coordinate results back to the original sectional structure.
	In practical scenarios, inconsistency may exist between the variation trend of low-dimensional segments and the dominant stretching direction of original sections. For instance, one sectional function may be monotonically increasing while the other is monotonically decreasing. In such cases, unified monotonicity (either fully increasing or fully decreasing) is required to guarantee the rationality of the mapping process. In this case study, coordinate correspondence is implemented along the dominant stretching direction of the original $z$-axis, where both the left and right sections vary from the minimum to the maximum $z$-values. The uniformly monotonically increasing subregions constitute the standardized generalized interval $[0,2]$.
	The two green subsegments correspond to the two terminals of the normalized interval, as indicated by the red and blue regions in the subgraph of Fig. \ref{fig-5-18}(b), while the black region represents the integrated projection of the two complete sections. This subgraph characterizes the distribution of subsegments before evolution. Notably, since each point in the subregions has a one-to-one correspondence with the global sectional point set, repeated PCA dimensionality reduction is avoided for local subsegments. Instead, the coordinates of local points are directly projected onto the low-dimensional segments according to their positional relationships in the global section. This processing strategy is consistently adopted for all subsequent cases, where PCA is only performed on global regions, and local fragments are projected based on global positional constraints.
	Consistent with the processing of the Rössler system in the previous section, the generalized coordinates of pre-evolutionary sections are established, and all post-evolutionary points are projected onto the same coordinate system to obtain the evolutionary relationship diagram in the main graph of Fig. \ref{fig-5-18}(b). The results indicate that both subsegments overlap within the interval $[0,1]$ after evolution, corresponding to the domain of $CPS_{B}4$.
	Similarly, backward evolution is conducted on the right section, yielding two indigo linear fragments shown in Fig. \ref{fig-5-18}(c). As illustrated in the subgraph of Fig. \ref{fig-5-18}(d), their low-dimensional projections are located at the central region of the global segment. The post-evolutionary low-dimensional points are further projected to generate the updated generalized $v$-coordinates, and the main graph of Fig. \ref{fig-5-18}(d) presents the corresponding evolutionary relationship. The two fragments overlap within the interval $[1,2]$, which corresponds to the right section in the high-dimensional phase space.
	By synthesizing the evolutionary results of the two sections, the integrated mapping relationship shown in Fig. \ref{fig-5-18}(e) is obtained. It is observed that both the $[0,1]$ and $[1,2]$ intervals can be partitioned into two distinct symbolic domains.
	Furthermore, the two sections possess nearly identical structural morphology, segment length, and fully symmetric evolutionary behaviors, which enables the combination of the left and right sections into a unified domain. In the low-dimensional projection space, the $[1,2]$ interval is completely translated and superimposed onto the $[0,1]$ interval, leading to the coincidence of all corresponding sectional fragments. The optimized overlapping correspondence is presented in the subgraph of Fig. \ref{fig-5-18}(f), which provides a clear foundation for binary symbolic allocation. The symbolic partition boundary is determined by the unique extreme point in the curve, and point sets on both sides of the boundary are assigned different symbolic states, as visualized in the main graph of Fig. \ref{fig-5-18}(f).
	
\begin{figure*}[htbp]  
		\begin{minipage}{0.48\linewidth}
			\centerline{\includegraphics[width=8cm]{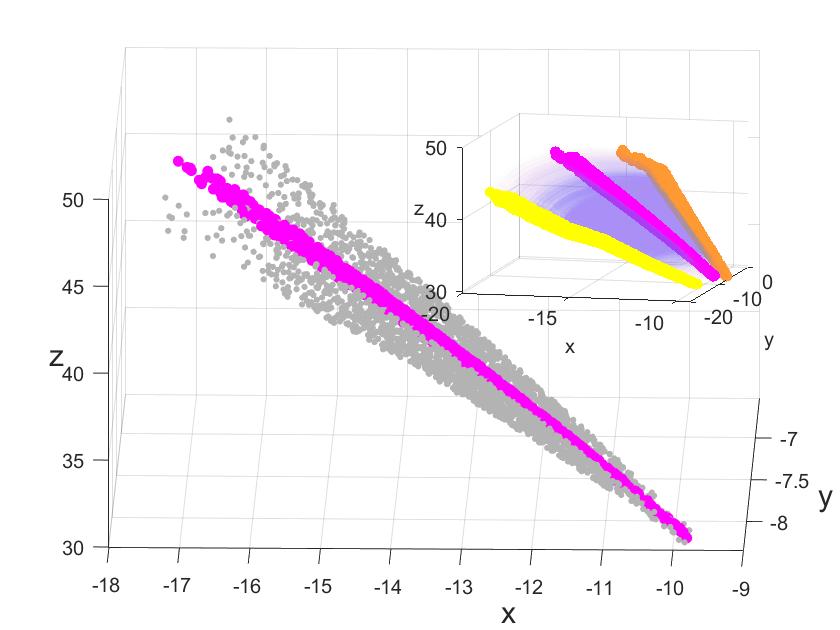}}
			\centerline{(a)}
		\end{minipage}
		\hfill
		\begin{minipage}{0.48\linewidth}
			\centerline{\includegraphics[width=8cm]{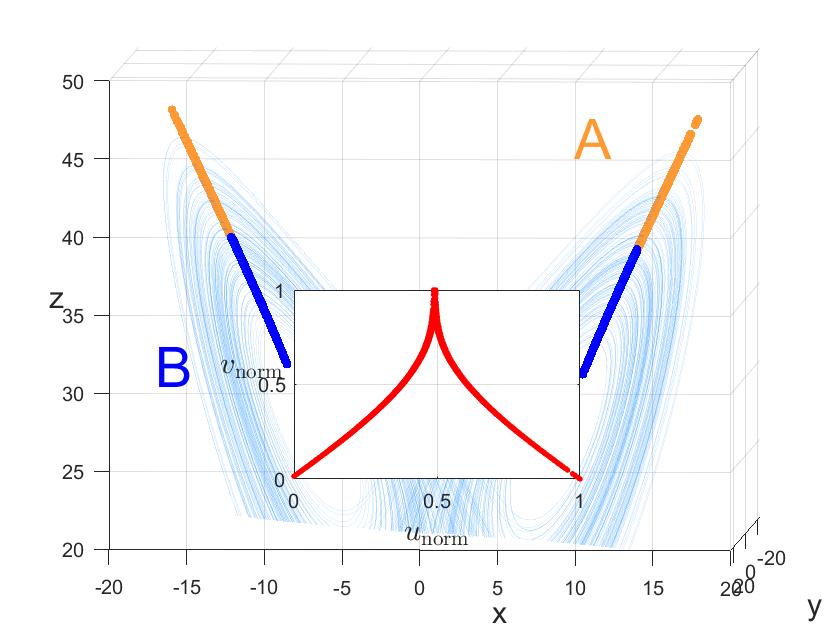}}
			\centerline{(b)}
		\end{minipage}
		
		\caption{ The final refinement of Poincar é's structure (a) involves performing OP interpolation on $CPS_ {B} $and reconstructing it
			(b) The generalized coordinate simplification and symbol division of the refined Poincar é section mapping are presented.} 
		\label{fig-5-19} 
	\end{figure*}

	The boundaries of the left and right sections in the above figure can also be finely constructed via the local interpolation method. Eq. \eqref{eq:a_1} is adopted to realize refined regional reconstruction of the original domains. The endpoints of each section are determined using its two neighboring CPSs, and interpolation is performed with $M=5$, as illustrated in the subgraph of Fig. \ref{fig-5-19}(a). After interpolation, consistent OP extreme points are extracted, and the refined sectional results are presented by the magenta segments in the main graph, while the gray segments denote the original CPSs. The right domain is processed following the same interpolation procedure to obtain the refined counterpart section.
	Based on the two newly refined sections, low-dimensional projection and generalized coordinate reconstruction are implemented, and the final evolutionary relationship is displayed in the subgraph of Fig. \ref{fig-5-19}(b). The system states are classified into two categories according to the critical points for symbolic assignment, and the symbolic results are mapped back to the original sectional domains, as shown in the main graph. The obtained results are consistent with those in Fig. \ref{fig-5-18}(f). The optimized sections exhibit nearly linear profiles and are closer to the ideal Poincaré section state.

	The completion of symbolic partitioning for the Lorenz chaotic system demonstrates that the proposed OP construction-based symbolic partitioning method is applicable to chaotic systems involving scroll coupling and evolutionary behaviors. Similar to the Rössler system, the Lorenz system possesses a relatively single chaos generation mechanism and can be regarded as an approximate ideal chaotic model. On this basis, this study further extends the proposed method and generalizes the two-symbol chaotic partitioning strategy to more complex and universal chaotic systems.

	\subsection{Lü system}

	This study constructs generalized Poincaré sections to analyze complex chaotic systems. The investigations on the Rössler and Lorenz systems reveal that coupling and self-folding represent nonlinear regional overlapping behaviors generated by two distinct chaotic mechanisms. Since overlapping dynamical regions cannot be fully characterized by a single symbolic state, the construction of accurate symbolic partition boundaries is indispensable for chaotic symbolic analysis. In practical chaotic systems, these two evolutionary mechanisms generally coexist simultaneously. Such multi-mechanism nonlinear evolution enables the generation of more than two symbolic states, which constitutes a complex multi-symbol partitioning scenario. To further validate the effectiveness of the proposed OP-based Poincaré section method in handling sophisticated chaotic dynamics, this paper conducts symbolic partitioning on the Lü and Chen systems with complex evolutionary mechanisms.
	The Lü system\cite{HAN2004221} is a three-dimensional autonomous chaotic system belonging to the generalized Lorenz system family, which serves as a bridge connecting the Lorenz and Chen systems.

	\begin{equation}
		\begin{cases}
			\dot{x} = a(y - x) \\
			\dot{y} = -xz + cx \\
			\dot{z} = xy - bz
		\end{cases}
		\label{eq:lu}
	\end{equation}
	
\begin{figure*}[htbp]  
		\begin{minipage}{0.48\linewidth}
			\centerline{\includegraphics[width=8cm]{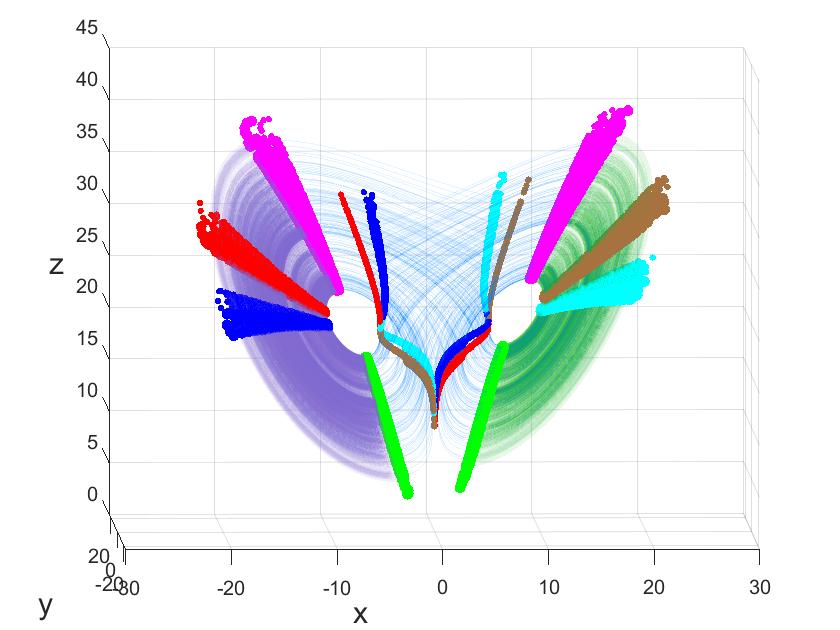}}
			\centerline{(a)}
		\end{minipage}
		\hfill
		\begin{minipage}{0.48\linewidth}
			\centerline{\includegraphics[width=8cm]{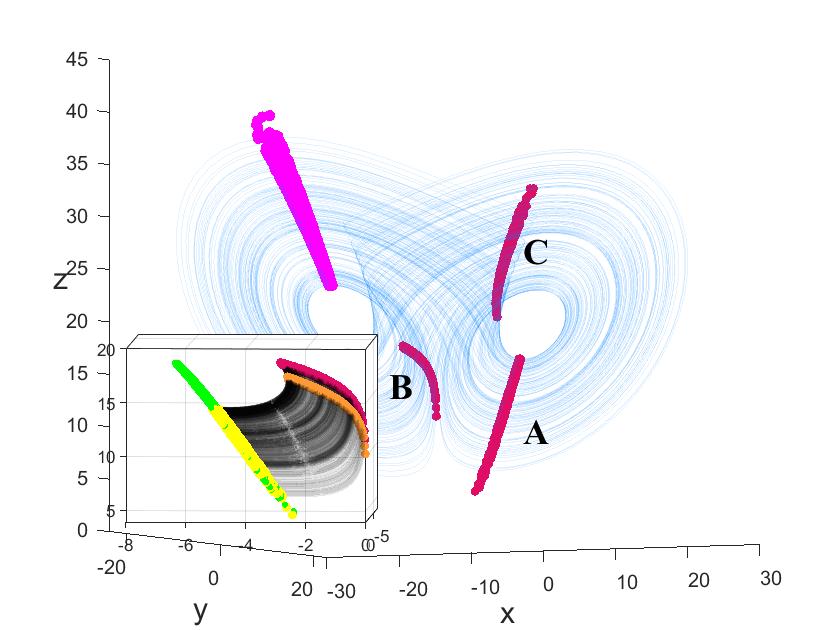}}
			\centerline{(b)}
		\end{minipage}
		\vfill
		\begin{minipage}{0.48\linewidth}
			\centerline{\includegraphics[width=8cm]{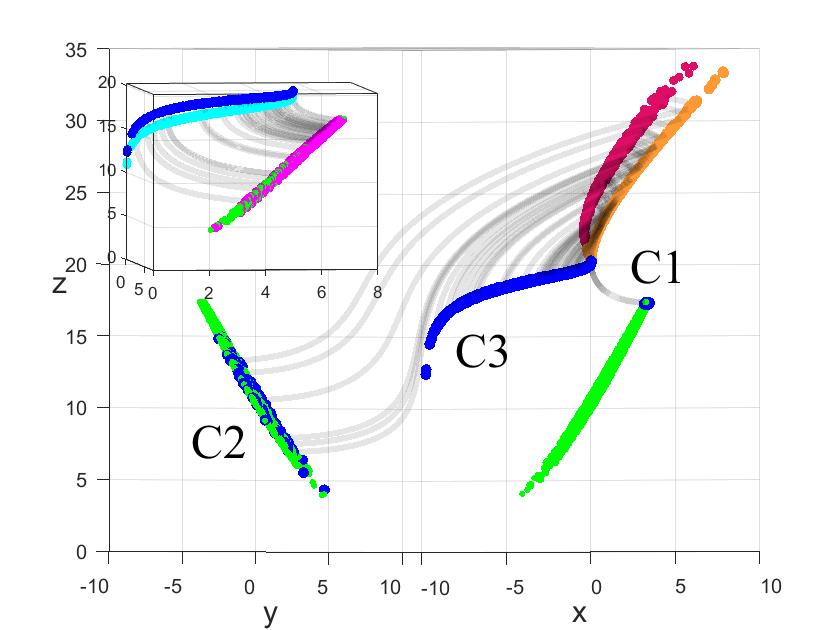}}
			\centerline{(c)}
		\end{minipage}
		\hfill
		\begin{minipage}{0.48\linewidth}
			\centerline{\includegraphics[width=8cm]{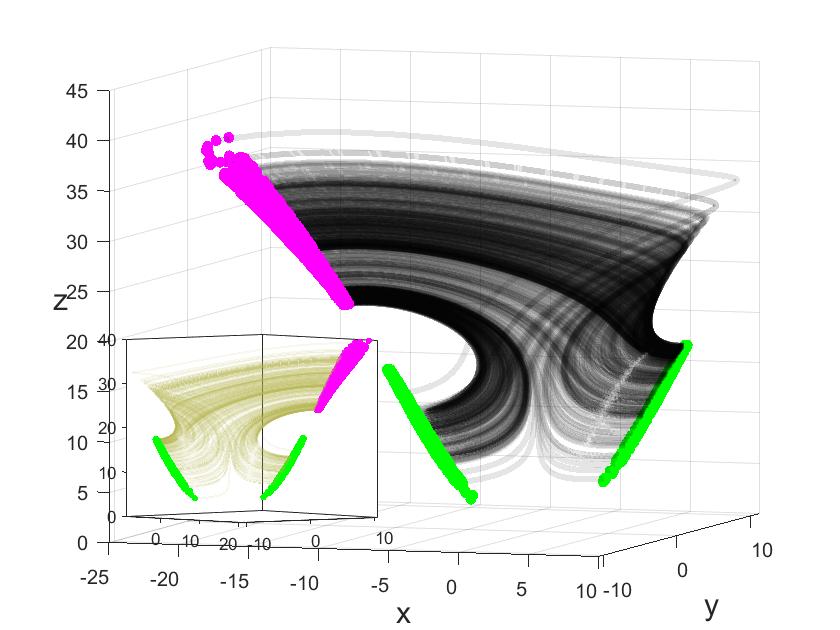}}
			\centerline{(d)}
		\end{minipage}
		\vfill
		\begin{minipage}{0.48\linewidth}			\centerline{\includegraphics[width=8cm]{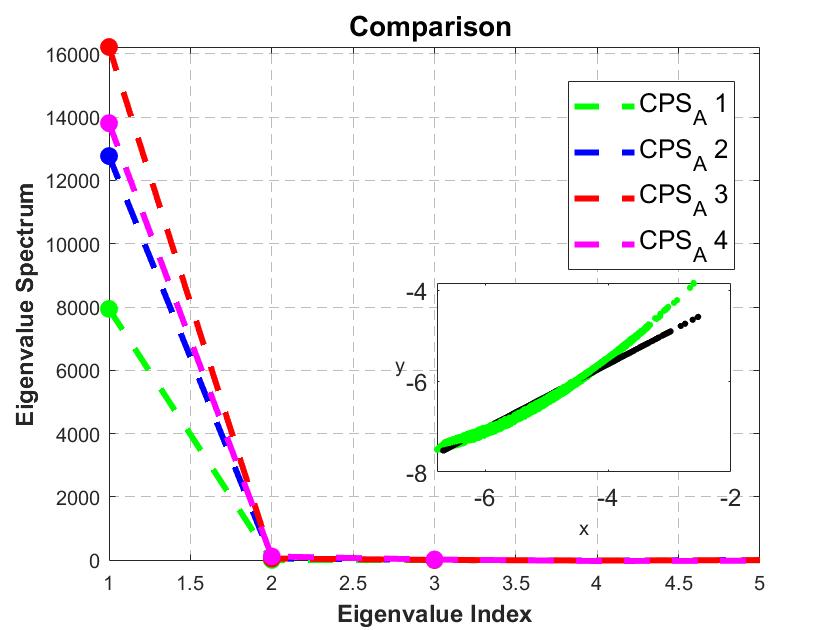}}
			\centerline{(e)}
		\end{minipage}
		\hfill
		\begin{minipage}{0.48\linewidth}
			\centerline{\includegraphics[width=8cm]{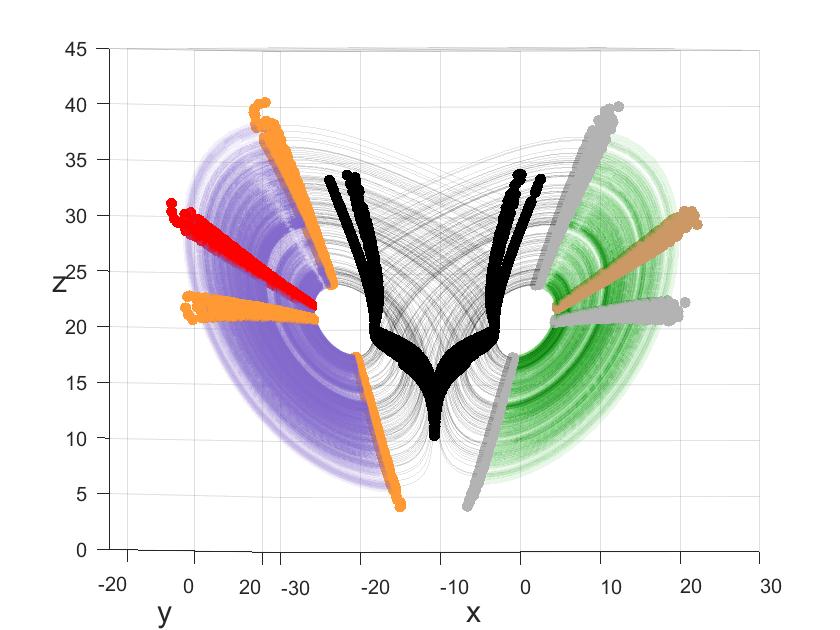}}
			\centerline{(f)}
		\end{minipage}

		\caption{CPS construction and effective Poincaré section screening of the Lü chaotic system via the OP method. (a) Initial CPS division and preliminary scroll judgment; (b) Non-equivalent divergence process of the scroll corresponding to $CPS_{A}$; (c) Further evolutionary process of $CPS_{A}3$; (d) Analysis of divergence and coupling processes between $CPS_{A}$ and $CPS_{B}$; (e) PCA analysis of coarse segments in $CPS_{A}$; (f) Construction of the final Poincaré section by extracting the simplest CPSs from $CPS_{A}$ and $CPS_{B}$} 
		\label{fig-5-30} 
	\end{figure*}

	The typical chaotic parameters are set as 
	a=36, b=3, c=20,
	. The system also generates a dual-scroll attractor, which features a more compact attractor structure, and trajectories stay for a shorter time within each scroll.
	Fig. \ref{fig-5-30}(a) shows the structure of a two-coupled-scroll chaotic attractor. Consistent with the evolutionary parameters of the Lorenz system, this attractor is obtained by evolving a steady-state initial point for 
	800000
	steps with a time step of 
	0.05
	. Similar to the Rössler system, through preliminary classification based on ordinal patterns (OPs), non-monotonic OPs in three different directions are extracted. After distinguishing between local maxima and minima, six initial Candidate Poincaré Sections (CPSs) are obtained, which are marked with different colors in the figure for distinction. Each initial CPS can diverge into at least two distinct clustered regions, so these CPSs are obviously rough-grained.
	In this paper, these coarse CPSs are further divided according to their clustered regions, and a total of 16 refined CPSs are obtained. The analysis of evolutionary order and equivalence process for these 16 CPSs is relatively complex, due to the large number of CPSs and the possible coexistence of coupling and divergence processes. Therefore, it is necessary to pre-judge whether coupling and divergence exist in this region. Consistent with the previous cases, the equivalence process is first judged to determine the number of scrolls, and then coupling process analysis is carried out. To classify these CPSs into scroll categories, the 16 CPSs first need to be sorted by the number of sampling points, and the results are listed in the following table:
	
	\begin{table}[h]%
		
		\centering  
		\caption{Different Types of Poincaré Sections and Their Quantitative Statistics}.

		\scalebox{0.4}[0.4]{
			\resizebox{\textwidth}{!}{%
				\begin{tabular}{ccccc}
					\hline
					
					\hline
					CPS type & \multicolumn{4}{c}{number} \\ 
					\hline
					$CPS_A$ &	$3603$  & $3602$ & $3602$ & 	$3602$ \\
					$CPS_B$ &	$3564$  & $3564$ & $3564$ & 	$3564$\\ 
					$CPS_C$ & $2105$ &-&-&-\\ 
					$CPS_D$  &$2084$&-&-&-\\   
					$CPS_E$  &$2012$&-&-&-\\ 
					$CPS_F$	& 	$2001$ &-&-&-\\ 
					$CPS_G$ & $1021$&-&-&-\\ 
					$CPS_H$  &$1003$&-&-&-\\ 
					$CPS_I$	& 	$938$ &-&-&-\\ 
					$CPS_J$	& 	$910$ &-&-&-\\ 
					\hline

				\end{tabular}
			}
		}
	\end{table}
	
	Based on the above statistical results, all CPSs can be classified into four categories. Among them, $CPS_A$ and $CPS_B$ contain the largest number of CPSs, with four sections for each category, which correspondingly correspond to two independent chaotic scrolls. The remaining eight CPSs belong to different categories with varying sampling point quantities, all of which possess fewer point sets than $CPS_A$ and $CPS_B$.
	Specifically, the four CPSs of $CPS_A$ satisfy linear equivalence throughout the evolutionary process. The equivalent evolutionary region of $CPS_A$ is illustrated by the purple area in Fig. \ref{fig-5-20}(a), with the evolutionary sequence following green, blue, red, and magenta. According to this evolutionary order, the corresponding sections are named $CPS_{A}1$, $CPS_{A}2$, $CPS_{A}3$, and $CPS_{A}4$, respectively.
	Similarly, the four CPSs of $CPS_B$ are mutually linearly equivalent during system evolution. The equivalent region of $CPS_B$ is represented by the green area in Fig. \ref{fig-5-30}(a), with the evolutionary direction following green, cyan, brown, and magenta. In accordance with this sequence, the corresponding CPSs are defined as $CPS_{B}1$, $CPS_{B}2$, $CPS_{B}3$, and $CPS_{B}4$. Consistent with the scroll characteristics of the Lorenz chaotic system, these CPSs form two spatially separated local equivalent evolutionary domains. Therefore, the investigated Lü system is preliminarily identified as a dual-scroll chaotic system with at least two independent scroll structures.

\begin{figure*}[htbp]  
		\begin{minipage}{0.48\linewidth}
			\centerline{\includegraphics[width=8cm]{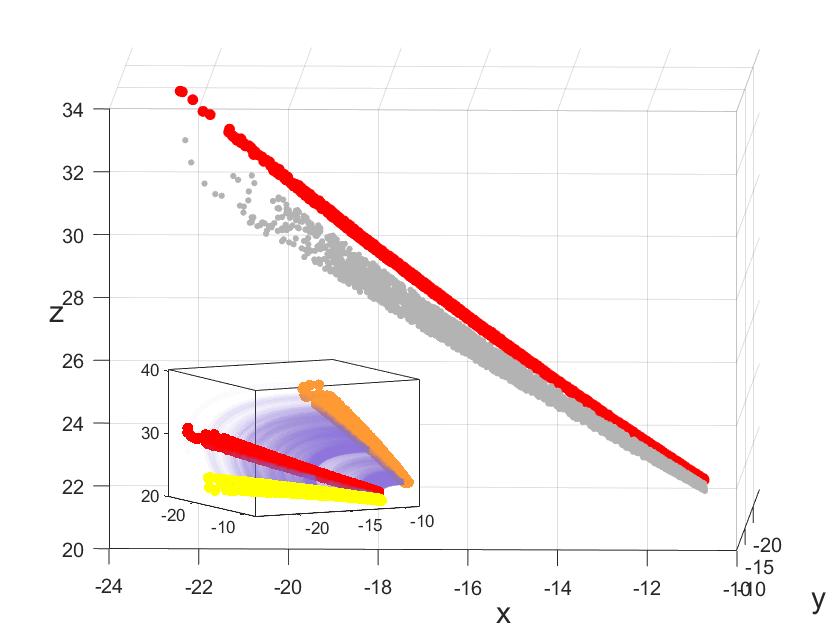}}
			\centerline{(a)}
		\end{minipage}
		\hfill
		\begin{minipage}{0.48\linewidth}
			\centerline{\includegraphics[width=8cm]{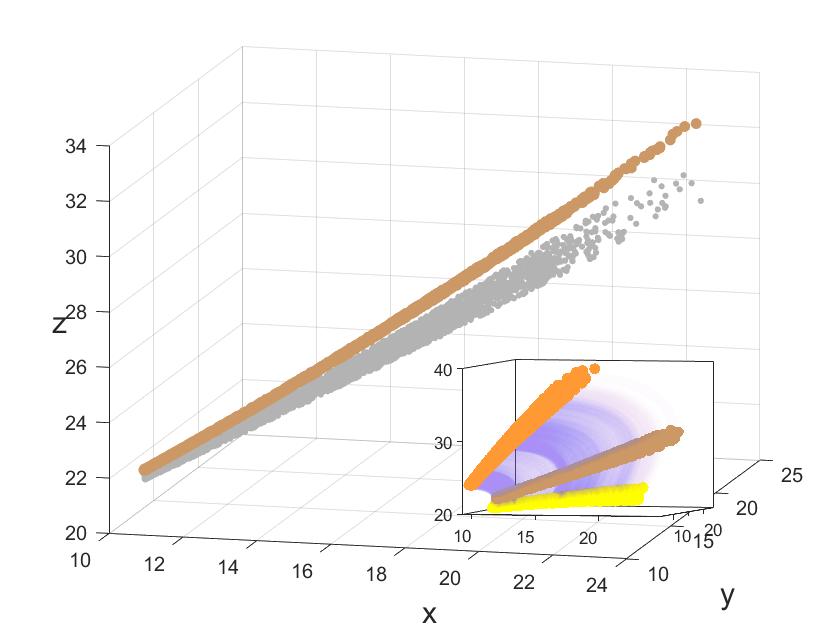}}
			\centerline{(b)}
		\end{minipage}
		\vfill
		\begin{minipage}{0.48\linewidth}
			\centerline{\includegraphics[width=8cm]{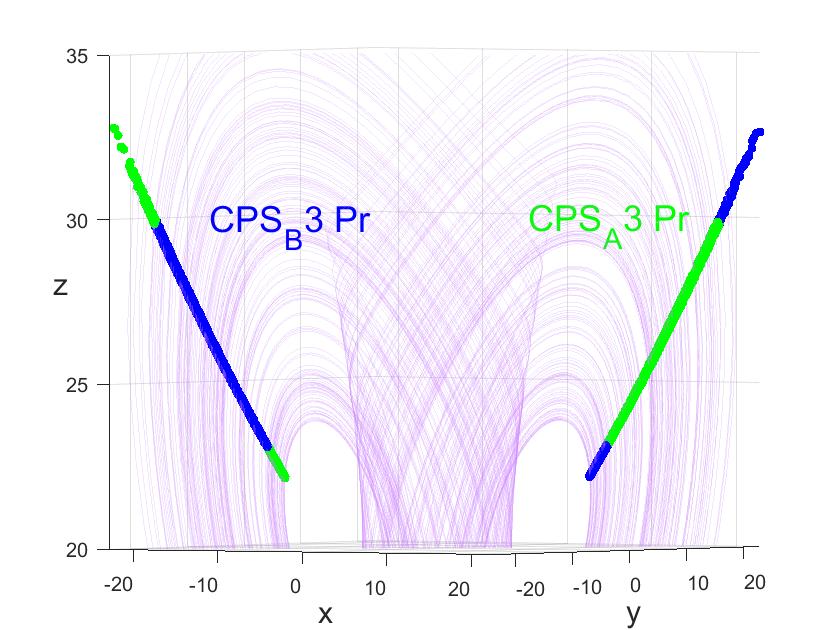}}
			\centerline{(c)}
		\end{minipage}
		\hfill
		\begin{minipage}{0.48\linewidth}
			\centerline{\includegraphics[width=8cm]{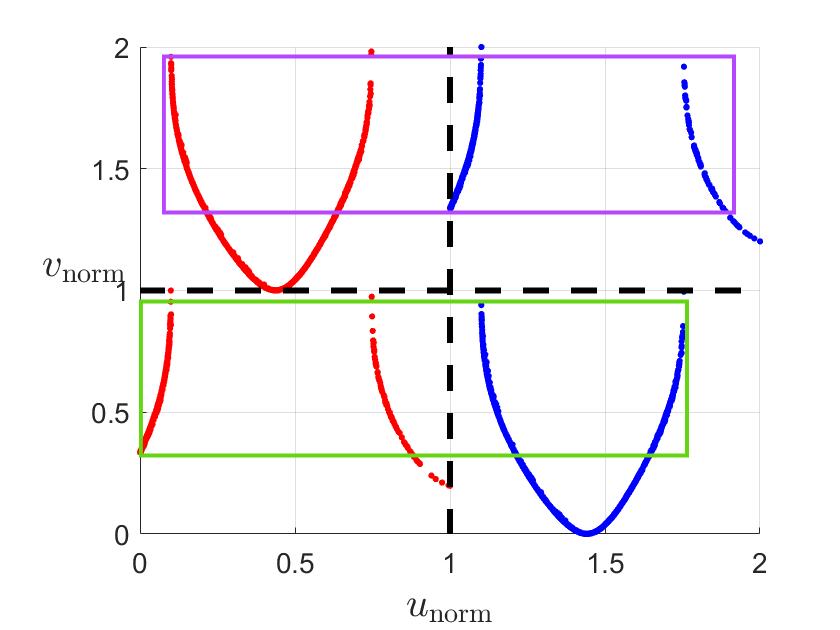}}
			\centerline{(d)}
		\end{minipage}
		\vfill
		\begin{minipage}{0.48\linewidth}
			\centerline{\includegraphics[width=8cm]{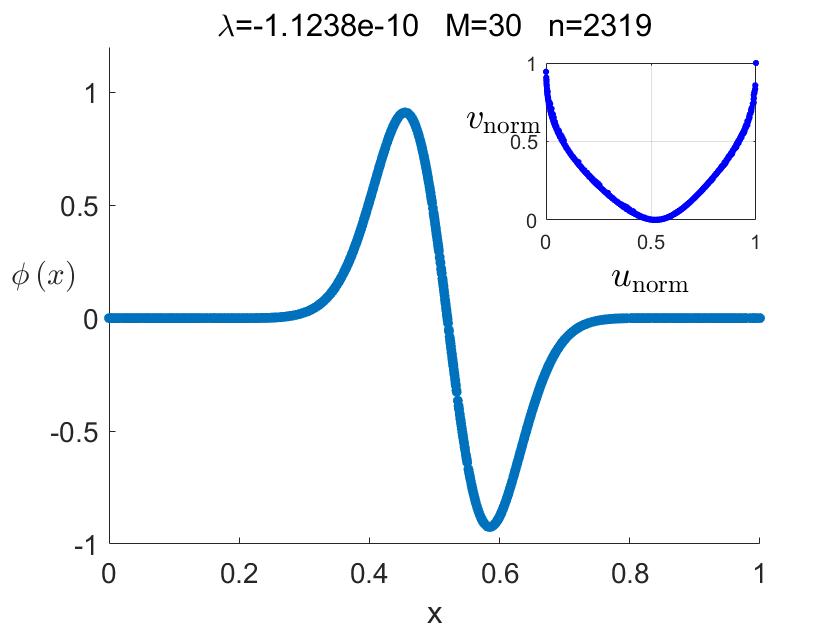}}
			\centerline{(e)}
		\end{minipage}
		\hfill
		\begin{minipage}{0.48\linewidth}
			\centerline{\includegraphics[width=8cm]{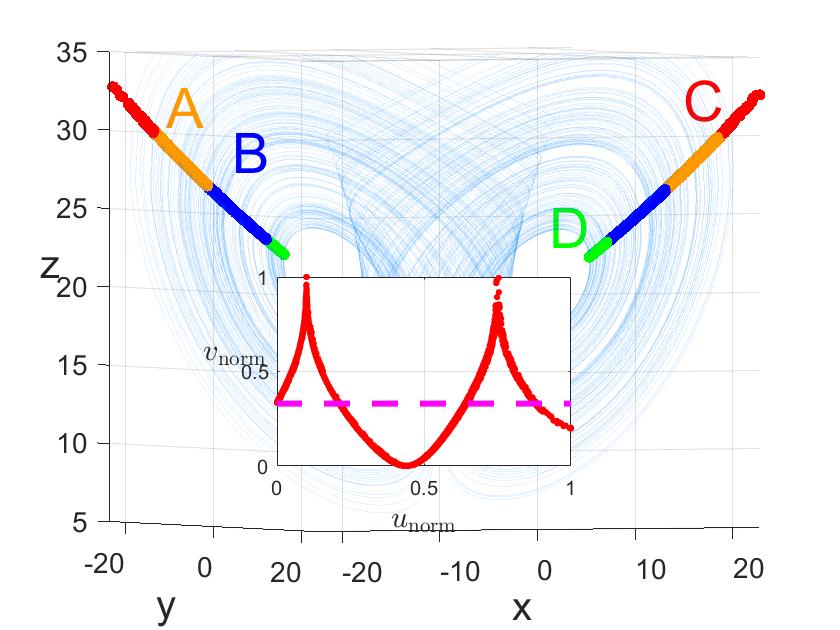}}
			\centerline{(f)}
		\end{minipage}
		
		\caption{Backward evolutionary decoupling of two components of effective sections, low-dimensional region construction, and generalized coordinate establishment for symbolic partitioning. (a) Structural refinement of $CPS_{A}3$ via OP interpolation; (b) Structural refinement of $CPS_{B}3$ via OP interpolation; (c) Pre-image fragments of $CPS_{A}3$ and $CPS_{B}3$ after backward evolutionary decoupling; (d) Symbolic partition boundary determination via generalized coordinate analysis of the final Poincaré section; (e) Symbolic partitioning of the longest fragment of refined $CPS_{A}3\text{Pr}$ based on GKA; (f) Symmetry analysis of $CPS_{A}3$ and $CPS_{B}3$, as well as generalized coordinate simplification and symbolic partitioning of Poincaré section mapping.
		} 
		\label{fig-5-31} 
	\end{figure*}

	The remaining CPSs exhibit distinct sampling point quantities without equal distribution, indicating that they do not satisfy the equivalence evolutionary criterion. Accordingly, these CPSs are identified as transitional sections involved in the coupling and divergence processes between the two scrolls. To verify the participation of these transitional CPSs in coupling–divergence dynamics, the inlet and outlet sections (i.e., the initial and terminal CPSs) of the two scrolls are determined following the analytical framework of previous cases. According to the multi-scroll theoretical model and prior case analyses, pure dual-scroll coupling inevitably induces divergence behaviors. Specifically, if the terminal CPS of each scroll can evolve into the initial CPSs of both scrolls, the initial section of either scroll originates from the terminal sections of two different scrolls.
	By analyzing the divergence dynamics of individual scrolls, all remaining transitional CPSs are proven to be invalid. For intuitive analysis and presentation, the eight transitional CPSs in Fig. \ref{fig-5-30}(a) are matched with the categorized labels in the statistical table.In the upper $z$-direction region of the figure, the red, blue, cyan, and brown CPSs correspond to $CPS_{J}$, $CPS_{H}$, $CPS_{G}$, and $CPS_{I}$, respectively. Meanwhile, the brown, cyan, blue, and red CPSs in the lower $z$-direction region are defined as $CPS_{E}$, $CPS_{C}$, $CPS_{D}$, and $CPS_{F}$, respectively.
	The divergence process is analyzed starting from the terminal section $CPS_{A}4$ of the purple scroll region. One-step evolution of $CPS_{A}4$ yields three distinct CPS components, as illustrated by the three thick red segments in Fig. \ref{fig-5-30}(b). Segment A constitutes a local component of $CPS_{B}1$, demonstrating that partial trajectories of the $CPS_{A}$ scroll diverge into the $CPS_{B}$ scroll. Segment B belongs to the transitional region of $CPS_{C}$. Further evolution of this transitional component, as shown in the subgraph, reveals that it evolves into $CPS_{E}$ in one step and subsequently into $CPS_{A}1$ in the next step, verifying that the divergent trajectories of the $CPS_{A}$ scroll can return to its own initial region.
	Segment C located in the $CPS_{G}$ region presents more complex dynamics and requires further evolutionary analysis, as depicted in Fig. \ref{fig-5-30}(c). After one-step evolution, Segment C evolves into $CPS_{I}$ (orange segment) and further diverges into three new thick segments C1, C2, and C3. Among them, C1 corresponds to $CPS_{B}1$ and C2 corresponds to $CPS_{A}1$, while C3 remains in the transitional region of $CPS_{D}$ and requires continuous evolution. As shown in the subgraph, C3 evolves into $CPS_{F}$ in one step and finally converges to $CPS_{A}1$ after another step of evolution. Although the divergence process of the Lü system is more intricate than that of the Lorenz system, it still conforms to the fundamental dual-scroll coupling mechanism. After eliminating all transitional invalid CPSs, the simplified one-step divergence diagram is obtained, as displayed in Fig. \ref{fig-5-30}(d).
	Similarly, backward evolution is performed on the terminal section $CPS_{B}4$ of the green scroll region. Excluding transitional regions, $CPS_{B}4$ can diverge into two initial sections $CPS_{A}1$ and $CPS_{B}1$ via one-step evolution. Combining the main graph and subgraph results, $CPS_{A}1$ is verified to be jointly coupled by the terminal sections $CPS_{A}4$ and $CPS_{B}4$ of the two scrolls. Therefore, the final effective Poincaré section is composed of valid CPSs extracted from both scroll regions.
	Since internal evolution within each individual scroll satisfies the equivalence principle, CPS screening is essential to achieve the simplest structural configuration of the final Poincaré section. CPS screening is implemented separately for the two scroll regions, with the $CPS_{A}$ group taken as an example. Isomap analysis is first conducted to obtain the eigenvalue distribution of the geodesic distance matrix $D_G$. The Isomap eigenvalue decay curves are plotted in Fig. \ref{fig-5-30}(e). The maximum eigenvalue of $CPS_{A}1$ is 7940, which is significantly lower than those of the other three $CPS_{A}$ components. This result is opposite to the characteristics of the Rössler and Lorenz systems, indicating that the other three CPSs undergo obvious structural stretching, while $CPS_{A}1$ possesses a distinct local structural morphology.
	PCA dimensionality reduction is further adopted for structural evaluation. For $CPS_{B}1$, structural reconstruction based solely on the first principal component exhibits poor fitting performance with obvious bending deformation in the original section, as indicated by the black segment in the subgraph. In contrast, the other three $CPS_{A}$ sections achieve high-precision reconstruction via single-principal-component fitting. Among these three valid sections, $CPS_{A}3$ with the highest first principal component contribution rate is selected as the optimal component, marked by the red section in Fig. \ref{fig-5-30}(f), while the remaining equivalent CPSs in the $CPS_{A}$ group are colored in orange.
	The same screening procedure is applied to the green scroll region, and $CPS_{B}3$ with the maximum PCA principal component contribution rate is determined as the optimal component (brown section in the figure), with other equivalent $CPS_{B}$ sections marked in gray and transitional invalid CPSs marked in black.
	After completing CPS classification and screening, return mapping analysis is performed on the optimized final Poincaré sections via backward evolution, which captures evolutionary fragments derived from different sectional segments. Nevertheless, the two optimized coarse segments exhibit excessive width, leading to low sectional precision. To address this issue, OP interpolation refinement is implemented prior to return mapping analysis. For $CPS_{A}3$, interpolation endpoints are selected from its adjacent equivalent sections $CPS_{A}2$ and $CPS_{A}4$, as illustrated in the subgraph of Fig. \ref{fig-5-31}(a). Extreme points are extracted to optimize the sectional structure, yielding the refined red section in the main graph, which presents a nearly linear morphology and is closer to the ideal Poincaré section compared with the original gray segment.
	Similarly, $CPS_{B}3$ is refined by interpolating the endpoints of its neighboring sections $CPS_{B}2$ and $CPS_{B}4$. Extreme point re-extraction and structural optimization are performed, as shown in the subgraph of Fig. \ref{fig-5-31}(b), and the refined brown section is obtained in the main graph.
	
Previous analyses demonstrate that the divergence of scroll-terminal CPSs and their subsequent re-entry into distinct scrolls inevitably induces coupling interactions between the two scroll regions. Following the decoupling framework established for the Lorenz chaotic system, identical decoupling analysis is performed on the dual-segment Poincaré section of the Lü chaotic system. First, decoupling is implemented for $CPS_{A}3$, and the evolved pre-image segment is denoted as $CPS_{A}3\text{Pr}$, which is visualized as the green region in Fig. \ref{fig-5-31}(c). The decoupling process yields three distinct linear fragments, among which one fragment locates on $CPS_{B}3$ and the other two belong to $CPS_{A}3$; the two fragments on $CPS_{A}3$ can be regarded as a single linear segment with a sparse middle region. Similarly, decoupling of $CPS_{B}3$ produces the evolved pre-image segment $CPS_{B}3\text{Pr}$, represented by the blue region in the figure. Three analogous fragments are obtained, with one fragment distributed on $CPS_{A}3$ and the remaining two on $CPS_{B}3$. The pre-image evolutionary fragments are clearly distributed on the Poincaré section without spatial overlap or frequent cross-distribution, providing qualified regional conditions for generalized coordinate construction.
Prior to generalized coordinate establishment, PCA dimensionality reduction is separately performed on $CPS_{A}3$ and $CPS_{B}3$ to construct low-dimensional equivalent line segments, adopting the same correspondence strategy utilized for the Lorenz system. Notably, PCA processing of $CPS_{A}3$ reveals a monotonically decreasing correspondence between the dominant direction of the original high-dimensional region and the low-dimensional projected segment. To guarantee consistent monotonicity, the low-dimensional sequence is inverted and subsequently normalized, achieving the same regional correspondence effect as the two subgraphs in Fig. \ref{fig-5-18}(a). The two low-dimensional segments are normalized to the intervals $[0,1]$ and $[1,2]$ respectively, forming a unified long segment and establishing the generalized coordinate system spanning $[0,2]$.
Based on the pre-evolutionary generalized coordinates, all discrete points of $CPS_{A}3\text{Pr}$ and $CPS_{B}3\text{Pr}$ as well as their one-step evolutionary points are projected onto the unified coordinate system. The mapping relationship of $CPS_{A}3\text{Pr}$ and its evolutionary points on the $[0,1]$ interval is constructed, as illustrated in Fig. \ref{fig-5-31}(d). Four independent symbolic subregions exist within this interval, marked by green rectangular frames in the figure, corresponding to four distinct forward local evolutionary domains and thus requiring four independent symbolic states for classification. Similarly, the mapping of $CPS_{B}3\text{Pr}$ and its evolutionary points on the $[1,2]$ interval presents four local folding-to-unification evolutionary processes, marked by blue-green rectangular frames, which also necessitates four distinct symbolic classifications.
After determining the total number of symbolic states and initial partition boundaries, two local folded evolutionary mappings are identified in the domains $[1,2]\times[0,1]$ and $[0,1]\times[1,2]$. Although the two terminals of each mapping correspond to different symbolic states, these local mappings exhibit continuous stretching-folding dynamics with inherent critical points. Therefore, generalized Koopman analysis (GKA) is adopted for both domains to acquire accurate symbolic partition boundaries. Taking the mapping in the $[1,2]\times[0,1]$ domain as an example, affine transformation is implemented via bidirectional normalization of pre- and post-evolutionary data, completing linear transformation including stretching, compression, and translation. The processed mapping result is displayed in the subgraph of Fig. \ref{fig-5-31}(e).
Koopman matrix construction is performed on the low-dimensional local segment using Gaussian basis functions with equally spaced sampling. The number of basis functions starts at $M=10$ and increases incrementally by 10. Valid leading eigenvalues (VLEZ) are obtained when $M=30$, as shown in the main graph. The zero-crossing point of the eigenvalue curve is determined as the optimal symbolic partition boundary. The unimodal forward low-dimensional segment in the $[1,2]$ interval corresponds to the long green segment on $CPS_{B}3$ in Fig. \ref{fig-5-31}(c). Likewise, GKA is applied to the local unimodal mapping in the $[0,1]\times[1,2]$ domain to identify the symbolic boundary for the long blue segment on $CPS_{A}3$ in Fig. \ref{fig-5-31}(c). In contrast, the mappings in $[0,1]\times[0,1]$ and $[1,2]\times[1,2]$ exhibit discontinuous intermediate regions, where symbolic classification can be simply realized by assigning different symbolic states on both sides of the discontinuity.
Synthesizing the above analytical results, a total of eight independent symbolic regions are obtained, with four symbolic partitions allocated for each of $CPS_{A}3\text{Pr}$ and $CPS_{B}3\text{Pr}$. The two sectional structures and their evolutionary dynamics are perfectly symmetric, enabling the fusion of the left and right sections. In the low-dimensional projection space, the $[1,2]$ interval is fully translated and superimposed onto the $[0,1]$ interval, leading to complete coincidence of corresponding sectional fragments, as demonstrated in the subgraph of Fig. \ref{fig-5-31}(f). The overlapping correspondence is significantly optimized: the region above the magenta dashed line contains four overlapping subregions for four-symbol allocation. For the discontinuous intermediate region below the dashed line, GKA is further performed to determine accurate symbolic partition boundaries. Through symmetric fusion, the original eight symbolic states are simplified to four, as visualized in the main graph of Fig. \ref{fig-5-31}(f).
Although the Lü chaotic system investigated in this work belongs to a high-dimensional chaotic case, the effective low-dimensional region of its evolved sectional domain is one-dimensional. Nevertheless, the effective low-dimensional regions of Poincaré sections for more complex chaotic attractors can present two-dimensional or even higher-dimensional characteristics. Accordingly, the proposed OP construction method is further extended and applied to chaotic systems with more intricate structural and dynamical features in subsequent analyses.

	\subsection{Chen system}
	
\begin{figure*}[htbp]  
		\begin{minipage}{0.48\linewidth}
			\centerline{\includegraphics[width=8cm]{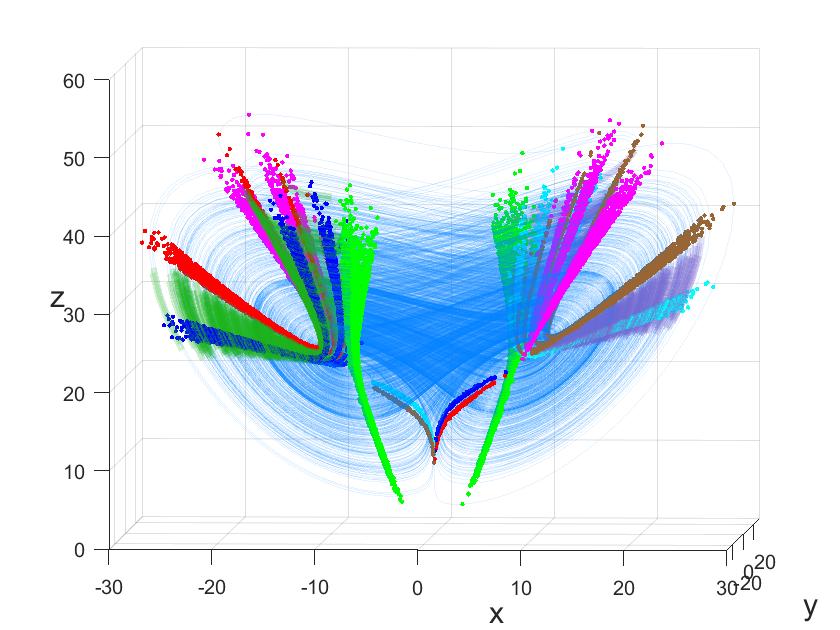}}
			\centerline{(a)}
		\end{minipage}
		\hfill
		\begin{minipage}{0.48\linewidth}
			\centerline{\includegraphics[width=8cm]{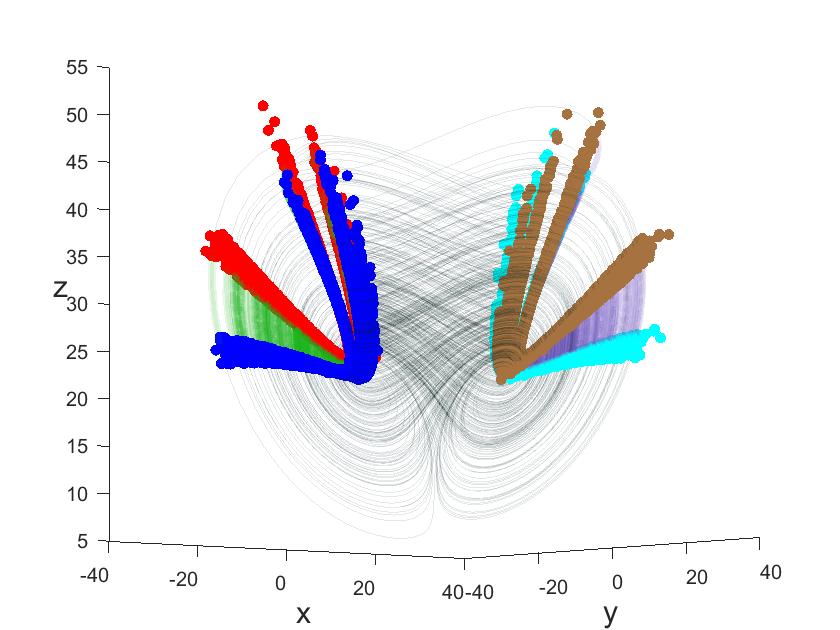}}
			\centerline{(b)}
		\end{minipage}
		\vfill
		\begin{minipage}{0.48\linewidth}
			\centerline{\includegraphics[width=8cm]{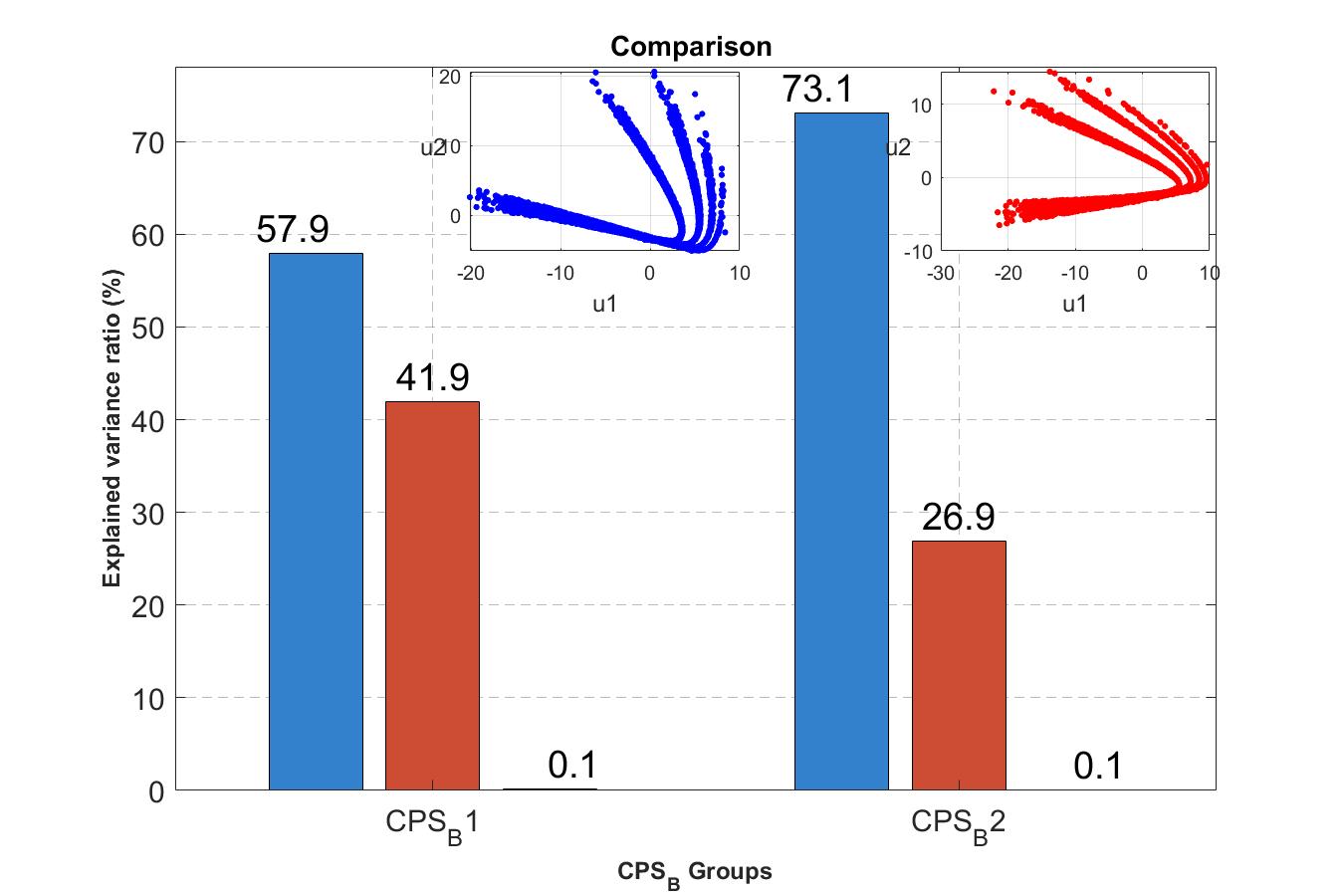}}
			\centerline{(c)}
		\end{minipage}
		\hfill
		\begin{minipage}{0.48\linewidth}
			\centerline{\includegraphics[width=8cm]{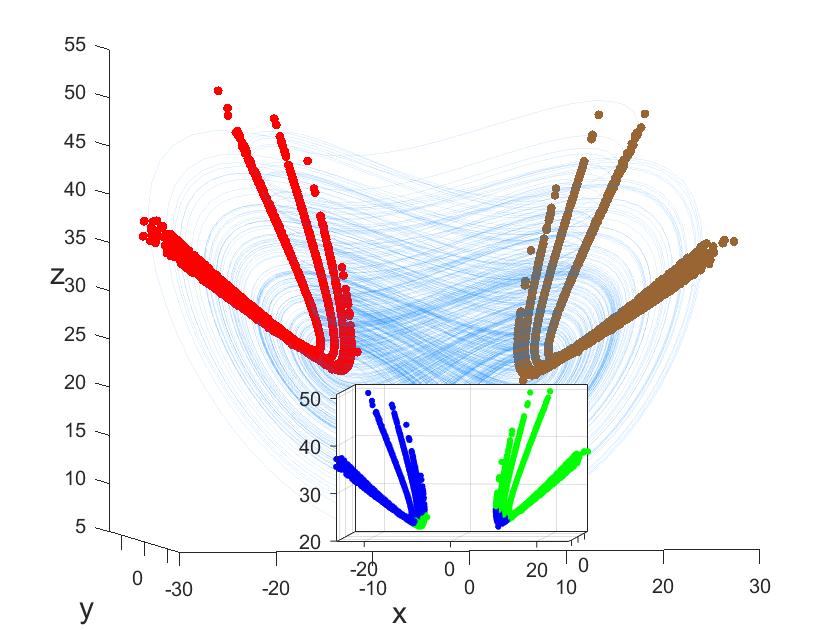}}
			\centerline{(d)}
		\end{minipage}
		
		\caption{CPS construction and effective Poincaré section screening of the Chen chaotic system via the OP method. (a) Initial CPS division and preliminary scroll judgment; (b) Presentation of four valid CPSs in $CPS_{A}$ and $CPS_{B}$; (e) PCA analysis of two sectional segments in $CPS_{A}$; (f) Construction of the final Poincaré section and its backward evolutionary fragments by extracting the optimal simplest CPSs from $CPS_{A}$ and $CPS_{B}$.} 
		\label{fig-5-33} 
	\end{figure*}

Both the Lorenz and Lü systems yield well-structured, linear-segment Poincaré sections with simple topological rules. However, Poincaré sections constructed for more complex chaotic systems exhibit high-dimensional and intricate morphological features. Such complex sectional structures can also emerge in typical three-dimensional chaotic systems. For instance, the Poincaré section of the Duffing oscillator investigated in Chapter 4 presents sophisticated fractal characteristics, which is constructed based on the system oscillation period. To generalize the proposed OP method for sectional construction of universal complex three-dimensional chaotic systems, this section further investigates the typical and highly complex Chen chaotic system.
Proposed by Chen Guanrong in 1999\cite{1999YET}, the Chen system is a three-dimensional autonomous chaotic system and serves as one of the three representative systems of the generalized Lorenz system family, together with the Lorenz and Lü systems. It is the first class of chaotic systems constructed via anti-control strategy, which is transformed from the original Lorenz system through state feedback modification. The dynamical equations of the Chen system are given as follows:
	\begin{equation}
		\begin{cases}
			\dot{x} = a(y - x) \\
			\dot{y} = (c - a)x - xz + cy \\
			\dot{z} = xy - bz
		\end{cases}
		\label{eq:chen}
	\end{equation}
	Chaos is generated when the system parameters are set to 
	$a=35,\; b=3,\; c=28$
	. The core characteristics of the attractor include stronger positive feedback, more complex topological structure, and a more compact and twisted attractor morphology.
	Fig. \ref{fig-5-33}(a) shows the structure of the Chen chaotic attractor. Consistent with the sampling time and frequency adopted for the Lü system, this attractor is obtained by evolving a steady-state initial point for 
	800000
	steps with a time step of 
	0.05
	. Similar to the aforementioned continuous chaotic systems, through preliminary classification based on ordinal patterns (OPs), non-monotonic OPs in three different directions are extracted. After distinguishing between local maxima and minima, six initial Candidate Poincaré Sections (CPSs) are obtained, which are marked with different colors in the figure for distinction. Each initial CPS can diverge into at least two distinct clustered regions, so these CPSs are obviously rough-grained.
	In this paper, these coarse CPSs are further divided according to their clustered regions, and a total of 12 refined CPSs are obtained. For multi-scroll chaotic systems, the determination of evolutionary order and equivalence process analysis for these 12 CPSs is relatively complex, due to the existence of coupling processes. Consistent with the previous cases, the equivalence process is first judged to determine the number of scrolls, and then coupling process analysis is carried out. To classify these CPSs into scroll categories, the 12 CPSs first need to be sorted by the number of sampling points, and the results are listed in the following table:
	\begin{table}[h]%
		
		\centering  
		\caption{Different Types of Poincaré Sections and Their Quantitative Statistics}.

		\scalebox{0.3}[0.3]{
			\resizebox{\textwidth}{!}{%
				\begin{tabular}{ccc}
					\hline
					
					\hline
					CPS type & \multicolumn{2}{c}{number} \\ 
					\hline
					$CPS_A$ &	$7105$  & $7105$\\
					$CPS_B$ &	$7094$  & $7094$\\ 
					$CPS_C$ & $6802$ & $6802$\\ 
					$CPS_D$  & $6788$ & $6788$ \\  
					$CPS_E$  & $359$ & $359$\\ 
					$CPS_F$	& $349$ & $349$ \\
					\hline

				\end{tabular}
			}
		}
	\end{table}
	
Only two types of CPSs with the maximum magnitude are identified in the Chen system, among which $CPS_A$ corresponds to the local purple scroll region. The evolutionary domain formed by these two $CPS_A$ segments is extremely narrow, which stems from the prominent nonlinear coupling and folding characteristics of the system. Its evolutionary sequence follows brown to cyan, and the two segments are denoted as $CPS_{A}1$ and $CPS_{A}2$, respectively. Similarly, $CPS_B$ corresponds to the local green scroll region with a narrow evolutionary domain, and its evolutionary sequence follows blue to red, defined as $CPS_{B}1$ and $CPS_{B}2$.
$CPS_C$ corresponds to the green and magenta sectional segments adjacent to the $CPS_A$ side, labeled $CPS_{C}1$ and $CPS_{C}2$, while $CPS_D$ covers the green and magenta sections near the $CPS_B$ side, denoted as $CPS_{D}1$ and $CPS_{D}2$. The red and blue sections in the middle domain belong to the $CPS_E$ type, namely $CPS_{E}1$ and $CPS_{E}2$, and the cyan and brown middle sections are classified as the $CPS_F$ type, including $CPS_{F}1$ and $CPS_{F}2$.
Although the magnitudes of $CPS_C$ and $CPS_D$ are second only to those of $CPS_A$ and $CPS_B$, they are verified as invalid transitional CPSs located in the coupling region between the two scrolls. For instance, $CPS_{D}1$ and $CPS_{F}2$ couple with $CPS_{B}1$ after one-step evolution. All other transitional CPSs can diverge and eventually enter the two local scroll regions of $CPS_A$ and $CPS_B$ after multi-step evolution.
	
	Only four valid CPSs are presented in Fig. \ref{fig-5-33}(b). Given the inherent coupling behavior between the two scrolls, a unique optimal CPS is extracted from each scroll to construct the final Poincaré section. For complex two-dimensional sectional structures, the establishment of generalized coordinates and linear equivalence judgment is considerably challenging. Nevertheless, if the evolution of one CPS fully converges to another CPS without additional coupling or divergence, the evolutionary correspondence between the two CPSs can be defined as an equivalence process. In this case, Isomap and PCA are not mandatory for low-dimensional linear segment construction; however, PCA analysis is still performed to evaluate the dimensionality and morphological characteristics of the low-dimensional structure.
	First, PCA analysis is conducted on $CPS_{B}1$ and $CPS_{B}2$, and the corresponding principal component contribution rates are illustrated in Fig. \ref{fig-5-33}(c). The results demonstrate that $CPS_{B}2$ possesses a higher first principal component contribution rate than $CPS_{B}1$. The two subgraphs located directly above the bar charts present the low-dimensional corresponding regions of $CPS_{B}1$ and $CPS_{B}2$, respectively. The low-dimensional region of $CPS_{B}2$ (marked in red) exhibits a narrower bandwidth and a coarser linear morphology, indicating superior PCA dimensionality reduction performance. Accordingly, $CPS_{B}2$ is selected as one component of the final Poincaré section. Similarly, PCA analysis is performed on $CPS_{A}1$ and $CPS_{A}2$, and $CPS_{A}2$ is determined as the other optimal component of the final sectional structure.
	Based on the selected optimal components, the final Poincaré section is constructed, as shown in Fig. \ref{fig-5-33}(d). Consistent with the analytical procedure for the Lü system, backward evolution is implemented to obtain the decoupled pre-image sections $CPS_{A}2\text{Pr}$ and $CPS_{B}2\text{Pr}$, which are visualized in the subgraph. The green two-dimensional fragment denotes $CPS_{B}2\text{Pr}$, while the blue two-dimensional fragment represents $CPS_{A}2\text{Pr}$. Compared with all previous cases, the pre-image regions of the Chen system exhibit significantly more complex topological features, characterized by intrinsic two-dimensional fractal structures and multi-scale insufficient folding behaviors analogous to two-dimensional chaotic mappings. Therefore, Koopman analysis is adopted to realize the preliminary symbolic partition of these intricate sectional regions.
	
	\begin{figure*}[htbp]  
		\begin{minipage}{0.48\linewidth}
			\centerline{\includegraphics[width=8cm]{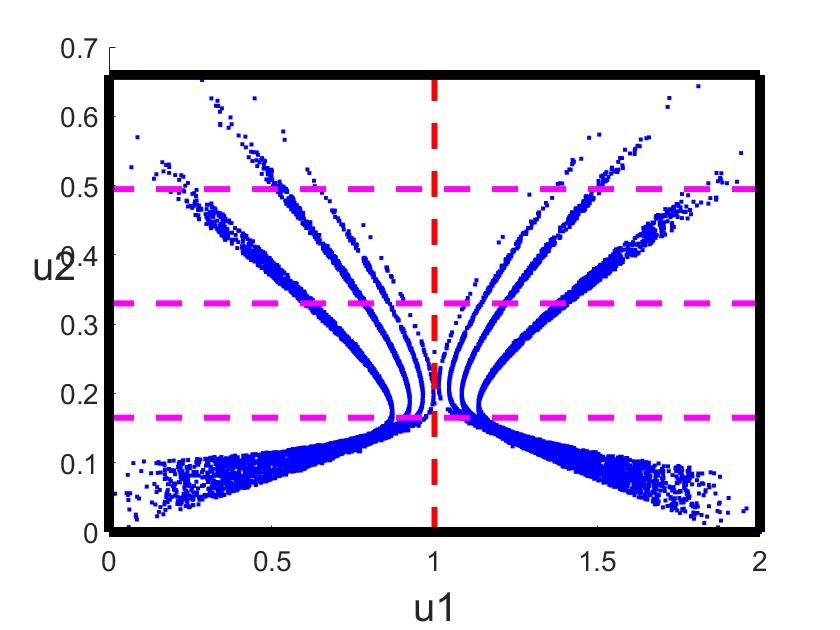}}
			\centerline{(a)}
		\end{minipage}
		\hfill
		\begin{minipage}{0.48\linewidth}
			\centerline{\includegraphics[width=8cm]{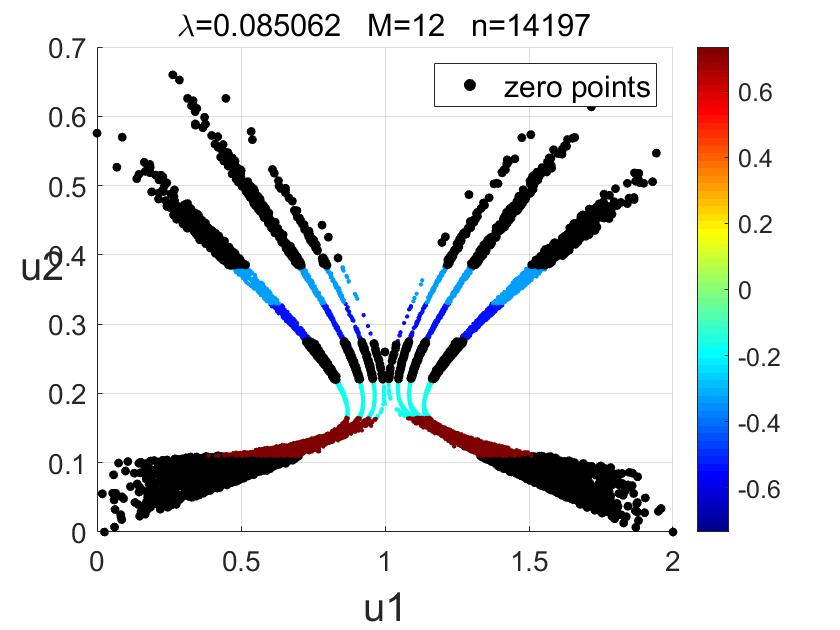}}
			\centerline{(b)}
		\end{minipage}
		\vfill
		\begin{minipage}{0.48\linewidth}
			\centerline{\includegraphics[width=8cm]{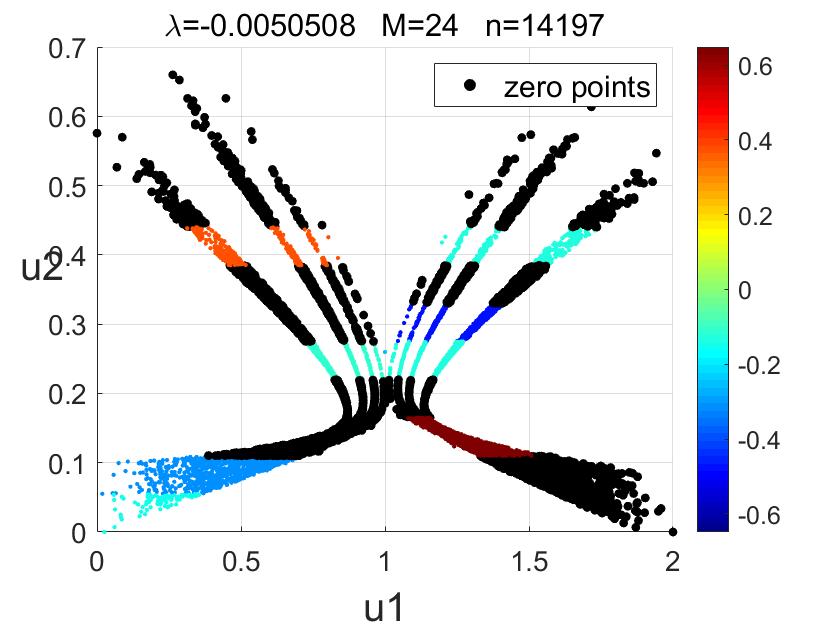}}
			\centerline{(c)}
		\end{minipage}
		\hfill
		\begin{minipage}{0.48\linewidth}
			\centerline{\includegraphics[width=8cm]{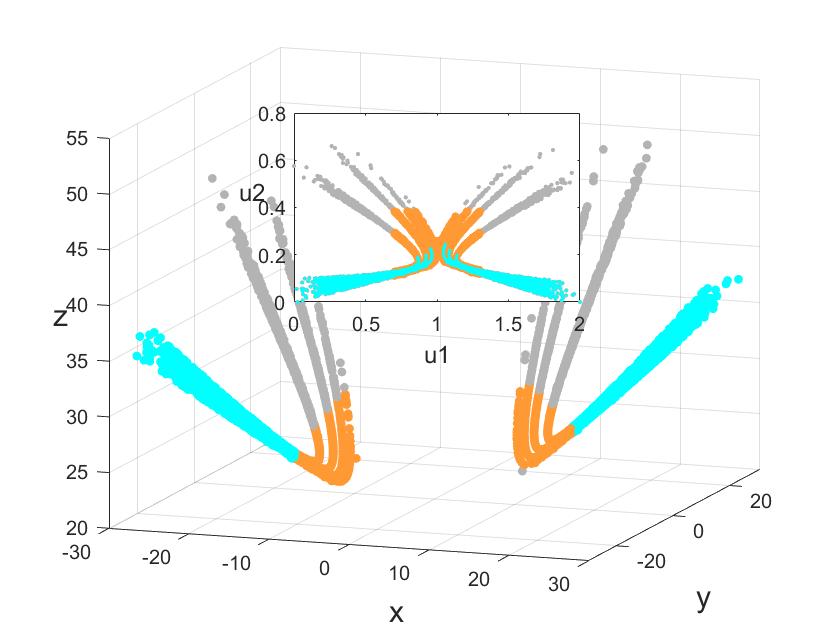}}
			\centerline{(d)}
		\end{minipage}
		\vfill
		\begin{minipage}{0.48\linewidth}
			\centerline{\includegraphics[width=8cm]{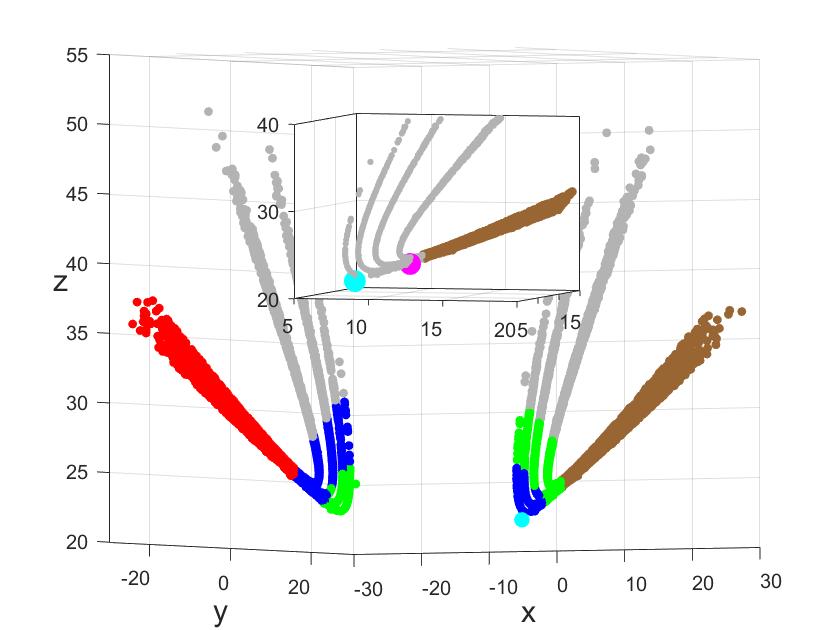}}
			\centerline{(e)}
		\end{minipage}
		\hfill
		\begin{minipage}{0.48\linewidth}
			\centerline{\includegraphics[width=8cm]{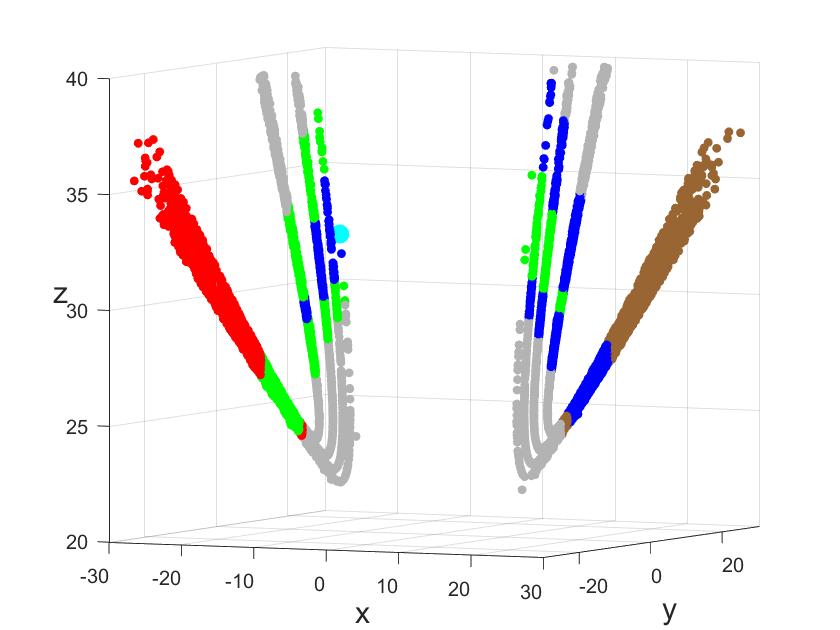}}
			\centerline{(f)}
		\end{minipage}
		\caption{Localization of symbolic partition boundaries for the Chen chaotic system: backward evolutionary decoupling of two effective sectional components, low-dimensional region construction, and generalized coordinate establishment for symbolic partitioning. (a) Low-dimensional normalized generalized coordinates of $CPS_{A}2$ and $CPS_{B}2$ and generalized rectangular window basis preprocessing; (b) Characteristic Lyapunov exponent (CLE) obtained via Koopman analysis under locally symmetric rectangular windows for $CPS_{A}2$ and $CPS_{B}2$; (c) CLE extracted from separated $CPS_{A}2$ and $CPS_{B}2$ via Koopman analysis; (d) Initial symbolic partition boundaries obtained through CLE localization; (e) One-step backward evolution analysis of the initial boundary regions; (f) Two-step forward evolution analysis of $CPS_{A}2\text{M}$ and $CPS_{B}2\text{M}$.
		} 
		\label{fig-5-34} 
	\end{figure*}
	
\begin{figure*}[htbp]  
		\begin{minipage}{0.48\linewidth}
			\centerline{\includegraphics[width=8cm]{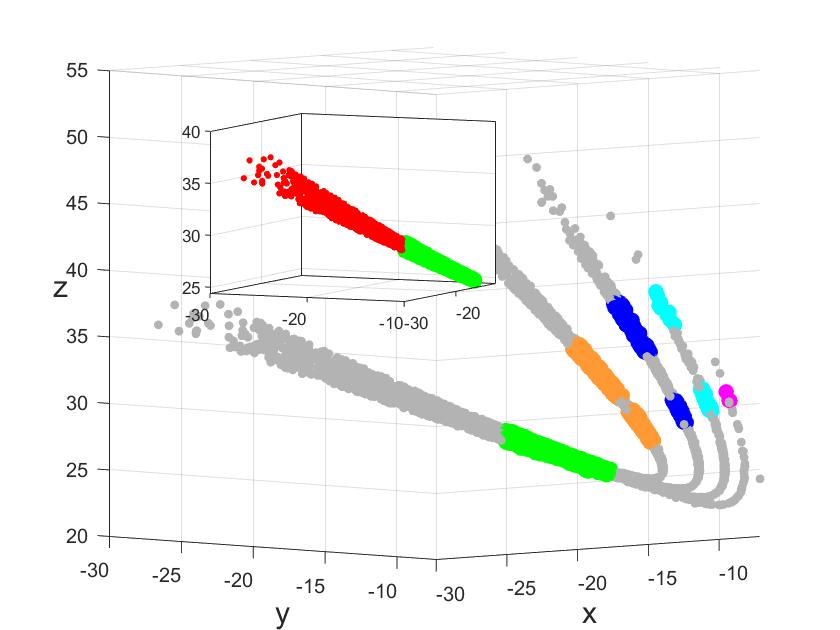}}
			\centerline{(a)}
		\end{minipage}
		\hfill
		\begin{minipage}{0.48\linewidth}
			\centerline{\includegraphics[width=8cm]{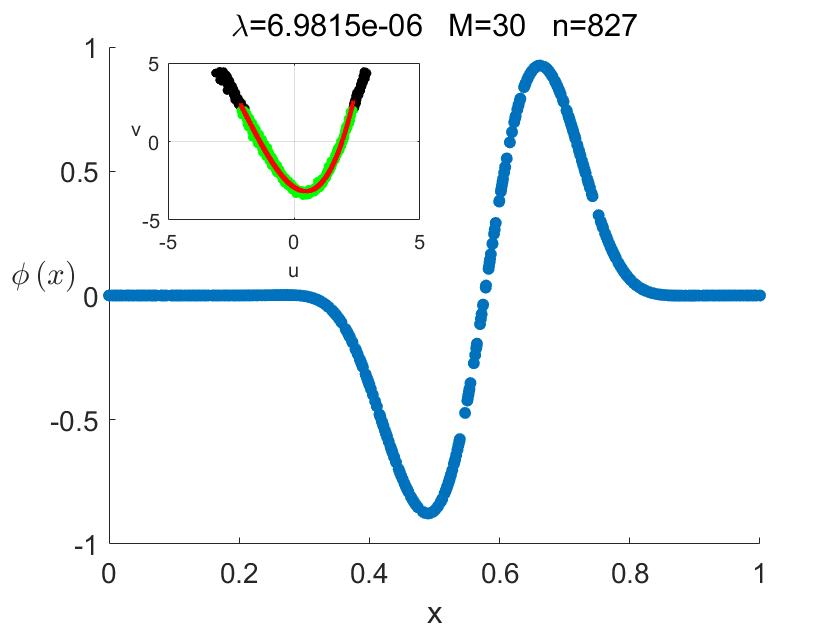}}
			\centerline{(b)}
		\end{minipage}
		\vfill
		\begin{minipage}{0.48\linewidth}
			\centerline{\includegraphics[width=8cm]{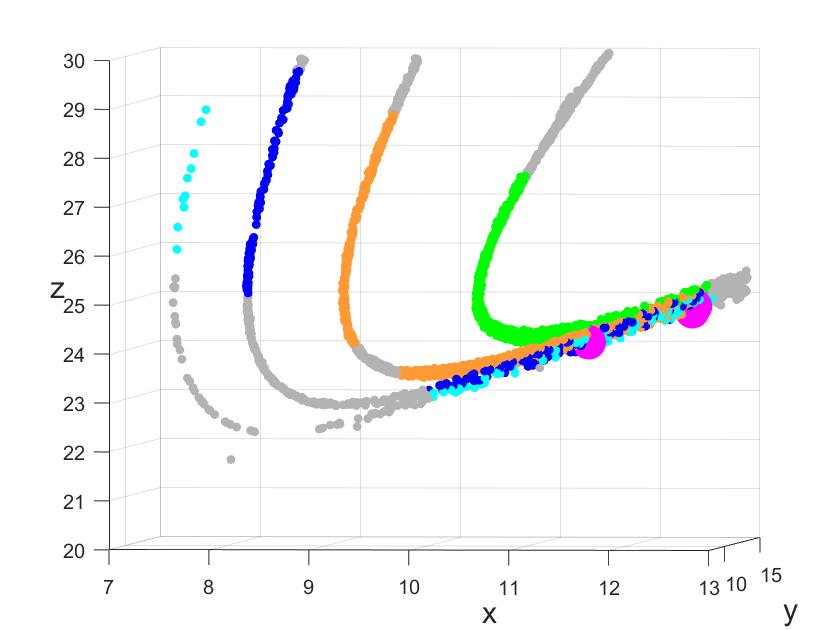}}
			\centerline{(c)}
		\end{minipage}
		\hfill
		\begin{minipage}{0.48\linewidth}
			\centerline{\includegraphics[width=8cm]{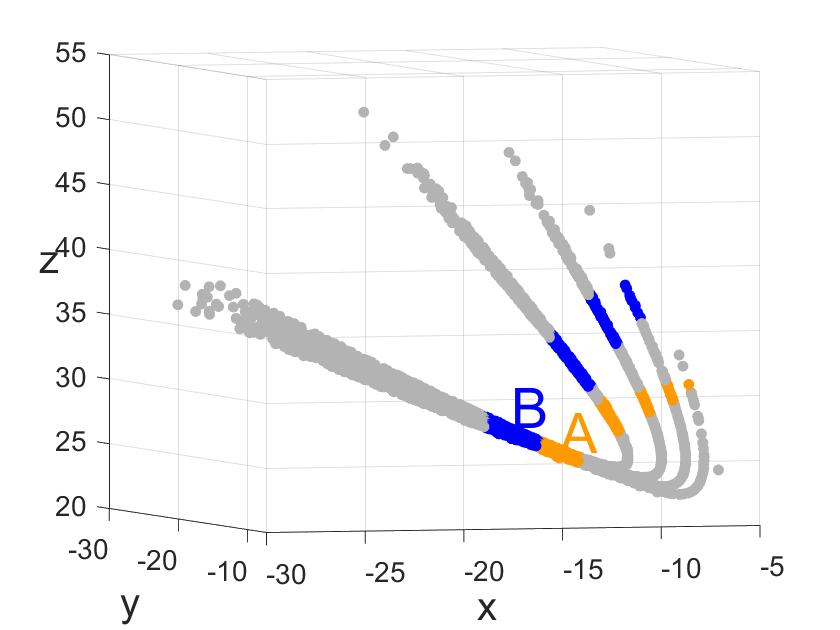}}
			\centerline{(d)}
		\end{minipage}	
		\caption{Symbolic partitioning analysis of $CPS_{B}2\text{MPr}_{A}$. (a) Classification of $CPS_{B}2\text{MPr}_{A}$ into four coarse segments; (b) Refined symbolic partition boundary of the green fragment of $CPS_{B}2\text{MPr}_{A}$ via GKA analysis; (c) One-step evolutionary presentation of four classified coarse segments of $CPS_{B}2\text{MPr}_{A}$; (d) Visualization of final symbolic partition boundaries for $CPS_{B}2\text{MPr}_{A}$.
		} 
		\label{fig-5-35x} 
	\end{figure*}

After confirming the existence of coupling overlap in the Chen system, this study further examines whether folding behaviors similar to those observed in the Lü chaos system emerge. Distinct from the one-dimensional linear segments in previous cases, the effective Poincaré section of the Chen system presents a two-dimensional structure with unequal lengths along the two principal directions. Therefore, normalization is implemented along the first principal direction with the maximum length, and the second principal component is proportionally scaled following the same normalization ratio. Specifically, the principal direction of the left sectional region is normalized to $[0,1]$, while that of the right region is scaled to $[1,2]$. The normalized low-dimensional regions are illustrated in Fig. \ref{fig-5-34}(a).
Koopman analysis is performed on the two-dimensional sectional domain, where the pink line denotes the dominant partition direction for generalized rectangular basis construction without breaking global continuity, and the red dashed line further separates the coupled domains for refined partitioning. Without considering the red dashed boundary, Koopman analysis is conducted with the basis number starting from $M=3$ and increasing incrementally until valid characteristic Lyapunov exponents (CLEs) are acquired. Reliable CLE results are obtained at $M=12$, as displayed in Fig. \ref{fig-5-34}(b). The positive and negative oscillations in the middle domain verify the existence of local folding behavior. This result fully incorporates the coupling interaction between paired domains during the folding process. Accordingly, prior to partitioning the low-dimensional domains of the two CPSs, all symmetric subregions covered by generalized square basis wave packets are merged to eliminate redundant local division.
For comparison, if coupling effects are neglected and the two low-dimensional CPS domains are simply separated by the red dashed line in Fig. \ref{fig-5-34}(a), valid CLEs can be obtained at $M=24$, as shown in Fig. \ref{fig-5-34}(c). Although bidirectional oscillations are observed and confirm the existence of coupled evolution, the localized oscillatory features are far less distinct than those in Fig. \ref{fig-5-34}(b). Based on the localized oscillation positions in Fig. \ref{fig-5-34}(b), reasonable spatial truncation is performed on the critical domain. The truncated orange local region in the subgraph of Fig. \ref{fig-5-34}(d) evolves into the cyan domain after one-step forward iteration. Partial spatial overlap remains between the original orange region and the evolved cyan region, and the topological structure of the cyan domain is still intricate. Further truncation is implemented on the local part of the cyan region close to the normalized coordinate of 1, which moderately reduces the boundary range of the orange domain without destroying global continuity, thereby verifying the rationality of the secondary truncation operation.
After truncation, the optimized low-dimensional section is mapped back to the original high-dimensional domain, as presented in the main graph. The orange pre-image region and its one-step evolved cyan region achieve nearly complete non-overlapping distribution, and the cyan region forms a simple topological structure composed of two coarse segments. The two cyan segments belong to different Poincaré sectional components and are distinguished in Fig. \ref{fig-5-34}(e). The red coarse segment located on $CPS_{B}2$ is defined as $CPS_{B}2\text{M}$, and the brown coarse segment on $CPS_{A}2$ is denoted as $CPS_{A}2\text{M}$.
The corresponding pre-image domains are also divided into two independent parts. One-step forward evolution of the $CPS_{B}2\text{M}$ pre-image generates the green region in the figure, which diverges to both $CPS_{A}2$ and $CPS_{B}2$ components. This phenomenon demonstrates that scroll coupling occurs within this local domain. Similarly, the blue region derived from one-step evolution of the $CPS_{A}2\text{M}$ pre-image also diverges across $CPS_{A}2$ and $CPS_{B}2$, fully retaining the dynamical characteristics of dual-scroll decoupling and coupling evolution.
An isolated black point is observed at the bottom of the $z$-direction on the $CPS_{A}$ side, which is not included in the truncated local pre-image domain in Fig. \ref{fig-5-34}(d). This omission arises from excessive truncation during cyan region optimization, which removes partial marginal features of the orange pre-image domain and manifests as a single discrete point in this case. To compensate for this omission, forward evolution is performed on this isolated point, as illustrated in the subgraph of Fig. \ref{fig-5-34}(e). The original discrete point is marked in cyan, and its one-step evolutionary state is marked in pink, which lies exactly on the extension line of the brown $CPS_{A}2\text{M}$ segment. Since the pink point belongs to the $CPS_{A}2\text{M}$ domain, the isolated pre-image point should be classified as part of the blue $CPS_{A}2\text{M}$ pre-image region.
Nevertheless, the one-step pre-image regions of $CPS_{A}2\text{M}$ and $CPS_{B}2\text{M}$ are inappropriate for symbolic partitioning. These domains exhibit obvious structural bending and local fractal overlap, which hinders quantitative analysis and violates the criterion of selecting Poincaré sections with the simplest topology in this study. To address this issue, secondary backward evolution is performed on the two pre-image regions, yielding the optimized results in Fig. \ref{fig-5-34}(f). All fractal subregions of the green and blue domains evolve into non-overlapping coarse segments that fully satisfy the optimal structural criteria for Poincaré section selection. The cyan point reserved from prior truncation is also updated in the secondary pre-image evolution and remains adjacent to the blue $CPS_{A}2$ pre-image domain.
The two-step pre-image domains of $CPS_{A}2\text{M}$ and $CPS_{B}2\text{M}$ are finally adopted as valid boundary domains for symbolic partitioning. Taking the backward evolutionary process of $CPS_{B}2\text{M}$ as the research object, its two-step pre-image domain is defined as $CPS_{B}2\text{MPr}$, which is further categorized into $CPS_{B}2\text{MPr}_{A}$ (local subdomain on the $CPS_{A}2$ side) and $CPS_{B}2\text{MPr}_{B}$ (local subdomain on the $CPS_{B}2$ side).
Symbol partitioning is first implemented for $CPS_{B}2\text{MPr}_{B}$. Affected by the inherent fractal characteristics of the Poincaré section, this domain can be further decomposed into five independent coarse segments, as shown in Fig. \ref{fig-5-35x}(a). The green segment presents continuous linear morphology, while the orange, cyan, and blue segments contain internal discontinuities and can be regarded as sparsely distributed local segments. For the continuous green segment, the two-step evolutionary states are both standard coarse segments, enabling PCA dimensionality reduction, generalized coordinate construction, and generalized Koopman analysis (GKA).
The subgraph in Fig. \ref{fig-5-35x}(b) presents the generalized coordinates based on the first principal component of the green segment before and after evolution, forming a unimodal chaotic mapping structure. The wide black regions at both ends interfere with curve fitting and are therefore eliminated, retaining only the evolutionary domain with $v<2$. Although the reserved green mapping region still has finite bandwidth, it can be accurately fitted by a cubic polynomial function: $v=0.1729u^{3}+1.0298u^{2}-1.0716u-2.9099$, as shown by the orange fitting curve, which approximates a standard one-dimensional chaotic mapping. Affine transformation is further performed via bidirectional normalization to eliminate coordinate range differences between pre- and post-evolutionary states.
Koopman analysis is conducted on the normalized unimodal mapping using Gaussian basis functions with uniformly distributed grid points. Spectral decomposition is performed with the basis number starting from $M=10$ and increasing by 10 iteratively until valid leading eigenvalues (VLEZs) are obtained. In this case, qualified VLEZs are captured at $M=30$, as illustrated in the main graph. The symbolic partition boundary of the green segment is determined by the zero-crossing position of the VLEZ oscillation, corresponding to the continuous coarse segment at the bottom of the $z$-direction in Fig. \ref{fig-5-35x}(d).
For the remaining sparse segments, valid partition boundaries naturally lie at internal discontinuities after two-step evolution, eliminating the need for refined GKA localization. Each coarse segment supports binary symbolic partitioning and shares a consistent folding direction, as verified by the one-step evolutionary results of all segments in Fig. \ref{fig-5-35x}(c). The bent and closely distributed domains at the bottom of the $z$-direction correspond to the lower regions in Fig. \ref{fig-5-35x}(a) and exhibit consistent dynamical behaviors, which can be integrated into a unified coarse segment. Similar to the Hénon mapping, no new symbolic states are generated during local refinement. To maintain consistency with the folding characteristics of the green segment, the other three coarse segments are assigned identical symbolic rules, and the final unified symbolic partitioning results are displayed in Fig. \ref{fig-5-35x}(d).

\begin{figure*}[htbp]  
		\begin{minipage}{0.48\linewidth}
			\centerline{\includegraphics[width=8cm]{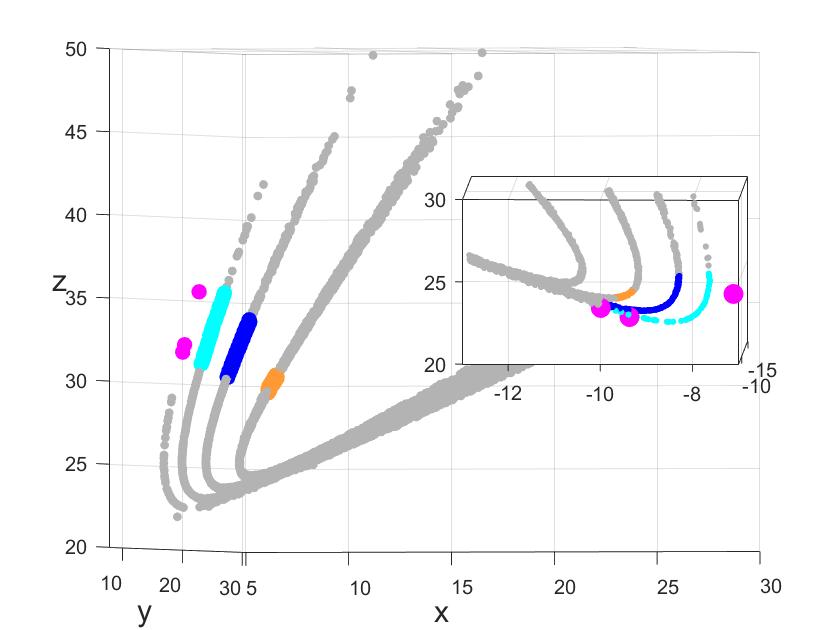}}
			\centerline{(a)}
		\end{minipage}
		\hfill
		\begin{minipage}{0.48\linewidth}
			\centerline{\includegraphics[width=8cm]{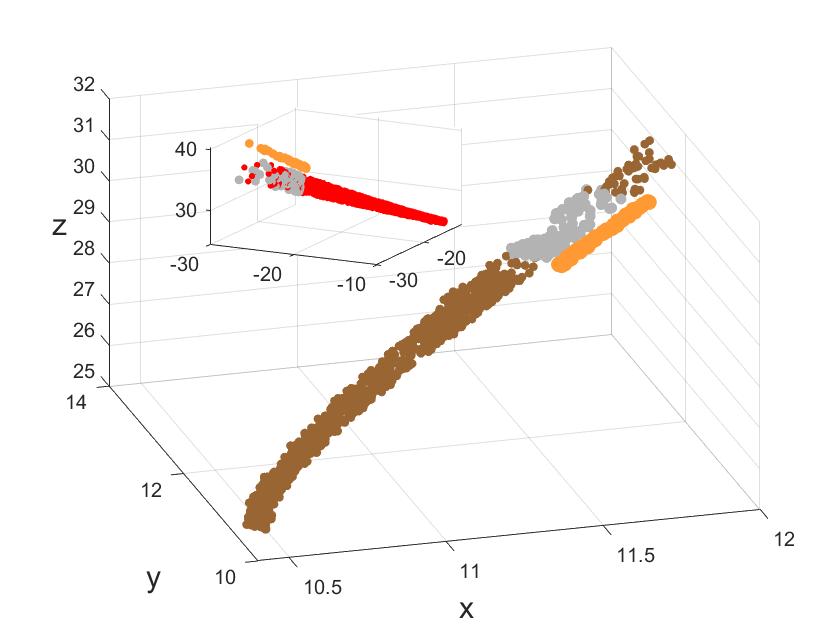}}
			\centerline{(b)}
		\end{minipage}
		\vfill
		\begin{minipage}{0.48\linewidth}
			\centerline{\includegraphics[width=8cm]{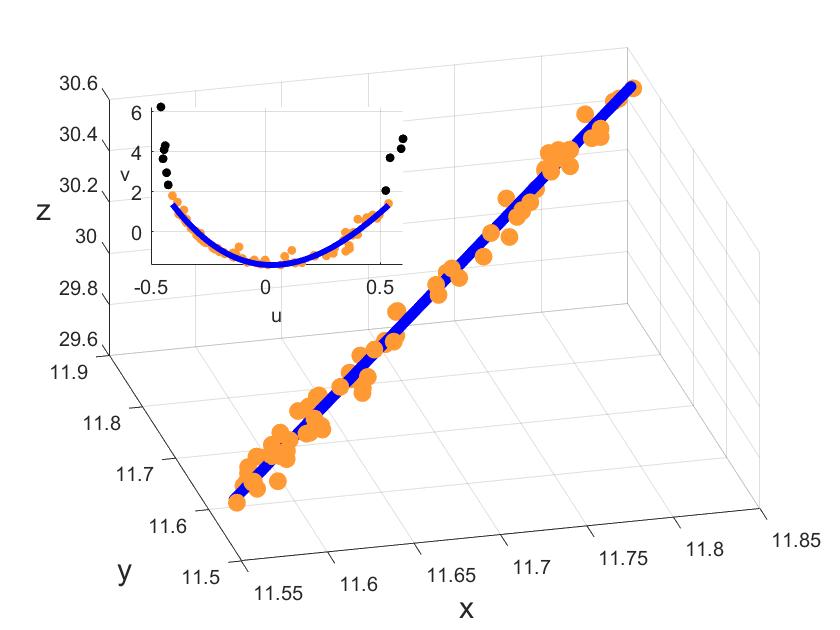}}
			\centerline{(c)}
		\end{minipage}
		\hfill
		\begin{minipage}{0.48\linewidth}
			\centerline{\includegraphics[width=8cm]{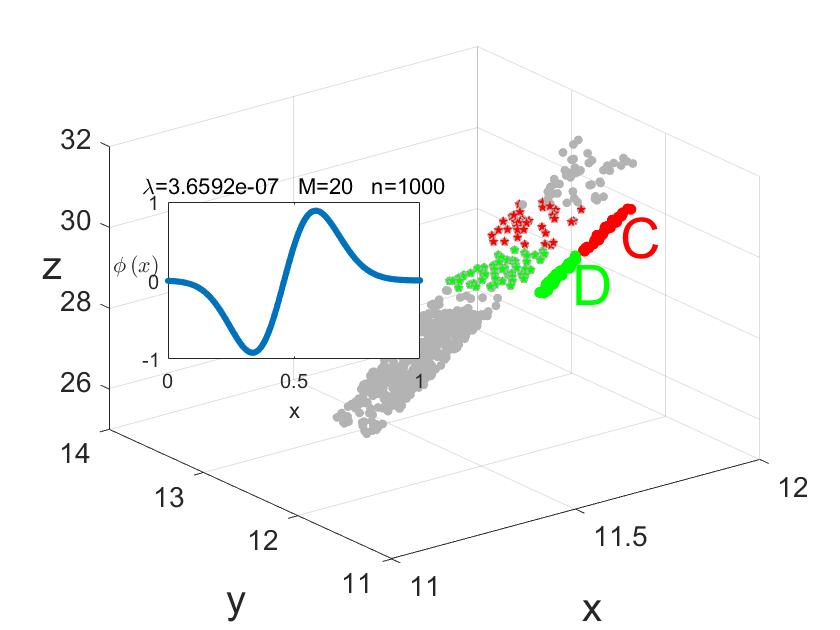}}
			\centerline{(d)}
		\end{minipage}
		\vfill
		\begin{minipage}{0.48\linewidth}
			\centerline{\includegraphics[width=8cm]{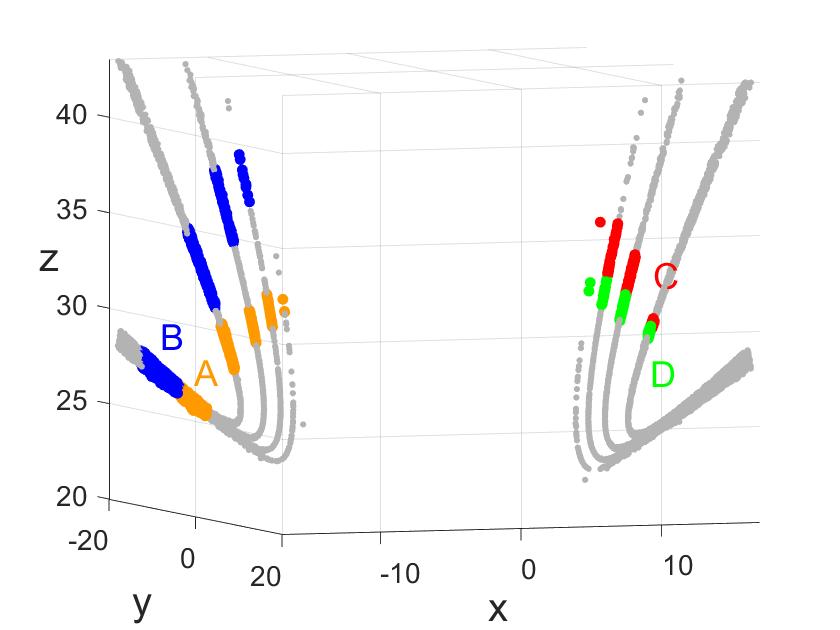}}
			\centerline{(e)}
		\end{minipage}
		\hfill
		\begin{minipage}{0.48\linewidth}
			\centerline{\includegraphics[width=8cm]{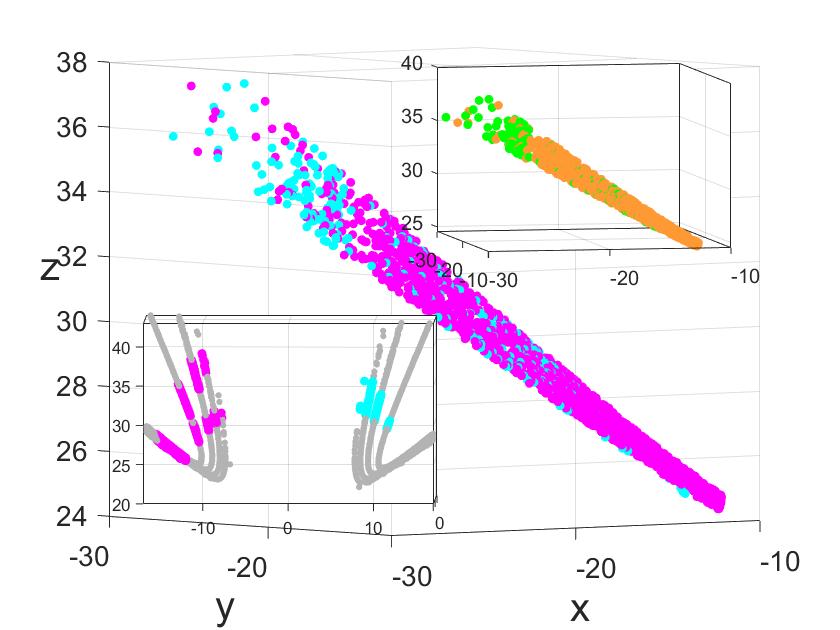}}
			\centerline{(f)}
		\end{minipage}
		\caption{
			Symbolic partitioning analysis of $CPS_{B}2\text{MPr}_{B}$. (a) Classification of $CPS_{B}2\text{MPr}_{B}$ into four coarse segments and presentation of one-step evolutionary regions; (b) OP refinement and reconstruction of the orange segment of $CPS_{B}2\text{MPr}_{B}$ and its two-step evolutionary segment; (c) Generalized coordinate presentation and interpolation fitting of the orange segment after PCA dimensionality reduction; (d) Visualization of symbolic partition regions for $CPS_{B}2\text{MPr}_{A}$ and $CPS_{B}2\text{MPr}_{B}$; (e) Refined symbolic partition boundaries of the orange fragment via GKA analysis; (f) Two-step evolution of symbolic regions for regional overlap verification and symbolic quantity determination on $CPS_{B}2\text{MPr}_{B}$.
		} 
		\label{fig:chen_de} 
	\end{figure*}

	Subsequently, symbolic partitioning is performed for $CPS_{B}2\text{MPr}_{A}$. As illustrated in Fig. \ref{fig:chen_de}(a), this domain can also be decomposed into four coarse segments, which possess distinct structural features compared with $CPS_{B}2\text{MPr}_{B}$. Only the purple segment presents a sparse distribution, while the other three segments exhibit continuous morphological characteristics. Nevertheless, consistent with $CPS_{B}2\text{MPr}_{B}$, mild bending occurs in all segments after one-step evolution except for the indistinct red segment. Numerical tests on the bilateral positions of each bent segment verify that the evolutionary directions remain consistent with those of the original pre-evolutionary segments, including the bilateral endpoints of the magenta and orange regions. Accordingly, if only one unique partition boundary exists for each local region after two-step evolution, these domains can be classified via dual-symbol partitioning rules.
	The blue and cyan segments are analogous to the green coarse segment of $CPS_{B}2\text{MPr}_{B}$. PCA dimensionality reduction is capable of generating standard unimodal mappings for these two segments, and generalized Koopman analysis (GKA) is further implemented to obtain valid leading eigenvalues (VLEZs) and determine accurate symbolic partition boundaries. In contrast, the orange coarse segment suffers from an excessively short spatial length and relatively large bandwidth, which invalidates effective PCA dimensionality reduction. The first principal component contribution rate of the original orange segment is only 82.6\%, and that of its two-step evolutionary segment is 80.4\%. In comparison, the first principal component contribution rates of the remaining two segments exceed 98\% before and after evolution, guaranteeing reliable low-dimensional mapping performance.
	Furthermore, the orange segment contains merely 90 sampling points, which is insufficient for constructing a valid Koopman approximation matrix for spectral feature analysis. Therefore, this study focuses on the targeted symbolic partitioning analysis of the special orange segment.

	To reduce the structural width and optimize the morphological characteristics of the orange segment, structural refinement is implemented via the ordinal pattern (OP) interpolation method consistent with previous cases, followed by re-extraction of extreme point positions. The refined results are illustrated in Fig. \ref{fig:chen_de}(b). In the main graph, the gray fragment represents the original unrefined segment located on $CPS_{A}2\text{M}$, while the red fragment denotes the optimized segment after OP interpolation and fitting, achieving a first principal component contribution rate of 99.7\%. The orange segment in the subgraph corresponds to the two-step evolved structure after interpolation refinement, where the gray unrefined segment on $CPS_{B}2\text{M}$ yields a first principal component contribution rate of 99.8\%.
	After PCA dimensionality reduction, generalized coordinates based on the first principal component of the low-dimensional structure are obtained, as shown in the subgraph of Fig. \ref{fig:chen_de}(c), presenting a sparse unimodal evolutionary mapping. The black sparse regions at both ends of the mapping interfere with curve fitting and are therefore removed to improve fitting accuracy. Only the evolutionary data satisfying $v<2$ are retained, leaving 80 valid discrete points, as highlighted by the orange region. Although the reserved data still exhibit finite bandwidth, they can be accurately fitted by a cubic polynomial function with the expression: $v=-4.9575u^{3}+14.4374u^{2}-0.7430u-1.6517$. The fitted smooth unimodal mapping is represented by the blue curve.
	Different from the sparse discrete independent variables adopted for the green segment in prior cases, uniform interpolation with 1000 equidistant sampling points is performed on the two endpoints of the orange segment to construct continuous independent variables for curve fitting, corresponding to the blue fitting curve in the subgraph. The uniformly interpolated point set on the PCA low-dimensional structure is mapped back to the original high-dimensional domain via Eq. \eqref{eq:A_PCA_return2}. The main graph shows that the sparse orange point set is uniformly distributed around the ideal dense linear segment, realizing effective structural regularization.
	Following the acquisition of the smooth unimodal mapping, data normalization is implemented to complete affine transformation. Generalized Koopman analysis (KA) is subsequently conducted on the normalized unimodal mapping. Gaussian basis functions with equidistant grid points covering the pre-evolutionary domain are adopted to construct the Koopman approximation matrix. Spectral decomposition is performed with the basis number starting from $M=10$ and increasing by 10 iteratively until valid leading eigenvalues (VLEZs) are captured. In this case, qualified VLEZs are obtained at $M=20$, as demonstrated in the subgraph of Fig. \ref{fig:chen_de}(d).
	The symbolic partition boundary determined by the zero-crossing point of VLEZ oscillations corresponds to the blue segmented region in Fig. \ref{fig:chen_de}(c). The remaining 80 sparse discrete points are assigned symbolic states according to their proximity to the labeled points on the fitted blue linear segment, forming the dense segmented region with refined symbolic partitioning in the main graph of Fig. \ref{fig:chen_de}(d). This OP-interpolated refined region is mapped back to the original domain, yielding the final symbolic partitioning results for the original coarse orange segment on $CPS_{A}2\text{M}$, which is visualized on the coarse point set of the gray segmented region in the figure.
	Combining the GKA results of the blue and cyan segments and the folding correspondence characteristics of the sparse magenta segment, the complete symbolic partitioning of $CPS_{B}2\text{MPr}_{A}$ is achieved, as shown by the red and green local regions in Fig. \ref{fig:chen_de}(e). The symbolic states of this domain are defined as C and D, which differ from the symbolic states A and B of $CPS_{B}2\text{MPr}_{B}$. This discrepancy verifies the existence of intrinsic coupling interactions between $CPS_{B}2\text{MPr}_{A}$ and $CPS_{B}2\text{MPr}_{B}$, which induces distinct symbolic partition rules for the two local domains.
	As illustrated in the lower-left subgraph of Fig. \ref{fig:chen_de}(e), $CPS_{B}2\text{MPr}_{B}$ is defined as the coarse symbolic region AB (magenta), and $CPS_{B}2\text{MPr}_{A}$ is defined as the independent coarse symbolic region CD (cyan). Two-step evolution of the two symbolic regions generates the unified $CPS_{B}2\text{M}$ domain in the main graph, where discrete points belonging to different symbolic states overlap mutually. Further refined verification confirms pairwise overlapping and coupling evolution among the four symbolic subregions. For instance, points from symbolic region A and symbolic region D produce coupled overlap after two-step evolution, as displayed in the upper-right subgraph of Fig. \ref{fig:chen_de}(f). Similar coupling and overlapping behaviors are also observed between region A and region C, region B and region C, as well as region B and region D. Consequently, all four symbolic states are mutually independent and uniquely defined.

	\begin{figure*}[htbp]  
		\begin{minipage}{0.48\linewidth}
			\centerline{\includegraphics[width=8cm]{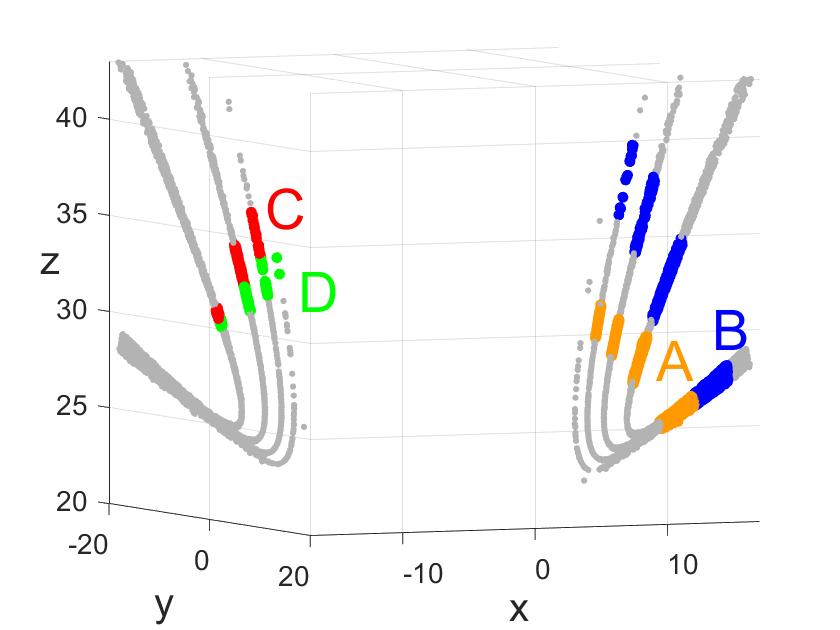}}
			\centerline{(a)}
		\end{minipage}
		\hfill
		\begin{minipage}{0.48\linewidth}
			\centerline{\includegraphics[width=8cm]{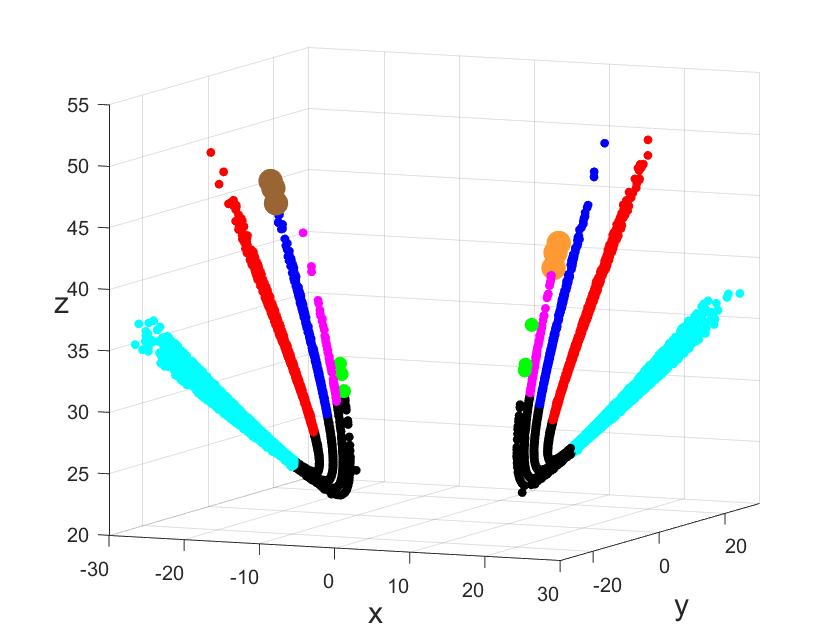}}
			\centerline{(b)}
		\end{minipage}
		\vfill
		\begin{minipage}{0.48\linewidth}
			\centerline{\includegraphics[width=8cm]{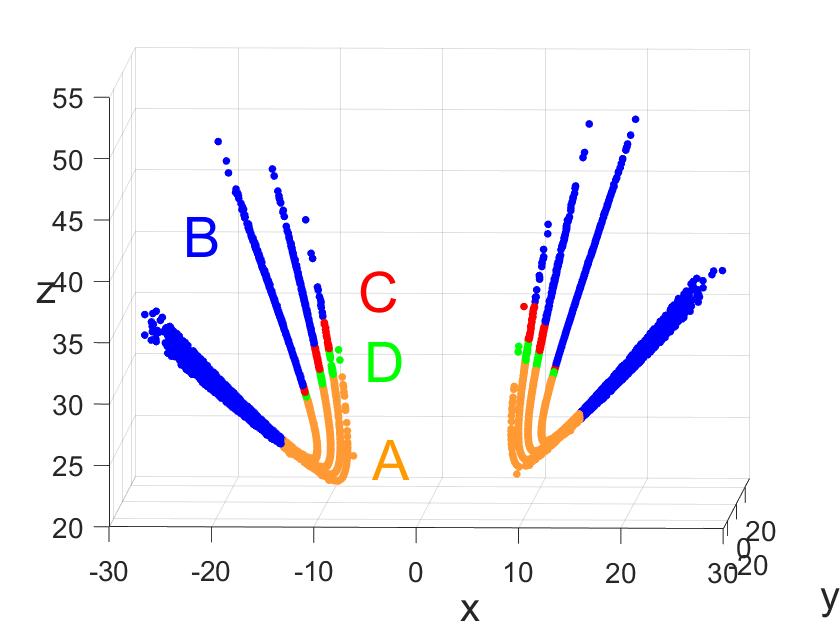}}
			\centerline{(c)}
		\end{minipage}
		\hfill
		\begin{minipage}{0.48\linewidth}
			\centerline{\includegraphics[width=8cm]{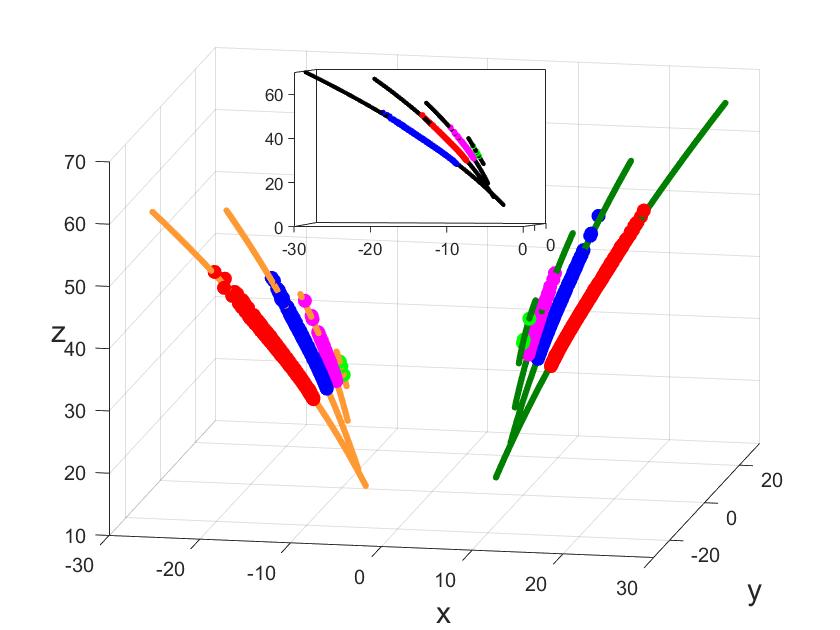}}
			\centerline{(d)}
		\end{minipage}
		\caption{Final symbolic partitioning analysis of Poincaré sections for the Chen chaotic system. (a) Visualization of symbolic partition regions for $CPS_{A}2\text{MPr}$; (b) Evolution analysis of residual Poincaré sectional regions for potential additional symbolic partition domains; (c) Presentation of the final symbolic partitioning results of Poincaré sections for the Chen chaotic system; (d) Fitting curve analysis of local sparse point sets on Poincaré sections for symmetry evaluation and final symbolic quantity determination.
		} 
		\label{fig_chen_eve} 
	\end{figure*}
	
The above procedures elaborate the refined symbolic partitioning of \(CPS_{B}2\text{Pr}\). Similarly, identical analytical steps are implemented for \(CPS_{A}2\text{Pr}\), and the corresponding results are presented in Fig. \ref{fig_chen_eve}(a). The symbolic distribution of \(CPS_{A}2\text{Pr}\) exhibits central symmetry with respect to the partition pattern of \(CPS_{B}2\text{MPr}\) shown in Fig. \ref{fig:chen_de}(e). The one-step evolutionary results of \(CPS_{A}2\text{Pr}\) and \(CPS_{B}2\text{Pr}\) form the black transitional regions in Fig. \ref{fig_chen_eve}(b), while their two-step evolution generates the symmetric cyan coarse segments displayed in the same figure. Since valid symbolic partitioning is performed on the two-step pre-image local domains, further partitioning of the one-step pre-image regions is unnecessary. Following the fundamental principle that all local domains on the Poincaré section must be assigned valid symbolic states, the black transitional regions are reasonably assigned the same symbolic label as the adjacent symbolic A regions.
The cyan symmetric segments are verified to originate from coupled evolutionary processes via backward iteration, whereas the one-step backward evolution of other linear segments strictly maps to the corresponding coarse sectional segments with linear equivalence. Specifically, the green coarse segment on \(CPS_{B}2\) evolves backward in one step and maps to the magenta segment on \(CPS_{A}2\), as marked by the orange coarse points. These orange points on \(CPS_{A}2\) further evolve backward and converge to the upper end of the cyan segment on \(CPS_{B}2\). Meanwhile, the entire magenta segment on \(CPS_{A}2\) backward iterates to the upper local domain of the blue segment.
According to this backward iteration rule, the internal segments undergo a spiral ascending process along the \(z\)-direction: the segments on \(CPS_{B}2\) first map backward to the adjacent outer segments of \(CPS_{A}2\), and further iteration drives the trajectories to the upper domains of the outer segments on \(CPS_{B}2\), until they finally converge to \(CPS_{B}2\text{M}\). Similarly, backward evolution initiated from \(CPS_{B}2\) yields identical spiral iteration behavior, where trajectories alternately evolve outward and upward between the two scrolls and eventually converge to \(CPS_{A}2\text{M}\).
The above backward iteration processes are non-divergent and entirely linear. Accordingly, the red, blue, cyan, and green linear segments do not require refined symbolic partitioning, and no additional pre-image domains need to be extracted for boundary localization. Consequently, symbolic partitioning is exclusively required for the domains illustrated in Fig. \ref{fig_chen_eve}(a) and Fig. \ref{fig:chen_de}(e). All remaining transitional regions, such as the black evolutionary domains, are uniformly assigned the symbolic state of their neighboring symbolic B regions. The final symbolic allocation results are displayed in Fig. \ref{fig_chen_eve}(c), where the entire Poincaré section is completely partitioned into four independent symbolic states. This four-symbol partition theoretically requires identical sectional topology and fully symmetric evolutionary characteristics for the two sectional components.
Nevertheless, the two Poincaré sectional components present slight asymmetric features. As shown in Fig. \ref{fig_chen_eve}(b), the uppermost green point on \(CPS_{A}2\) is significantly higher than its counterpart on \(CPS_{B}2\), and the isolated black point at the bottom of the \(CPS_{A}2\) domain only appears on a single side of the attractor. These asymmetric phenomena are attributed to the sparse distribution of attractor trajectories caused by finite evolutionary steps rather than intrinsic structural asymmetry.
To validate the essential symmetry of the sectional structure, curve fitting is performed on all coarse segments. As illustrated in Fig. \ref{fig_chen_eve}(d), eight coarse segments covering the \(CPS_{A}2\) and \(CPS_{B}2\) domains are fitted separately. All segment morphologies can be quantitatively characterized by correlating the \(x\) and \(y\) coordinates with the \(z\)-component, yielding four quadratic polynomial fitting functions. The specific fitting equations are provided as follows:
	
	\begin{equation}
		\begin{cases}
			\begin{aligned}
				\hat{x} &= a_2 z^2 + a_1 z + a_0 \\
				\hat{y} &= b_2 z^2 + b_1 z + b_0
			\end{aligned}
		\end{cases}
		\label{eq:fitpoly}
	\end{equation}
	
	In the fitting functions, \(\hat{x}\) and \(\hat{y}\) denote the fitted coordinate values of the optimized curves. After acquiring all polynomial coefficients, interpolation is implemented along the \(z\)-component for structural quantitative analysis. Define the total length of the \(z\)-component as \(l(z)\), with \( \mathrm{min}(z) \) and \( \mathrm{max}(z) \) representing the minimum and maximum values of the original \(z\) sequence, respectively. The interpolation domain is extended to \( [\mathrm{min}(z)-0.8l(z),\mathrm{max}(z)+0.8l(z)] \), where 1000 equally spaced sampling points are generated to reconstruct the continuous fitted curves.
	The four fitted curves corresponding to the coarse segments on \(CPS_{B}2\) are marked in orange, while those for \(CPS_{A}2\) are marked in dark green. The two groups of fitted curves exhibit strict axial symmetry about \(x=0\) and \(y=0\) in the \(x\) and \(y\) directions. To further quantitatively verify and compare the symmetric characteristics, coefficient comparison is conducted for all fitted curves, and the detailed results are listed in the following table:
	
	\begin{table*}[t]%
		\centering
		
		\caption{Coefficient of fitting curve}.
		
		\resizebox{\textwidth}{!}{%
			\begin{tabular}{cccccccc}
				\cline{1-8}
				\multirow{2}{*}{Type}&\multirow{2}{*}{Local Subregion}& \multicolumn{6}{c}{Coefficient } \\
				\cline{3-8}
				&&  $a_{2}$ & $a_{1}$ &  $a_{0}$ & $b_{2}$ & $b_{1}$ &  $b_{0}$ \\
				\hline
				\multirow{4}{*}{$CPS_{A}2$}
				& PartA1 &	$0.0022$  & $0.2150$ & $3.298$  & $0.0042$ 
				& $0.1665$  & $3.4417$ \\
				& PartA2	&  	$0.0021$  & $0.1023$ & $5.0013$  & $0.0051$ 
				& $-0.0321$  & $6.7625$ \\
				& PartA3	&  $0.0028$  & $0.0015$ & $6.2913$  & $0.0029$ 
				& $0.0481$  & $5.1488$ \\
				& PartA4	& $0.0116$  & $-0.6362$ & $16.7597$  & $-0.0290$ 
				& $2.1981$  & $-31.8575$ \\
				\hline
				\multirow{4}{*}{$CPS_{B}2$}
				&  ParB1 &	$-0.0023$  & $-0.2101$ & $-3.3696$  & $-0.0032$ 
				& $-0.2384$  & $-2.2280$ \\
				& PartB2	&  	$-0.0020$  & $-0.1059$ & $-4.9205$  & $-0.0019$ 
				& $-0.1854$  & $-3.1317$ \\
				& PartB3	&  $-0.0029$  & $0.0045$ & $-6.3773$  & $-0.0040$ 
				& $0.0025$  & $-6.3432$\\
				& PartB4	& $0.0063$  & $-0.5521$ & $2.9341$  & $-0.0431$ 
				& $2.5642$  & $-46.6730$\\
				\hline
				\multirow{4}{*}{$CPS_{B}2S$}
				&  PartS1 &	$-0.0021$  & $-0.2212$ & $-3.1896$  & $-0.0037$ 
				& $-0.2031$  & $-2.8241$ \\
				& PartS2	&  	$-0.0018$  & $-0.1182$ & $-4.7187$  & $-0.0026$ 
				& $-0.1397$  & $-3.8808$ \\
				& PartS3	&  $-0.0017$  & $0.0706$ & $-5.1353$  & $-0.0047$ 
				& $0.0733$  & $-7.1731$\\
				& PartS4	& $0.0012$  & $-0.2266$ & $-2.2690$  & $-0.0166$ 
				& $0.8659$  & $-19.5492$\\
				
				\hline
				
			\end{tabular}
		}
	\end{table*}
	In the table, \(PartA1\), \(PartA2\), \(PartA3\), and \(PartA4\) correspond to the red, blue, magenta, and green coarse segments on the \(CPS_{A}2\) side, respectively, while \(PartB1\), \(PartB2\), \(PartB3\), and \(PartB4\) represent the red, blue, magenta, and green coarse segments on the \(CPS_{B}2\) side. As indicated by the fitting results, \(PartA1\) and \(PartB1\) possess coefficient values with similar magnitudes but opposite signs, satisfying approximate structural symmetry. For \(PartA2\) and \(PartB2\), the fitting coefficient \(b0\) in the \(y\)-direction shares the same sign, and the magnitude error is slightly larger than that between \(PartA1\) and \(PartB1\) but remains within a reasonable range, thereby also supporting approximate symmetry.
	The two core fitting coefficients \(a0\) and \(b0\) of \(PartA3\) and \(PartB3\) exhibit identical signs with minor deviations in magnitude. In contrast, all coefficients of \(PartA4\) and \(PartB4\) share consistent signs but present considerable magnitude discrepancies, indicating an asymmetric configuration. This asymmetry originates from the insufficient sampling points and localized sparse distribution of the \(PartA4\) and \(PartB4\) segments under finite iteration steps.
	Nevertheless, backward evolutionary analysis demonstrates that one-step backward iteration of \(PartA4\) and \(PartB4\) maps to the upper domains of \(PartB3\) and \(PartA3\), respectively. Accordingly, if \(PartA3\) and \(PartB3\) achieve dense distribution and complete symmetry under sufficient iterations, \(PartA4\) and \(PartB4\) will also evolve into fully symmetric structures. By extension, full symmetry of \(PartA1\) and \(PartB1\) guarantees the overall symmetry of the two Poincaré sectional components, each consisting of four coarse segments.
	Strict axial symmetry with respect to \(x=0\) and \(y=0\) can be realized by inverting the \(x\) and \(y\) components while retaining the original \(z\)-component for point sets on one sectional side. In this study, the point sets of all coarse segments on \(CPS_{A}2\) are processed via component inversion and superimposed on the original coarse segment point sets of \(CPS_{B}2\) to construct a new optimized sectional region denoted \(CPS_{B}2\text{S}\). The four refined coarse curves within \(CPS_{B}2\text{S}\) are defined as \(PartS1\), \(PartS2\), \(PartS3\), and \(PartS4\), as visualized in the subgraph.
	Quadratic polynomial fitting and uniform \(z\)-component interpolation are performed on the new segments, yielding four black fitted curves. The corresponding fitting coefficients are listed in the last four rows of the table. The results show that \(PartS1\), \(PartS2\), and \(PartS3\) achieve high consistency with the original symmetric curves of \(CPS_{B}2\text{S}\), with identical signs and approximate magnitudes for all coefficients. In comparison, \(PartS4\) presents a distinct difference from the original \(PartB4\).
	To evaluate the accuracy of different fitting schemes for dense attractor structures, morphological similarity analysis of adjacent segments is conducted. For both the \(PartA\) and \(PartB\) groups, adjacent segments among the first three parts differ only in the sign of a single coefficient, reflecting highly correlated topological characteristics. In contrast, \(PartA4\) and \(PartA3\) differ in three coefficient signs, and \(PartB4\) and \(PartB3\) also exhibit three sign discrepancies. However, \(PartS4\) only differs from \(PartS3\) in two coefficient signs. According to the similarity principle of adjacent coarse segments in chaotic sectional structures, the fitted curve of \(PartS4\) better approximates the dense structural morphology of the ideal \(PartB4\).
	Similarly, equivalent structural optimization can be implemented by inverting the point sets of \(CPS_{B}2\) coarse segments and superimposing them on the \(CPS_{A}2\) point sets, yielding consistent symmetric refinement results. In summary, the two Poincaré sectional components exhibit apparent asymmetry under finite evolutionary iterations, necessitating an eight-symbol partition scheme for accurate regional classification. Nevertheless, structural fitting and backward evolutionary verification confirm that the sectional structures will become completely dense and theoretically symmetric as the iteration number approaches infinity.
	Therefore, symmetric simplification of the two sectional components is scientifically valid for the Chen chaotic system, enabling reliable four-symbol symbolic partitioning. Furthermore, the proposed symbolic partitioning strategy retains robustness and applicability for chaotic systems with noisy sectional regions.

	\section{Conclusion}
	\label{sec:conclusion}
	
This chapter focuses on the topological partitioning problem of symbolic dynamics for continuous chaotic systems and systematically solves the fundamental challenges of symbolic modeling ranging from discrete mappings to continuous flows and from single-scroll to multi-scroll chaotic attractors. Continuous chaotic systems exhibit inherent local linearity in short-term dynamical evolution, such that folding structures cannot be directly captured via one-step iteration, which restricts traditional discrete mapping methods that rely solely on one-step stretching-folding characteristics. To address this limitation, this chapter introduces the ordinal pattern (OP) theory and establishes a complete methodological framework for Poincaré section construction and symbolic partitioning suitable for general continuous chaotic flows.
First, a simplified attractor model for continuous chaotic systems is established, and the theoretical rationality of constructing evolutionary equations based on ordinal pattern features is demonstrated. According to the topological complexity of chaotic attractors, continuous chaotic systems are classified into two typical categories, namely single-scroll and multi-scroll chaotic systems. Corresponding idealized processing models and systematic symbolic analysis frameworks are separately constructed for each category.
For single-scroll chaotic systems, an ordinal pattern based candidate Poincaré section (CPS) construction strategy is proposed. By introducing an equivalence judgment criterion, redundant topological information among multiple candidate sections is eliminated, and the topological simplification of effective CPSs is realized. On this basis, high-precision return mapping is established on the optimized Poincaré section to achieve reliable symbolic partition of chaotic sequences. In terms of practical verification, the Rössler system is adopted as a typical single-scroll prototype for comprehensive numerical analysis. The overall procedure includes initial CPS generation via OP interpolation, quantitative screening of valid CPSs through generalized coordinate transformation and linear equivalence discrimination combined with dimensionality reduction analysis, and accurate topological boundary localization of symbolic partitions via Koopman analysis (KA). The numerical results verify the high efficiency and precision of the proposed method for single-scroll chaotic configurations.
For multi-scroll chaotic systems, this chapter reveals the essential topological complexity that a single CPS is insufficient to completely characterize system dynamics. Multiple inequivalent CPSs are mutually correlated through inherent coupling evolution, and the final effective Poincaré section is reasonably constituted by multiple coupled CPS components. To elaborate this mechanism, the Lorenz system is investigated as a classical multi-scroll chaotic prototype. Two inequivalent CPSs are constructed to form a composite Poincaré section. A backward decoupling strategy combined with dimensionality reduction is adopted to build a unified generalized coordinate system, based on which global symbolic boundaries covering all local partition domains are determined. Furthermore, by exploiting the symmetric evolutionary characteristics of the Lorenz attractor, the generalized coordinate framework and symbolic quantity are effectively reduced, which significantly eliminates redundant symbolic encoding and improves modeling efficiency.
The Lü and Chen systems are further investigated as complex multi-scroll chaotic systems containing intertwined folding and coupling dynamical mechanisms. Different from simple multi-scroll superposition, their dynamical behaviors integrate intra-scroll stretching-folding evolution and inter-scroll nonlinear coupling interactions. The superposition of dual mechanisms brings greater challenges to symbolic dynamical modeling compared with conventional single-scroll or standard multi-scroll systems. For such composite chaotic systems, the OP-based CPS construction and generalized coordinate establishment framework is inherited and extended. Particularly for the Chen system, the essential influence of sectional symmetry on symbolic simplification is systematically explored. A local curve fitting strategy is introduced to refine sparse attractor structures under finite sampling conditions, realizing physically consistent symmetric simplification of symbolic partitions and improving the numerical robustness of topological boundaries.
Furthermore, the proposed OP-based Poincaré section construction method exhibits excellent generalization capability and numerical robustness. Stable and reliable symbolic analysis can be achieved for all investigated chaotic systems under noisy conditions with certain signal-to-noise ratios. To facilitate method reproducibility and technical promotion, this chapter presents the complete implementation procedure of the proposed framework in the appendix, taking the Rössler system as an example. The full technical pipeline, including OP structural refinement, topological equivalence discrimination, valid CPS screening, and Koopman-based boundary localization, is elaborated in detail, forming a self-consistent and executable algorithm loop.
Finally, to further improve the physical interpretability and numerical accuracy of symbolic partitioning, a generalized Koopman analysis framework is integrated for refined boundary correction. By globally linearizing the nonlinear evolutionary characteristics of chaotic systems via Koopman operator decomposition, the localization precision of symbolic topological boundaries is significantly optimized. The proposed symbolic dynamical method achieves innovations in both theoretical framework and practical implementation, providing a reliable and universal tool for topological classification and information encoding of complex chaotic systems in future research.

	\section*{Acknowledgements}
	This work was supported by the National Natural Science Foundation of
	China under Grants No.12375030.
	
	\appendix 

	\section*{Appendix A  Structural refinement adjustment of Rössler Poincaré sections via monotonic OP interpolation construction}\label{app1}
	
	\setcounter{equation}{0} 
	
	\renewcommand{\theequation}{A.\arabic{equation}}

\begin{figure*}[htbp]  
		\begin{minipage}{0.48\linewidth}
			\centerline{\includegraphics[width=8cm]{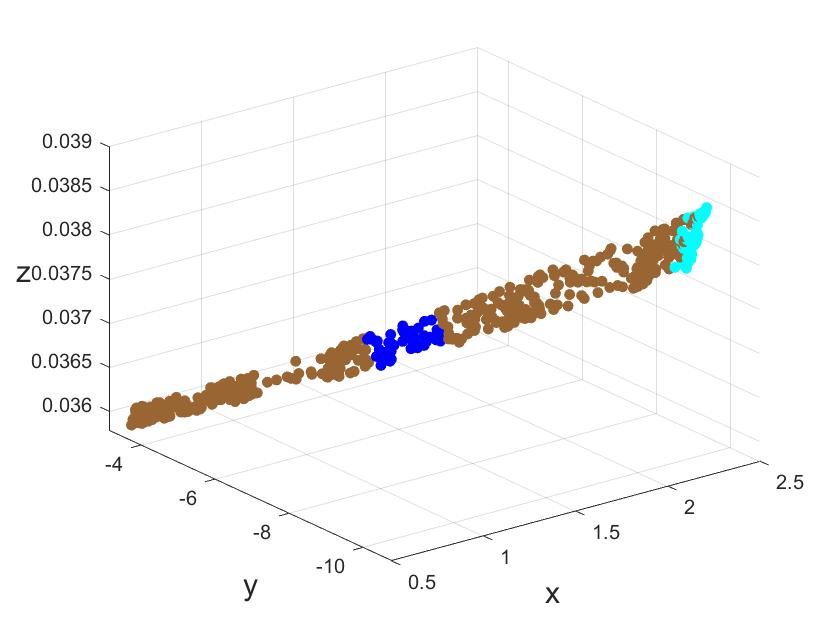}}
			\centerline{(a)}
		\end{minipage}
		\hfill
		\begin{minipage}{0.48\linewidth}
			\centerline{\includegraphics[width=8cm]{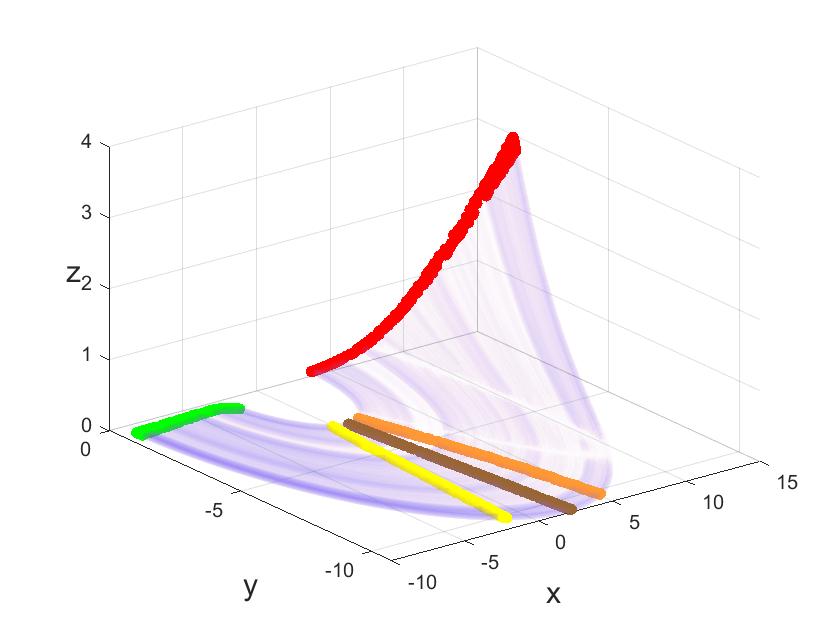}}
			\centerline{(b)}
		\end{minipage}
		\vfill
		\begin{minipage}{0.48\linewidth}
			\centerline{\includegraphics[width=8cm]{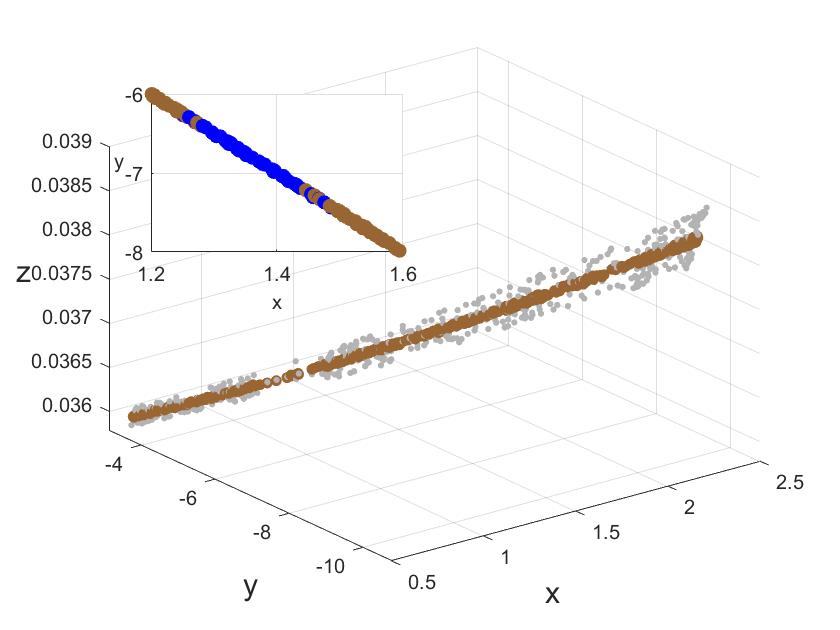}}
			\centerline{(c)}
		\end{minipage}
		\hfill
		\begin{minipage}{0.48\linewidth}
			\centerline{\includegraphics[width=8cm]{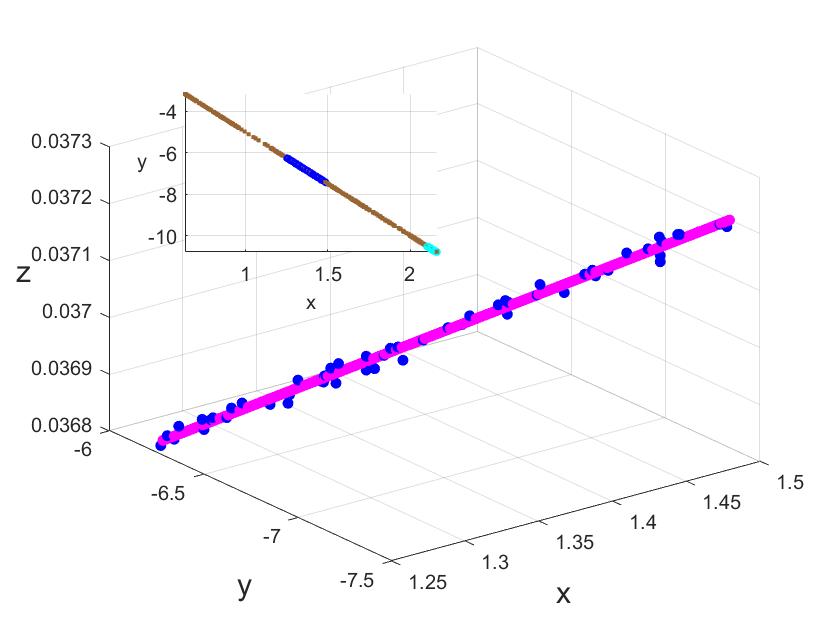}}
			\centerline{(d)}
		\end{minipage}
		\vfill
		\begin{minipage}{0.48\linewidth}
			\centerline{\includegraphics[width=8cm]{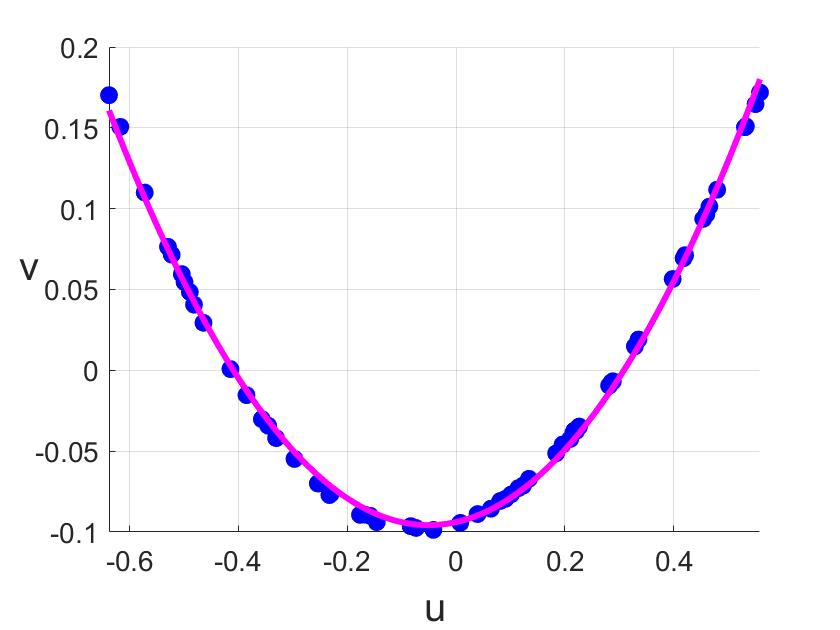}}
			\centerline{(e)}
		\end{minipage}
		\hfill
		\begin{minipage}{0.48\linewidth}
			\centerline{\includegraphics[width=8cm]{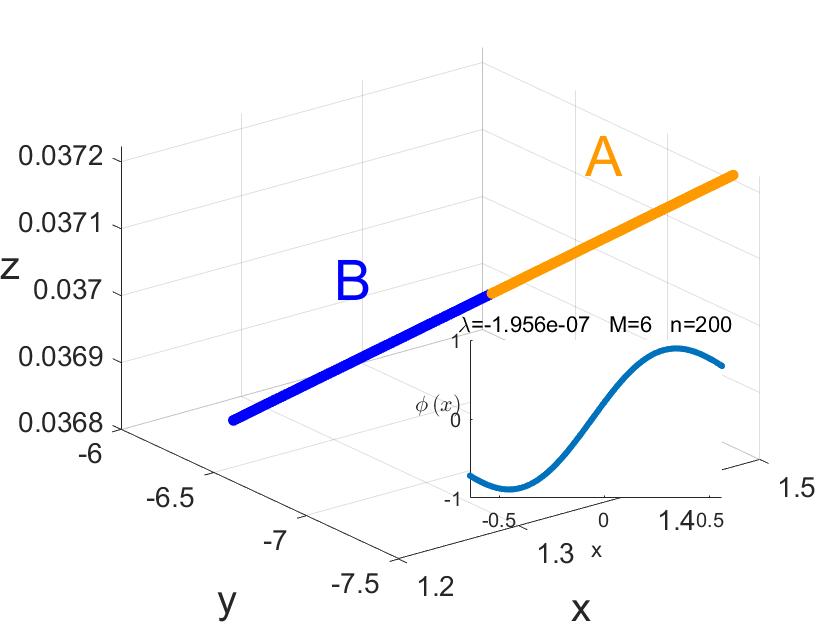}}
			\centerline{(f)}
		\end{minipage}

		\caption{Boundary optimization and refined reconstruction of CPS sections based on monotonic OP interpolation and non-monotonic topological correction}
		\label{fig-5-8} 
	\end{figure*}

	When extracting point sets corresponding to non-monotonic ordinal pattern (OP) regions, excessively large sampling intervals \(\tau\) commonly cause spatial deviation between the extracted representative points and the actual local extreme points of chaotic trajectories. As demonstrated in Fig. \ref{fig-5-1}(b), the four non-monotonic OP points obtained at \(\tau=50\) exhibit obvious positional offsets from the true extreme points of the original time series sampled at \(\tau=1\). To eliminate such systematic deviation, all sampling intervals adopted in this study are set equal to the numerical evolutionary step size, ensuring that the representative points of monotonic OP regions closely approximate the theoretical extreme positions.
	In the Rössler system case, a three-dimensional curved sectional region (i.e., CPS1) can be well captured under a fine sampling interval of \(\tau=0.01\), which serves as the optimal OP section for symbolic partitioning. Nevertheless, local refinement of symbolic boundaries for CPS1 leads to evident local broadening of sectional segments, as illustrated in the subgraph of Fig. \ref{fig-5-7}(d). This morphological distortion originates from the reduction of the principal-direction length during localization, while the spatial span of the point set in other three-dimensional directions remains unchanged, resulting in widened local regions and degraded linear morphological characteristics.
	In principle, reducing the sampling interval and increasing the number of sampling points can alleviate the broadening phenomenon. To further acquire refined, thin linear sectional regions with accurate extreme point positioning, this study proposes a monotonic OP-based regional interpolation strategy. First, the neighboring domains of the original coarse section are screened to verify whether the boundary trajectories form valid monotonic OP structures with the original section. Trajectory interpolation is subsequently implemented for qualified domains to correct non-monotonic OP regions toward the ideal extreme point set. The ideal section consists of tangent points of spiral trajectories, which theoretically form smooth continuous curves on the locally smooth surface of chaotic attractors.
	Structural verification indicates that the successive regional combination of CPS6, CPS1, and CPS2 satisfies linear OP constraints, providing valid domains for structural interpolation. However, direct interpolation generates an excessive number of sampling points, ranging from 140 for the shortest trajectory to 162 for the longest trajectory. To balance refinement accuracy and computational efficiency, an equal-division domain selection method based on trajectory point quantity is proposed.
	Specifically, the point quantities of the trajectory segments from CPS6 to CPS1 and from CPS1 to CPS2 are defined as \(T_1\) and \(T_2\), respectively. Equally divided sampling points \(N_s\) adjacent to the CPS1 domain are determined via uniform segmentation. Taking \(T_1=100\) as an example, the segmentation unit is calculated as \(100/M\). Since each trajectory point in the original time series possesses a unique sequential index \(N_{\text{CPS1}}\), the effective neighboring boundary indices on both sides of CPS1 are determined as \(N_{\text{CPS1}}-N_s\) and \(N_{\text{CPS1}}+N_s\).
	All segmented regions are uniformly divided into \(M\) equal parts, and the segmented positions closest to \(N_{\text{CPS1}}\) are extracted for structural reconstruction. According to the natural evolutionary sequence, the indices of CPS6, CPS1, and CPS2 are defined as \(N_1\), \(N_2\), and \(N_3\), respectively. The updated boundary indices of the reconstructed CPS1 neighboring regions are denoted as \(N_{1\text{new}}\) and \(N_{3\text{new}}\), which satisfy the following equations:
	\begin{equation}
		\begin{cases}
			N_{1new} = N_2 + \dfrac{M-1}{M}(N_3 - N_2), \\[10pt]
			N_{3new} = N_1 + \dfrac{1}{M}(N_2 - N_1),
		\end{cases}
		\quad k = 1,2,\dots,M-1
		\label{eq:a_1}
	\end{equation}
	The newly obtained indices \(N_{1\text{new}}\) and \(N_{3\text{new}}\) can be mapped to specific trajectory points in the evolutionary sequence. As illustrated in Fig. \ref{fig-5-8}(b), the yellow and orange regions correspond to the D1 and D3 domains determined by \(N_{1\text{new}}\) and \(N_{3\text{new}}\) under the equal-division parameter \(M=5\). In addition, the orange region presented in Fig. \ref{fig-5-5}(c) corresponds to the effective domain of \(N_{3\text{new}}\) when \(M=2\).
	
	In this study, the interpolation density is determined based on a unified principle: the interpolation point number of each trajectory is set to more than ten times the minimum number of points contained in the original trajectories. In the Rössler case, the number of discrete points for individual trajectories ranges from 56 to 61. Accordingly, a unified interpolation quantity of 560 points per trajectory is adopted for all trajectories, and spline interpolation is employed for structural refinement.
	Spline interpolation is selected primarily due to its excellent smoothness and numerical stability. In contrast to high-order global polynomial interpolation, spline interpolation effectively avoids spurious oscillations (Runge phenomenon) when processing unevenly distributed or scattered discrete points, which guarantees physically consistent and structurally reliable trajectory reconstruction.
	The original CPS1 consists of local maximum points in the \(y\)-direction. After spline interpolation, the same extreme-point extraction strategy is applied to the refined local domain, yielding the updated brown sectional region in Fig. \ref{fig-5-8}(c), while the original coarse section is plotted in gray. Comparative visualization verifies that the optimized sectional region becomes significantly thinner and more linear. Notably, since the interpolation is implemented only along individual trajectories without increasing the total number of trajectories, the total point quantity of the CPS section remains unchanged at 512. Each refined point can be strictly matched to its corresponding original trajectory point, establishing a one-to-one topological correspondence.
	Subsequently, the local blue region in Fig. \ref{fig-5-8}(a) is mapped to the refined sectional domain according to trajectory correspondence, as illustrated in the subgraph of Fig. \ref{fig-5-8}(c). The central part of the blue region fully covers the original domain, while partial coverage is observed at the boundaries. This incomplete boundary coverage originates from the finite width of the original coarse section, which induces inclined sectional boundaries during low-dimensional projection. Such geometric inclination prevents full boundary matching between the original projection and the refined narrow section.
	Nevertheless, the newly optimized section presents an ideal linear structure. All interior points can be completely enclosed by the two endpoints of the linear segment. As shown in the subgraph of Fig. \ref{fig-5-8}(d), the original 47 discrete points in the blue region form an optimized envelope containing 58 valid points under the linear endpoint constraint. One-step forward evolution is performed on these 58 points, and the evolved trajectories are mapped back to the reconstructed CPS1 domain, generating the cyan evolutionary region displayed in the subgraph.
	Both the blue pre-image region and the cyan evolutionary region exhibit standard linear morphological characteristics. The first principal component contribution rates reach 100\% and 99.9\%, respectively, which fully satisfies the prerequisite for generalized coordinate construction. Principal component projection is implemented for the two linear regions to establish low-dimensional generalized coordinates. The resulting scatter diagram in Fig. \ref{fig-5-8}(e) presents a typical one-dimensional unimodal chaotic mapping relationship, which provides a fundamental basis for subsequent symbolic partitioning.
	Generalized Koopman analysis (GKA) is further adopted to achieve refined local symbolic partitioning. Considering that the limited number of 58 evolutionary points may degrade the stability and accuracy of GKA spectral decomposition, secondary interpolation is performed on the effective projected variable \(Y\) along the \(u\)-direction. A total of 200 equally spaced interpolated points \(Y_{\text{interp}}\) are generated to regularize the mapping structure.
	Based on the evolutionary correspondence of the interpolated blue scatter points in Fig. \ref{fig-5-8}(e), polynomial fitting is implemented to establish the quantitative evolutionary equation. A quadratic polynomial is selected for optimal fitting, and the specific fitted expression is given as follows:
	\begin{equation}
		v_{p} = 0.4734v_{aff}^{2}+0.0750v_{aff}-0.0939
		\label{eq:5_20}
	\end{equation}
	The interpolated post-evolutionary cyan local region \(Yp_{\text{interp}}\) can be derived from the fitted curve. However, coordinate domain verification indicates that \(Y_{\text{interp}}\) and \(Yp_{\text{interp}}\) belong to inconsistent value ranges, which invalidates direct Koopman analysis. Therefore, affine transformation is adopted as a preprocessing procedure prior to spectral decomposition. Bidirectional affine mapping between \(Y_{\text{interp}}\) and \(Yp_{\text{interp}}\) is theoretically feasible, while both forward and backward mappings require linear stretching correction. In this study, unified normalization is performed to map both \(Y_{\text{interp}}\) and \(Yp_{\text{interp}}\) into the standard interval \([0,1]\), eliminating domain mismatch and realizing topological alignment.
	Standard Koopman analysis (KA) is subsequently conducted on the normalized unified domain. Valid leading eigenvalue zones (VLEZs) are successfully captured at a low basis number of \(M=6\), as illustrated in the subgraph of Fig. \ref{fig-5-8}(f). The zero-crossing positions of the VLEZ oscillations are identified as accurate symbolic partition boundaries for the normalized domain. After completing regional segmentation on the normalized coordinates, inverse affine transformation is implemented to back-project the partition boundaries to the original coordinate range of \(Y_{\text{interp}}\).
	Finally, the calibrated original partition boundaries are further mapped back to the newly reconstructed refined CPS1 section, completing the full pipeline of local symbolic boundary refinement. The detailed implementation procedures are elaborated in the following steps.
	
	\begin{equation}
		\boldsymbol{X} = \boldsymbol{Y} \boldsymbol{V}_k^\mathrm{T} + \boldsymbol{\mu}
		\label{eq:5_24}
	\end{equation}
	
	The backward mapping formula for reconstructing the original CPS region \(\boldsymbol{X}\) can be derived from Eqs. \eqref{eq:5_2}, \eqref{eq:5_3}, and \eqref{eq:5_4}. In this formula, \(\boldsymbol{V}_k^\mathrm{T}\) realizes the inverse reconstruction of the principal component \(Y\), and the term \(\mu\) compensates for the centralization offset to restore the original coordinate distribution. In this study, this inverse mapping mechanism is further generalized to the interpolated sectional region. The interpolated domain maintains consistent spatial positioning with the original CPS region, and its projection onto the \(u\)-direction coincides with that of the original domain, as illustrated in Fig. \ref{fig-5-8}(e). Accordingly, the inverse mapping principle is also applicable to Eq. \eqref{eq:5_24}, and the final derived backward reconstruction formula is presented as follows:

	\begin{equation}
		\boldsymbol{X}_{interp} = \boldsymbol{Y}_{interp} \boldsymbol{V}_k^\mathrm{T} + \boldsymbol{\mu}
		\label{eq:A_PCA_return2}
	\end{equation}
As depicted in Fig. \ref{fig-5-8}(b), the magenta region reconstructed via Eq. \eqref{eq:A_PCA_return2} exhibits significantly higher density and morphological regularity compared with the original blue scattered point set, while retaining identical spatial distribution positions. The partitioned \(Y_{\text{interp}}\) coordinates are inversely mapped to the reconstructed CPS1 section to generate the refined sectional region \(\boldsymbol{X}_{\text{interp}}\), as visualized in the main graph of Fig. \ref{fig-5-8}(f). A comprehensive structural improvement of the local Poincaré segment is observed in comparison with the unrefined subgraph in Fig. \ref{fig-5-7}(d).
The complete symbolic partitioning procedure is implemented in a hierarchical manner: symbolic boundaries are first assigned to the refined sparse point domain on the reconstructed CPS1 section, and the partition results are further mapped back to the original CPS1 region via one-to-one trajectory correspondence. Notably, the locally refined symbolic partitioning obtained via GKA analysis is completely consistent with the global symbolic segmentation results derived from full-domain KA analysis, which validates the accuracy, rationality and self-consistency of the proposed local refinement strategy.

%
%
%
%
%
%
%
%
%
%
%
%
%
%
\clearpage
	
	\bibliographystyle{unsrt}
	\bibliography{theXXX}

\end{document}